%% file: main.tex
 \documentclass[twocolumn]{aastex631}

\shorttitle{FEEDBACK: M16 Pillars}
\shortauthors{Karim et al.}

\def\hii{\hbox{{\rm H {\scriptsize II}}}}

\newcommand{\hone}{{\rm H}}
\newcommand{\htwo}{\ifmmode{{\rm H_2}} \else{H$_2$\/}\fi}
\newcommand{\nh}{\ifmmode{N(\hone)} \else{$N(\hone)$\/}\fi}
\newcommand{\nht}{\ifmmode{N(\htwo)} \else{$N(\htwo)$\/}\fi}
\newcommand{\ntot}{\ifmmode{N_\hone} \else{$N_\hone$\/}\fi}

\newcommand{\cii}{\ifmmode{{\rm [C\ II]}} \else{[C\ II]\/}\fi}
\newcommand{\oi}{\ifmmode{{\rm [O\ I]}} \else{[O\ I]\/}\fi}

\newcommand{\twcii}{\ifmmode{{\rm [^{12}C\ II]}} \else{[$^{12}$C\ II]\/}\fi}
\newcommand{\thcii}{\ifmmode{{\rm [^{13}C\ II]}} \else{[$^{13}$C\ II]\/}\fi}

\newcommand{\cp}{\ifmmode{{\rm C^{+}}} \else{C$^{+}$\/}\fi}
\newcommand{\twcp}{\ifmmode{{\rm ^{12}C^{+}}} \else{$^{12}$C$^{+}$\/}\fi}
\newcommand{\thcp}{\ifmmode{{\rm ^{13}C^{+}}} \else{$^{13}$C$^{+}$\/}\fi}

\newcommand{\co}{\ifmmode{{\rm CO(J=1-0)}} \else{CO(J=1-0)\/}\fi}

\newcommand{\twco}{\ifmmode{^{12}{\rm CO(J=1-0)}} \else{$^{12}$CO(J=1-0)\/}\fi}
\newcommand{\thco}{\ifmmode{^{13}{\rm CO(J=1-0)}} \else{$^{13}$CO(J=1-0)\/}\fi}
\newcommand{\ceighteeno}{\ifmmode{{\rm C^{18}O(J=1-0)}} \else{C$^{18}$O(J=1-0)\/}\fi}
\newcommand{\twcoA}{\ifmmode{^{12}{\rm CO}} \else{$^{12}$CO\/}\fi}
\newcommand{\thcoA}{\ifmmode{^{13}{\rm CO}} \else{$^{13}$CO\/}\fi}
\newcommand{\ceighteenoA}{\ifmmode{{\rm C^{18}O}} \else{C$^{18}$O\/}\fi}
\newcommand{\twcott}{\ifmmode{^{12}{\rm CO(J=3-2)}} \else{$^{12}$CO(J=3-2)\/}\fi}
\newcommand{\thcott}{\ifmmode{^{13}{\rm CO(J=3-2)}} \else{$^{13}$CO(J=3-2)\/}\fi}
\newcommand{\twcosf}{\ifmmode{^{12}{\rm CO(J=6-5)}} \else{$^{12}$CO(J=6-5)\/}\fi}

\newcommand{\sio}{\ifmmode{{\rm SiO(J=2-1)}} \else{SiO(J=2-1)\/}\fi}
\newcommand{\sioA}{\ifmmode{{\rm SiO}} \else{SiO\/}\fi}
\newcommand{\hcn}{\ifmmode{{\rm HCN(J=1-0)}} \else{HCN(J=1-0)\/}\fi}
\newcommand{\hcnA}{\ifmmode{{\rm HCN}} \else{HCN}\fi}
\newcommand{\hcop}{\ifmmode{{\rm HCO^{+}(J=1-0)}} \else{HCO$^{+}$(J=1-0)\/}\fi}
\newcommand{\hcopA}{\ifmmode{{\rm HCO^{+}}} \else{HCO$^{+}$\/}\fi}
\newcommand{\ntwohp}{\ifmmode{{\rm N_{2}H^{+}(J=1-0)}} \else{N$_{2}$H$^{+}$(J=1-0)\/}\fi}
\newcommand{\ntwohpA}{\ifmmode{{\rm N_{2}H^{+}}} \else{N$_{2}$H$^{+}$\/}\fi}
\newcommand{\cs}{\ifmmode{{\rm CS(J=2-1)}} \else{CS(J=2-1)\/}\fi}
\newcommand{\csA}{\ifmmode{{\rm CS}} \else{CS\/}\fi}
\newcommand{\jmton}[2]{\ifmmode{{\rm J=#1-#2}} \else{J=#1-#2\/}\fi}
\newcommand{\chisqdof}{\ifmmode{\chi^2 / {\rm dof}} \else{$\chi^2 / {\rm dof}$\/}\fi}

\newcommand{\vlsr}{\ifmmode{\rm V_{LSR}} \else{V$_{\rm LSR}$\/}\fi}
\newcommand{\nexpo}[2]{\ifmmode{#1 \times 10^{#2}}\else{$#1 \times 10^{#2}$}\fi}
\newcommand{\mjybmkms}{\ifmmode{{\rm mJy~beam^-1~\kms}} \else{mJy~beam$^{-1}$~\kms}\fi}
\newcommand{\mjybm}{\ifmmode{{\rm mJy~beam^{-1}}} \else{mJy~beam$^{-1}$}\fi}
\newcommand{\jybmkms}{\ifmmode{{\rm Jy~beam^{-1}~\kms}} \else{Jy~beam$^{-1}$~\kms}\fi}

\newcommand{\Tex}{\ifmmode{{T_{\rm ex}}} \else{$T_{\rm ex}$\/}\fi}
\newcommand{\losD}{\ifmmode{1740~{\rm pc}} \else{1740~pc}\fi}

\newcommand{\expo}[1]{\ifmmode{10^{#1}}\else{$10^{#1}$}\fi}

\newcommand\pdeg           {$.\kern-.25em ^{^\circ}$}
\newcommand\hh{\mbox{$^{\mathrm h}$}}
\newcommand\mm{\mbox{$^{\mathrm m}$}}

\newcommand\degree {\arcdeg}
\newcommand\dV{\ifmmode{\,{\Delta V}}\else{{$\Delta V$}}\fi}
\newcommand\kms{\ifmmode{\,{\rm km~s^{-1}}}\else{{${\rm km~s^{-1}}$}}\fi}

\newcommand{\shelf}{Ridge}
\newcommand{\wrt}{with respect to}

\graphicspath{{./}}

\begin{document}

\title{SOFIA FEEDBACK Survey: The Pillars of Creation in \cii\ and Molecular Lines}

\author[0000-0001-8844-5618]{Ramsey L. Karim}
\affiliation{University of Maryland, Department of Astronomy, Room 1113 PSC Bldg. 415, College Park, MD 20742-2421, USA}

\author[0000-0002-7269-342X]{Marc W. Pound}
\affiliation{University of Maryland, Department of Astronomy, Room 1113 PSC Bldg. 415, College Park, MD 20742-2421, USA}

\author[0000-0003-0306-0028]{Alexander G.G.M. Tielens}
\affiliation{University of Maryland, Department of Astronomy, Room 1113 PSC Bldg. 415, College Park, MD 20742-2421, USA}

\author[0000-0003-4260-2950]{Maitraiyee Tiwari}
\affiliation{University of Maryland, Department of Astronomy, Room 1113 PSC Bldg. 415, College Park, MD 20742-2421, USA}
\affiliation{Max-Planck Institute for Radioastronomy, Auf dem Hügel, D-53121 Bonn, Germany}

\author[0000-0002-0915-4853]{Lars Bonne}
\affiliation{SOFIA Science Center, USRA, NASA Ames Research Center, M.S. N232-12, Moffett Field, CA 94035, USA}

\author[0000-0003-0030-9510]{Mark G. Wolfire}
\affiliation{University of Maryland, Department of Astronomy, Room 1113 PSC Bldg. 415, College Park, MD 20742-2421, USA}

\author[0000-0003-3485-6678]{Nicola Schneider}
\affiliation{I. Physikalisches Institut, Universität zu Köln, Zülpicher Str. 77, D-50937 Köln, Germany}

\author[0000-0002-7640-4998]{Ümit Kavak}
\affiliation{SOFIA Science Center, USRA, NASA Ames Research Center, M.S. N232-12, Moffett Field, CA 94035, USA}

\author[0000-0002-8876-0690]{Lee G. Mundy}
\affiliation{University of Maryland, Department of Astronomy, Room 1113 PSC Bldg. 415, College Park, MD 20742-2421, USA}

\author[0000-0003-2555-4408]{Robert Simon}
\affiliation{I. Physikalisches Institut, Universität zu Köln, Zülpicher Str. 77, D-50937 Köln, Germany}

\author[0000-0002-1708-9289]{Rolf Güsten}
\affiliation{Max-Planck Institute for Radioastronomy, Auf dem Hügel, D-53121 Bonn, Germany}

\author[0000-0001-7658-4397]{Jürgen Stutzki}
\affiliation{I. Physikalisches Institut, Universität zu Köln, Zülpicher Str. 77, D-50937 Köln, Germany}

\author[0000-0003-4516-3981]{Friedrich Wyrowski}
\affiliation{Max-Planck Institute for Radioastronomy, Auf dem Hügel, D-53121 Bonn, Germany}

\author{Netty Honingh}
\affiliation{I. Physikalisches Institut, Universität zu Köln, Zülpicher Str. 77, D-50937 Köln, Germany}

\begin{abstract}
    We investigate the physical structure and conditions of photodissociation regions (PDRs) and molecular gas within the Pillars of Creation in the Eagle Nebula using SOFIA FEEDBACK observations of the \cii\ 158~$\mu$m line.
    These observations are velocity resolved to 0.5~\kms\ and are analyzed alongside a collection of complimentary data with similar spatial and spectral resolution: the \oi\ 63~$\mu$m line, also observed with SOFIA, and rotational lines of CO, \hcnA, \hcopA, \csA, and \ntwohpA.
    Using the superb spectral resolution of SOFIA, APEX, CARMA, and BIMA, we reveal the relationships between the warm PDR and cool molecular gas layers in context of the Pillars' kinematic structure.

    We assemble a geometric picture of the Pillars and their surroundings informed by illumination patterns and kinematic relationships and derive physical conditions in the PDRs associated with the Pillars. 
    We estimate an average molecular gas density $n_{\htwo} \sim \nexpo{1.3}{5}$~cm$^{-3}$ and an average atomic gas density $n_{\hone} \sim \nexpo{1.8}{4}$~cm$^{-3}$ and infer that the ionized, atomic, and molecular phases are in pressure equilibrium if the atomic gas is magnetically supported.
    We find pillar masses of 103, 78, 103, and 18~$M_{\odot}$ for P1a, P1b, P2, and P3 respectively, and evaporation times of $\sim$1--2~Myr.
    The dense clumps at the tops of the pillars are currently supported by the magnetic field.
    Our analysis suggests that ambipolar diffusion is rapid and these clumps are likely to collapse within their photoevaporation timescales.
\end{abstract}


\section{Introduction} \label{sec:intro}
\input{introduction}

\section{Observations}\label{sec:observations}
 \input{observations}

\section{Results} \label{sec:results}
\input{results_features}

\section{Photodissociation Regions} \label{sec:pdr}
\input{analysis_pdrs}

\section{Mass and Physical Conditions} \label{sec:physconds}
\input{analysis_conditions}

\input{analysis_pressure.tex}

\section{Discussion} \label{sec:discussion}
\input{discussion}

\section{Conclusion} \label{sec:conclusion}
\input{conclusion}

\begin{acknowledgements}
We thank the anonymous referee for detailed and constructive comments which improved the quality of this paper.
We thank E. Tarantino for many illuminating discussions on the structure and properties of \hii\ regions and A. Dittmann for the conversation about rotational support.

This work is based on observations made with the NASA/DLR Stratospheric Observatory for Infrared Astronomy (SOFIA).
SOFIA is jointly operated by the Universities Space Research Association, Inc. (USRA), under NASA contract NAS2-97001, and the Deutsches SOFIA Institut (DSI) under DLR contract 50 OK 0901 to the University of Stuttgart.
Financial support for the SOFIA Legacy Program, FEEDBACK, at the University of Maryland was provided by NASA through award SOF070077 issued by USRA.
The FEEDBACK project is supported by the BMWI via DLR, project Nos. 50 OR 2217 (FEEDBACK-plus). Support for CARMA observations was provided by NSF AST-2932160.

This work is based in part on observations made with the NASA/ESA/CSA James Webb Space Telescope.
The data were obtained from the Mikulski Archive for Space Telescopes at the Space Telescope Science Institute, which is operated by the Association of Universities for Research in Astronomy, Inc., under NASA contract NAS 5-03127 for JWST. These observations are associated with program \#2739.
The authors acknowledge the team led by Klaus Pontoppidan for developing their observing program with a zero-exclusive-access period.
This research made use of Regions, an Astropy package for region handling \citep{Bradley2022zndo...7259631B}.
This research has made use of the VizieR catalogue access tool, CDS, Strasbourg, France.

The spectral line cubes for CO(\jmton{1}{0}), CO(\jmton{6}{5}), \hcnA, \hcopA, \csA, \ntwohpA, and \oi\ are available from the Digital Repository at the University of Maryland: \url{http://hdl.handle.net/1903/30441}.
The SOFIA \cii\ observations are made public through the NASA/IPAC Infrared Science Archive (IRSA).
\end{acknowledgements}

\vspace{5mm}
\facilities{SOFIA(upGREAT), APEX(LAsMA, SEPIA660), CARMA, BIMA, JWST(NIRCam), Herschel(PACS, SPIRE), Spitzer(IRAC)}

\software{astropy \citep{astropy2013, astropy2018, astropy2022},
    Spectral Cube (\url{https://doi.org/10.5281/zenodo.3558614}),
    regions (\url{https://doi.org/10.5281/zenodo.7259631}),
    pvextractor (\url{https://github.com/radio-astro-tools/pvextractor}),
    numpy \citep{numpy_harris2020array},
    scipy \citep{2020SciPy-NMeth},
    pandas \citep{pandas_McKinney_2010, pandas_McKinney_2011},
    matplotlib \citep{matplotlib_Hunter:2007},
    pdrtpy \citep{Pound2023AJ....165...25P}}

\appendix

\section{Measured Main Beam Temperatures and Spectra} \label{sec:measurements}
\input{measurements}

\section{Systematic Velocity and [CII] Background} \label{sec:systematic-cii-background}

\input{systematic_background}

\section{Kinematics and Geometry of P1a} \label{sec:geomdyn}
\input{kinematic_modeling}

\bibliography{main}{}
\bibliographystyle{aasjournal}

\end{document}

%% file: introduction.tex
The Eagle Nebula, also known as M16, is a well-studied \hii\ region lying above the Galactic midplane at $(l,~b) = (16\arcdeg9540,~+0\arcdeg7934)$, $(\alpha,\,\delta)_{J2000} = (18\hh18\mm48\fs0,~$-$13\arcdeg48\arcmin24\arcsec)$.
The \hii\ region is illuminated by a $\sim$2~Myr old stellar cluster, NGC~6611 \citep{1993AJ....106.1906H}, born from the giant molecular cloud (GMC) W~37 \citep{2016RAA....16...56Z}.
Filaments primarily extended in the Galactic $b$ direction, parallel to the larger scale elongation of W~37, lie within W~37 near the location of M16 \citep{2012A&A...542A.114H, 2019A&A...627A..27X}.

The \hii\ region associated with M~16 is powered by the early-type stars in the NGC~6611 cluster, a $\sim2 \times 10^4~M_\odot$ cluster whose most massive member is estimated to be $\sim80~M_{\odot}$ \citep{1993AJ....106.1906H, 2012A&A...542A.114H}.
The cluster, at a heliocentric distance of $\sim$2~kpc, comprises a number of pre-main sequence (PMS) stars as well as stars surrounded by circumstellar material, both signs of recent or ongoing star formation.
NGC~6611 members exhibit some age spread, as evidenced by several hundred PMS stars whose ages range from 0.25--1~Myr and a massive main sequence population about 2~Myr old lying alongside a handful of evolved massive stars, at least one of which is B2.5~I, which is around 6~Myr old \citep{1993AJ....106.1906H}.
\citet{2007A&A...462..245G, 2009A&A...496..453G, 2010A&A...521A..61G} and \citet{2013MNRAS.435.3058D} delve into the star formation history of the region and characterize two different populations of stars: a group of $>$8~Myr old stars which likely existed before the formation of the M16 \hii\ region, and a group of $\sim$1~Myr-old stars, broadly consistent with the $\sim$2~Myr age determination of NGC~6611 by other authors, whose formation event may have been triggered by a supernova shell's arrival $\sim$3~Myr ago.
We adopt the distance of $1740 \pm 130$~pc determined using GAIA parallax observations \citep{2019ApJ...870...32K} and a main sequence age of 2~Myr based on studies from the last two decades \citep{1999A&AS..134..525B, 2000A&A...358..886B, 2006A&A...457..265D, 2008A&A...489..459M, Stoop2023A&A...670A.108S}.

\begin{figure*}
    \centering
    \includegraphics[width=\textwidth]{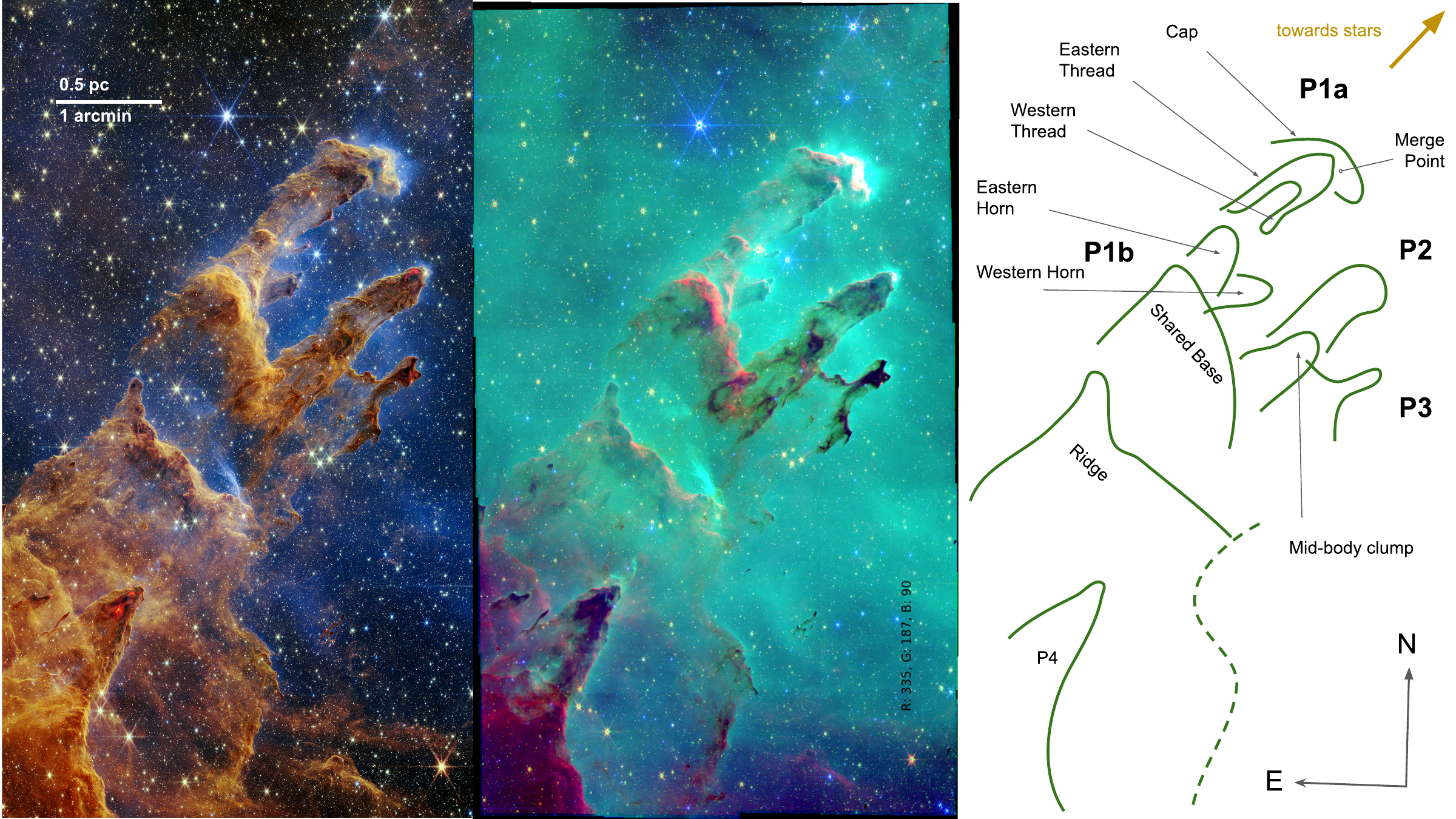}
    \caption{(\textit{Left}) JWST color composite prepared by Joseph DePasquale (STScI), Anton M. Koekemoer (STScI), Alyssa Pagan (STScI). Colors are Purple: F090W, Blue: F187N, Cyan: F200W, Yellow: F335M, Orange: F444W, Red: F470N. The image was obtained from \url{https://webbtelescope.org/contents/media/images/2022/052/01GF423GBQSK6ANC89NTFJW8VM}.
    (\textit{Center}) Three-color composite using JWST filters F090W (blue), F187N (green), and F335M (red). The stretches are nonlinear and limits have been adjusted. The 3.3~\micron\ band, in red, includes a PAH feature and therefore indicates far-ultraviolet illumination. Section~\ref{sec:obs-ancillary-data} includes a more detailed summary of the significance of these filters.
    (\textit{Right}) Schematic diagram of the Pillars on the sky at the same angular scale as the two images to the left. Features are marked with labels which will be used throughout the paper. P1a refers to the northern half of Pillar 1, including the Cap and Eastern and Western Threads. P1b refers to the southern half and includes the Eastern and Western Horns and part of the Shared Base. The dashed line to the south marks a boundary which is kinematically discontinuous with the \shelf\ despite its apparent continuity in the visible and IR images; this is discussed in Section~\ref{sec:shelf}.
    The three images are nearly RA-Dec aligned; ``up'' in the image is 2.3$^\circ$ east of north. The scale bar in the top-left corner shows $1\arcmin \approx0.5$~pc at 1740~pc line of sight.}
    \label{fig:pillars_jwst}
\end{figure*}

Within M16 lies an iconic pillar system whose Hubble Space Telescope (HST), and now JWST, images are well known to the public \citep{1996AJ....111.2349H}.
The three main Pillars, extending towards a handful of bright O5--7 stars $\sim3$~pc away in projection \citep{1993AJ....106.1906H}, are seen roughly parallel to the plane of the sky as in Figure~\ref{fig:pillars_jwst}.
Spectroscopic studies by \citet{1998ApJ...493L.113P} and \citet{2015MNRAS.450.1057M} have concluded that all the Pillars are inclined slightly towards or away from the observer, and the tallest pillar, called Pillar 1 (P1), is actually a superposition of two pillars: the northern half, called P1a, is actually behind the cluster and the southern half, called P1b, is in between the cluster and the observer (see Figure~\ref{fig:pillars_jwst}).
\citet{1998ApJ...493L.113P} detects coherent molecular gas flows along the line of sight, which they conclude are projected motions along the pillar.
They find that the observed radial velocity gradient along the pillar implies that the dynamical timescale for dissipation, $\sim10^5$~yr, may be shorter than their estimated photoevaporation timescale $\sim10^7$~yr.
Both \citet{1998ApJ...493L.113P} and \citet{2015MNRAS.450.1057M} suggest that Pillars 2 and 3 (P2 and P3) are between the cluster and the observer and that P2 points away from the observer towards the cluster (throughout this paper, ``points'' refers to the orientation of the head; i.e., the head of P2 is farther from the observer than the tail).
The line-of-sight (LOS) velocity data collected by \citet{1998ApJ...493L.113P} and \cite{2015MNRAS.450.1057M} indicate that P3 points towards the observer, and the illumination data from the HST observations presented by \citet{1996AJ....111.2349H} and the MUSE observations analyzed by \citet{2015MNRAS.450.1057M} are consistent with P3 lying between the cluster and the observer.
Together, the velocity and illumination data imply that P3 does not point towards the cluster.
P3 might point towards a cluster member displaced towards the observer along the line of sight (Section~\ref{sec:star_g0}).

Due to the high spatial resolution of the JWST images, structures within the Pillar system are well-defined.
We use the JWST images throughout this study to measure the on-sky angular widths of features.


\subsection{The Photodissociation Region}
The photodissociation region (PDR) is the region of far-ultraviolet (FUV; 6--13.6~eV) illuminated neutral atomic and molecular gas just behind the ionization front (see \citealt{1985ApJ...291..722T} and \citealt{Wolfire2022ARA&A..60..247W} for detailed background).
Due primarily to photoelectric heating from small grains and polycyclic aromatic hydrocarbons (PAHs) \citep{bakes1994photoelectric}, the atomic region at the cloud surface is warmer ($T\sim100$~K) than the molecular gas ($T\sim10$~K) found deeper in the cloud.
Most of the cooling in the atomic PDR is done through a small handful of fine-structure lines, principally the 158~\micron\ \cii\ line (singly-ionized carbon), and at $n\gtrsim 10^4$~cm$^{-3}$, the 63~\micron\ and 145~\micron\ \oi\ lines (neutral oxygen) \citep{1985ApJ...291..722T, HT1999_revmodphys, Wolfire2022ARA&A..60..247W}.
Spectroscopically resolved observations of these lines, such as those we present here, allow us to probe the kinematics of the warmer, outer atomic region of the PDR.
Observations of \cii\ and \oi\ are presented by \cite{Schneider2012A&A...542L..18S, Schneider2021A&A...653A.108S} towards pillars and globules in Cygnus~X.

At $A_V \gtrsim 5$, carbon is found primarily in the molecular phase as carbon monoxide (CO).
From $A_V \sim 5$--10, FUV radiation still warms the predominantly molecular gas, which we can probe with low- to mid-J CO lines.
These lines, and particularly the bright \twco\ line, should be optically thick and trace this molecular PDR layer as opposed to the colder, less illuminated molecular gas $A_V \gtrsim 10$.

Molecular lines more sensitive to dense gas ($n \gtrsim 10^5$~cm$^{-3}$) will probe the denser, colder layers beyond the PDR which are not heated by FUV radiation.
By comparing observations of dense gas tracers, such as the \hcopA, \hcnA, and \csA\ observed towards the Pillars by \cite{1999A&A...342..233W}, with the PDR tracers introduced above, we can explore not only the molecular gas inventory of the Pillars but also how mass moves between phases and leaves the Pillars through bulk flows \citep{1998ApJ...493L.113P} or photoevaporation \citep{1996AJ....111.2349H, 2015MNRAS.450.1057M}.

We present a multi-wavelength analysis using both velocity-resolved and continuum observations tracing a variety of gas phases within and around the Pillars, from the cold, dense layers deep within the pillar heads to the warm, outer layers illuminated by the bright members of NGC~6611.
This study presents the first velocity-resolved \cii\ and \oi\ line observations of the Pillars of Creation which probe the conditions and kinematics of the FUV-illuminated PDR layer between the ionization front and the molecular gas within the Pillars.
We describe the Pillars and surrounding features in all tracers in Section~\ref{sec:results} and determine the location of major PDRs and discuss the illumination geometry in Section~\ref{sec:pdr}.
Our derivation of column densities, number densities, and pressures in the atomic and molecular phases of the gas is discussed in Section~\ref{sec:physconds}, where we also discuss pressure equilibrium between these phases and the ionized gas.
We discuss the photoevaporative timescale of the Pillars in Section~\ref{sec:discussion} and include a summary of our work and some closing remarks in Section~\ref{sec:conclusion}.

%% file: observations.tex
\subsection{SOFIA}

M16 was observed with upGREAT\footnote{upGREAT and GREAT were developed by the MPI für Radioastronomie and the KOSMA/Universität zu K\"oln, in cooperation with the MPI für Sonnensystemforschung and the DLR Institut f\"ur Planetenforschung.} on SOFIA on 9 flights between 2019 and 2022 from Palmdale, California and Tahiti in the \cii\ $^{3}$P$_{3/2} \rightarrow$ $^{3}$P$_{1/2}$ transition at 158~\micron\ and in the \oi\ 63~\micron\ line in parallel with the upGREAT receiver \citep{Risacher2018JAI.....740014R} onboard SOFIA.
Note that we did not use the large undersampled \oi\ map from the FEEDBACK mapping but instead, data from an earlier PI program (see below). 
An area of $\sim590$ arcmin$^2$ was mapped in the on-the-fly (OTF) mode and atmospheric calibration was done with the GREAT pipeline \citep{Guan2012A&A...542L...4G}.
A Fast Fourier Transform Spectrometer (FFTS) with 4~GHz instantaneous bandwidth and a frequency resolution of 0.244~MHz serves as a backend \citep{Klein2012A&A...542L...3K}.

The nominal angular resolution of the \cii\ and \oi\ data is 14.1\arcsec\ and 6\arcsec, respectively, but here we use a \cii\ data cube with a spatial resolution of 15.4\arcsec, a grid of 3.5\arcsec, and a spectral binning of 0.5~\kms.
The noise RMS in one channel is typically 1.0~K (Table~\ref{t-mapsummary}).
A 3rd order baseline was removed from the spectra, which were then averaged with 1/2 weighting (baseline noise).
Spectra are presented on a main beam brightness temperature scale $T_{\rm MB}$ using an average main beam efficiency of 0.65.
The forward efficiency is $\eta_{\rm f}$ = 0.97.
See \citet{Schneider2020PASP..132j4301S} for more observational details.
These \cii\ observations are made public through the NASA/IPAC Infrared Science Archive (IRSA)\footnote{\url{https://irsa.ipac.caltech.edu/Missions/sofia.html}}.

A smaller area of M16, covering mostly the Pillars of Creation, was observed during 2 flights in October 2016 (Cycle 5) with the GREAT receiver with two channels.
The 7-pixel HFA was tuned to the \oi\ 63 $\mu$m line and the single pixel L2 channel to the CO(\jmton{16}{15}) line at 1841.345506 GHz, which we do not discuss in this study.
We employed the Fast Fourier Transform Spectrometer backend AFFTS.
The center IF (intermediate frequency) was 1455~MHz for the \oi\ line and 1000~MHz for the CO line.
The map was obtained in beam-switched on-the-fly mapping mode.
The stepsize of the map was 2.4\arcsec\ which is at a higher sampling than the FEEDBACK large mapping.
The angular resolution of the \oi\ data is 6\arcsec.
All line intensities are reported as main beam temperatures scaled with main-beam efficiencies of 0.69 and 0.68 for \oi\ and CO, respectively, and a forward efficiency of 0.97.
From the spectra, a 3rd order baseline was removed and spectra were then averaged with 1/2 weighting (baseline noise).
We smoothed the \oi\ data to a resolution of 0.4~\kms.

\subsection{CARMA and BIMA}
We used the Combined Array for Research in Millimeter-wave Astronomy \cite[CARMA,][]{2006SPIE.6267E..13B} to map the Eagle Nebula pillars in four spectral lines that trace high H$_2$ volume density gas.
The observations were obtained with the 15-element array comprised of the six 10.4~m antennas and nine 6.1~m antennas.
Three E-array configuration tracks were obtained on August 8, September 8, and September 10, 2012,
and one D-array configuration track was obtained on November 12, 2012.
The antenna signals were transmitted to the eight-band spectral line correlator.
Four bands were used to observe the spectral lines \hcn, \hcop, \ntwohp, and \cs\ in 7.8 MHz bandwidths with spectral resolution 24 kHz/channel ($\Delta$V $\sim$0.08 \kms).
Four bands were used to measure continuum at $\sim 92$ GHz, each with 490 MHz 
bandwidth and 12.5 MHz/channel.
Combining the two sidebands gave a total continuum bandwidth of 3.84 GHz.
(Continuum emission was detected at the level of $\sim$ 10 mJy; it is not discussed in this study.)
The Eagle was observed with a 37-point hexagonal mosaic centered on $(\alpha,\,\delta)_{J2000} = (18\hh18\mm51\fs29, -13\arcdeg15\arcmin02.32\arcsec)$.
The map was sampled at the Nyquist interval of the 10.4~m antennas at the CS rest frequency because it is the highest frequency and thus smallest interval.
Finally, we also use in this paper the CO(\jmton{1}{0}), \thco\, and \ceighteeno\ archival Berkeley-Illinois-Maryland Array (BIMA) data of the Eagle pillars from \cite{1998ApJ...493L.113P} and \cite{2007Ap&SS.307..187P}. 

The data were reduced using the MIRIAD package \citep{1995ASPC...77..433S,2011ascl.soft06007S}.
After phase, amplitude, passband, and flux calibration, and flagging of bad data, visibilities were inverted onto a 0.5\arcsec\ spatial grid and 0.1 \kms\ channels using a robust weighting value of zero \citep{1995AAS...18711202B}.
The inverted images were deconvolved with the MIRIAD task {\it mossdi} which uses CLEAN algorithm of \citet{1984A&A...137..159S}.
Deconvolved maps were restored with a fitted 2D Gaussian beam.
Details of the observations are provided in Table \ref{t-mapsummary}.  

The \ntwohp\ line splits into a series of hyperfine transitions, the two strongest of which, (J, F1, F) $ = $ (1--0,~2--1,~2--1) and (1--0,~2--1,~3--2), lie $\sim$1~\kms\ from each other.
Since these are not well separated given the $\sim$1~\kms\ full-width at half-max (FWHM) of the lines, we only use spatial information from them and do not use their velocities.
Our observations include a transition of \ntwohpA\ (J, F1, F) $ = $ (1--0,~0--1,~1--2) which is well-separated from other transitions given the typical FWHM, so we use velocity information from this line.
The \hcn\ line is also split into hyperfine transitions, but the strong central transition is well separated from the others given the typical FWHM.

\subsection{APEX}
M16 was mapped on September 18-20, 2019, in good weather conditions (precipitable water vapor pwv = 0.6--0.9 mm) in the \thcott\ and \twcott\ transitions using the LAsMA spectrometer on the APEX\footnote{APEX, the Atacama Pathfinder Experiment is a collaboration between the Max-Planck-Institut für Radioastronomie, Onsala Space Observatory (OSO), and the European Southern Observatory (ESO).} telescope \citep{Gusten2006A&A...454L..13G}.
LAsMA is a 7-pixel single polarization heterodyne array that allows simultaneous observations of the two isotopomers in the upper ($^{12}$CO) and lower ($^{13}$CO) sideband of the receiver, respectively.

The array is arranged in a hexagonal configuration around a central pixel with a spacing of about two beam widths ($\theta_{\rm MB} = 18.2\arcsec$ at 345.8~GHz) between the pixels.
It uses a K mirror as de-rotator.
The backends are advanced Fast Fourier Transform Spectrometers \citep{Klein2012A&A...542L...3K} with a bandwidth of $2 \times 4$~GHz and a native spectral resolution of 61~kHz.
The mapping was done in total power on-the-fly mode using a reference position at $(\alpha,\,\delta)_{J2000} = (18\hh20\mm46\fs3, -13\arcdeg14\arcmin56\arcsec)$.

The mapped region, centered at $18\hh18\mm35\fs7, -13\arcdeg43\arcmin31.0\arcsec$ and of size $30\arcmin \times 22\arcmin$ at $-40$ deg angle (CCW against positive RA), was split into $2 \times 2$ tiles.
Each tile was scanned with a spacing of 9\arcsec\ (oversampling to 6\arcsec\ in scanning direction), resulting in a uniformly sampled map with high fidelity.
All spectra are calibrated in T$_{\rm MB}$ (main-beam efficiency $\eta_{\rm MB} = 0.68$ at 345.8~GHz).
All linear baselines were removed and all data re-sampled to 0.1~\kms\ spectral bins.
The final data cubes are built with a pixel size of 9.1\arcsec\ and an 18.2\arcsec\ beam after gridding.

M16 was mapped on May 3 and June 28, 2021, in reasonable weather conditions (pwv = 0.5--0.9 mm) in the \twcosf\ transition using the SEPIA660 receiver and the FFTS1 backend spectrometer on the APEX telescope.
The instrument was tuned so that the \twcosf\ line rest frequency 691.473~GHz lay in the upper sideband, whose bandwidth is 4~GHz and frequency resolution is 61~kHz.
The OTF-mapped region is a $4.3\arcmin \times 4.3\arcmin$ square centered at $18\hh18\mm51\fs2, -13\arcdeg50\arcmin0\arcsec$.
Spectra are calibrated in T$_{\rm MB}$ (main-beam efficiency $\eta_{\rm MB} = 0.45$).
The final cube is gridded to 4.5\arcsec\ pixels with a 9.6\arcsec\ beam and has a channel width of 0.25~\kms.

\subsection{Ancillary Data} \label{sec:obs-ancillary-data}
M16 was observed at 8~\micron\ using the InfraRed Array Camera (IRAC, \citealt{Fazio2004ApJS..154...10F}) on board \textit{Spitzer} as part of the GLIMPSE program \citep{Benjamin2003PASP..115..953B}, and at 70~\micron\ and 160~\micron\ using the Photodetector Array Camera and Spectrometer (PACS, \citealt{Poglitsch2010A&A...518L...2P}) and 250, 350, and 500~\micron\ using the Spectral and Photometric Imaging REceiver (SPIRE, \citealt{Griffin2010A&A...518L...3G}) aboard the \textit{Herschel} Space Observatory \citep{Pilbratt2010A&A...518L...1P} as part of the Hi-GAL Galactic plane survey \citep{Molinari2010A&A...518L.100M}.
The IRAC image was obtained from the GLIMPSE website\footnote{\url{https://irsa.ipac.caltech.edu/data/SPITZER/GLIMPSE/}} and the PACS and SPIRE observations were obtained from the Herschel Science Archive (HSA),
and they are accessible through the NASA/IPAC IRSA: \cite{glimpse_https://doi.org/10.26131/irsa210} and \cite{higal70_https://doi.org/10.26131/irsa27, higal160_https://doi.org/10.26131/irsa25, higal250_https://doi.org/10.26131/irsa24, higal350_https://doi.org/10.26131/irsa28, higal500_https://doi.org/10.26131/irsa26}.
Ultraviolet (UV) radiation, emitted in abundance from NGC~6611, excites large hydrocarbon molecules called polycyclic aromatic hydrocarbons, or PAHs, which fluoresce in the infrared (IR), and the 8~\micron\ filter covers a particularly strong feature in the PAH spectrum \citep{Tielens2008ARA&A..46..289T}.
Detailed studies by \cite{2007ApJ...666..321I} and \cite{Flagey2011A&A...531A..51F} discuss these and other \textit{Spitzer} mid-IR images.
Far-infrared (FIR) 70 and 160~\micron\ emission traces warm dust illuminated with FUV radiation from the stars, with 70~\micron\ relatively more sensitive to temperature than column density; M16 was studied in detail in the FIR by \cite{2012A&A...542A.114H}.

The Pillars were observed using JWST NIRCam \citep{Rieke2023PASP..135b8001R} as part of the Cycle 1 outreach campaign (PI: Pontoppidan, PID \#2739) and their data made publicly available via the Mikulski Archive for Space Telescopes (MAST): \dataset[10.17909/fbc0-1930]{\doi{10.17909/fbc0-1930}}.
We use images in the filters F090W, F187N, and F335M, which trace 0.9~\micron\ continuum and background emission from starlight, the 1.87~\micron\ Pa-$\alpha$ recombination line from ionized hydrogen, and the 3.3~\micron\ PAH feature.
The 0.9 and 1.87~\micron\ observations are particularly useful for locating areas of high near-IR (NIR) extinction towards the Pillars, and the 3.3~\micron\ observation highlights illuminated PDR surfaces at $\sim200$~AU resolution.

\begin{deluxetable*}{lrrccc}
\tablewidth{0pt} 
\tablecaption{Observations summary.\label{t-mapsummary}}
\tablecolumns{6}
\colnumbers
\tablehead{
\colhead{Species} &
\colhead{Frequency} &
\colhead{Beam Size} &
\colhead{Beam PA}  &
\colhead{dV} &
\colhead{RMS}    \\
\colhead{} & 
\colhead{(GHz)} &
\colhead{(\arcsec)} & 
\colhead{(degrees)} & 
\colhead{(\kms)}  &
\colhead{(K)}
} 
\startdata
Continuum    &   92\phantom{.0000000} & $9.25\times 5.94$ & --5.34 & \nodata & 0.0042 \\ 
\hcnA        &   88.6318470 & $11.31 \times 7.22$  & --6.57 & 0.10  & 0.52 \\ 
\hcopA       &   89.1885180 &  $11.30 \times 7.15$ & --7.30 & 0.10  & 0.56 \\ 
\ntwohpA     &   93.1735050 &  $10.59 \times 6.76$ & --6.48 & 0.10  & 0.64 \\ 
CS           &   97.9809680 &  $10.12 \times 6.45$ & --6.29 & 0.10  & 0.73 \\ 
\ceighteeno  &  109.7821600 & $13.73\times 7.42$   & --9.55 & 0.267 & 0.60 \\ 
\thco        &  110.2013530 &  $6.80 \times 4.29$  & --8.15 & 0.266 & 2.06 \\ 
\twco        &  115.2712040 &  $7.51\times4.39$    & --1.38 & 0.254 & 5.03 \\ 
\thcott      &  330.5879653 & 20.0                 & 0      & 0.111 & 0.68 \\
\twcott      &  345.7959899 & 19.2                 & 0      & 0.106 & 0.55 \\
\twcosf      &  691.4730000 & 9.6                  & 0      & 0.25  & 1.65 \\
\twcoA(\jmton{16}{15}) & 1841.345506\phantom{0} & 15.9 & 0  & \nodata & \nodata \\
\cii         & 1900.536900\phantom{0} & 15.4       & 0      & 0.5   & 1.0 \\
\oi          & 4744.777490\phantom{0} & 6.7        & 0      & 0.4   & 1.9 
\enddata
\tablecomments{Beam PA is the position angle of the elliptical beam, measured in degrees east of north. dV is the velocity bin width used in this study. RMS is the root-mean-squared noise within each velocity bin for the given beam and dV. The 92~GHz continuum and CO(\jmton{16}{15}) line were observed but are not used in this study.}
\end{deluxetable*}

%% file: results_features.tex
\begin{figure*}
    \centering
    \includegraphics[width=\textwidth]{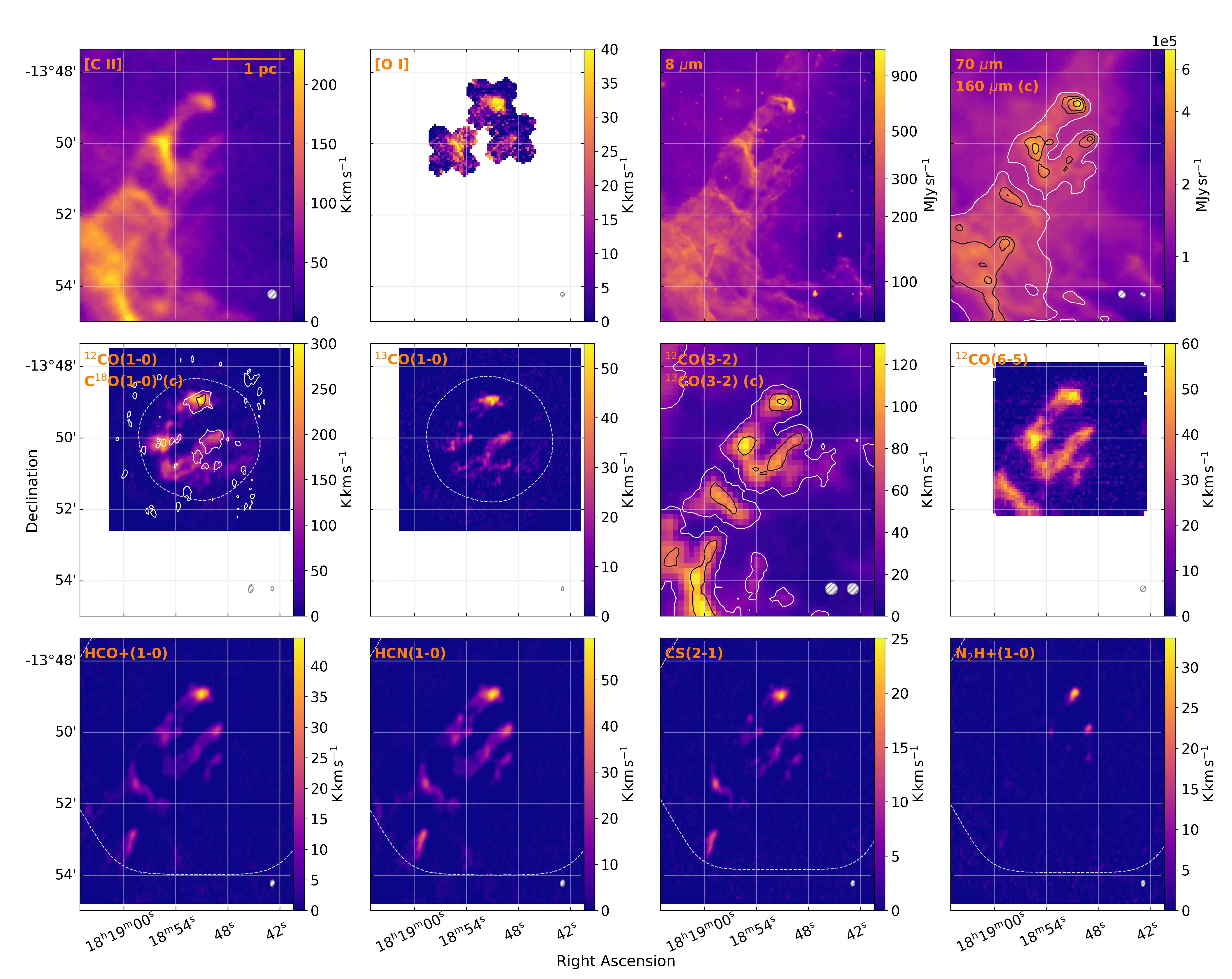}
    \caption{Integrated intensity and photometry centered on the Pillars. Contoured observations are marked with (c). The beam for each observation is shown in the lower right corner; the beam for the contoured observation, if present, is shown to the left of the color-scale observation's beam.
    All observations are shown at the same angular scale.
    \cii\ and \oi\ are integrated between $\vlsr=18$--27~\kms, as they tend to have longer blue wings, and molecular lines not otherwise specified are integrated between $\vlsr = 19$--27~\kms. \ntwohpA\ and \hcnA\ are integrated between $\vlsr = 12.6$--32~\kms\ \wrt\ the rest frequencies given in Table~\ref{t-mapsummary} in order to include several satellite lines for each species (see Section~\ref{sec:observations}). Contours are [0.5, 3.0]~K~\kms\ for \ceighteeno; [4, 24, 44]~K~\kms\ for \thcott; and [9, 18, 27, 45] $\times 10^5$~MJy~sr$^{-1}$ for the 160~\micron\ image (the last contour spacing is intentionally uneven to increase visibility towards P1a). The contour colors are chosen to increase contrast with the image and have no further significance. The dashed white contour on the interferometric molecular line observations marks 50\% gain from the primary beam mosaic pattern; outside this contour the sensitivity falls off (noise increases). Note the BIMA CO(\jmton{1}{0}) maps are smaller than the CARMA maps of the other species and so cover the southeast \shelf\ with very low sensitivity. Nonetheless it is weakly detected.}
    \label{fig:moment_panels}
\end{figure*}

We present a rich collection of observations of the Pillars of Creation alongside publicly available archival data.
Integrated intensity and continuum maps in Figure~\ref{fig:moment_panels} show the Pillars in a variety of tracers, from warm atomic gas (\cii\ and \oi) and UV illuminated gas (8~\micron\ PAH and 70~\micron\ dust continuum) to warm (CO) and dense (\hcopA\ and \csA) molecular gas.

\begin{figure}
    \centering
    \includegraphics[width=\linewidth]{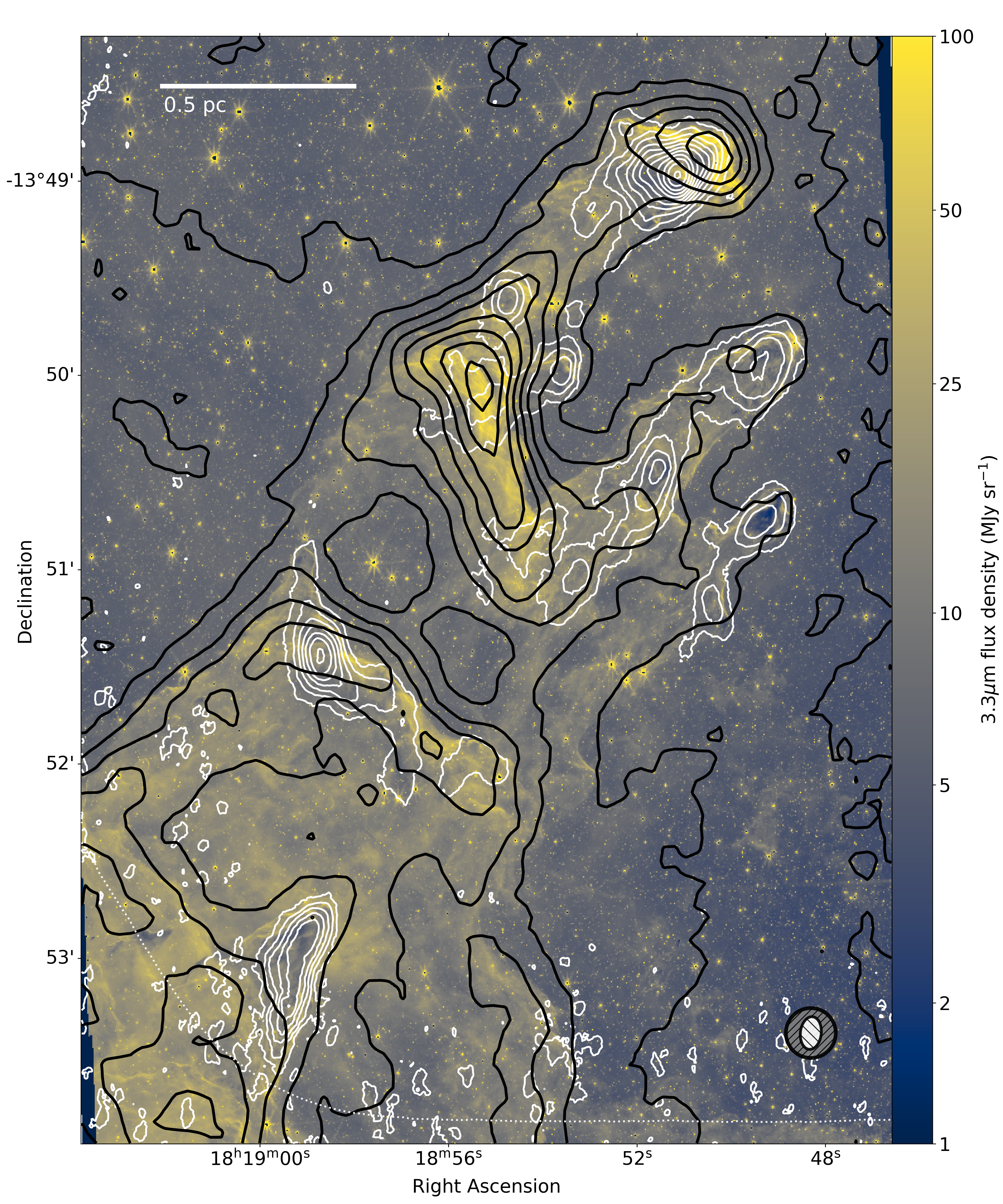} 
    \caption{Integrated \cii\ (black) and \csA\ (white) line intensities between $\vlsr = 19$--27~\kms\ overlaid on the F335M image. \cii\ integrated intensities are marked at the [25, 50, 75, 100, 125, 150, 175, 200, 225]~K~\kms\ levels, and \csA\ at [0.8, 3.2, 5.6, 8, 10.4, 12.8, 15.2, 17.6, 20, 22.4, 24.8]~K~\kms. The dotted white line marks the 50\% gain contour for the \csA\ observations; see the caption of Figure~\ref{fig:moment_panels} for more detail. The \cii\ beam is shown in grey in the lower right corner, and the \csA\ beam superimposed in white. Note the spatial offsets between the two lines towards the Cap and the Threads, among other locations. The large-scale \cii\ emission indicated by the lowest two contours is the $\vlsr \sim 25~\kms$ background discussed in detail in Appendix~\ref{sec:systematic-cii-background}.}
    \label{fig:cii_cs_f335m}
\end{figure}

This data set is ripe for comparison between tracers of different physical conditions as well as between our velocity-resolved line maps and extremely high spatial resolution JWST images.
Comparing integrated \cii\ intensity to the F335M image in Figure~\ref{fig:cii_cs_f335m} reveals myriad relationships between emission features both strong and weak, from the bright emission along the Pillars to the faint legs extending from the base of the Pillars down to the southeast (near $\alpha,\,\delta = 18\hh18\mm56\fs,\,-13\arcdeg51\arcmin20\arcsec$ in Figure~\ref{fig:cii_cs_f335m}),
while the \csA\ integrated intensity map overlaid in that figure unveils the density structure beneath.

\begin{figure*}
    \centering
    \includegraphics[width=\textwidth]{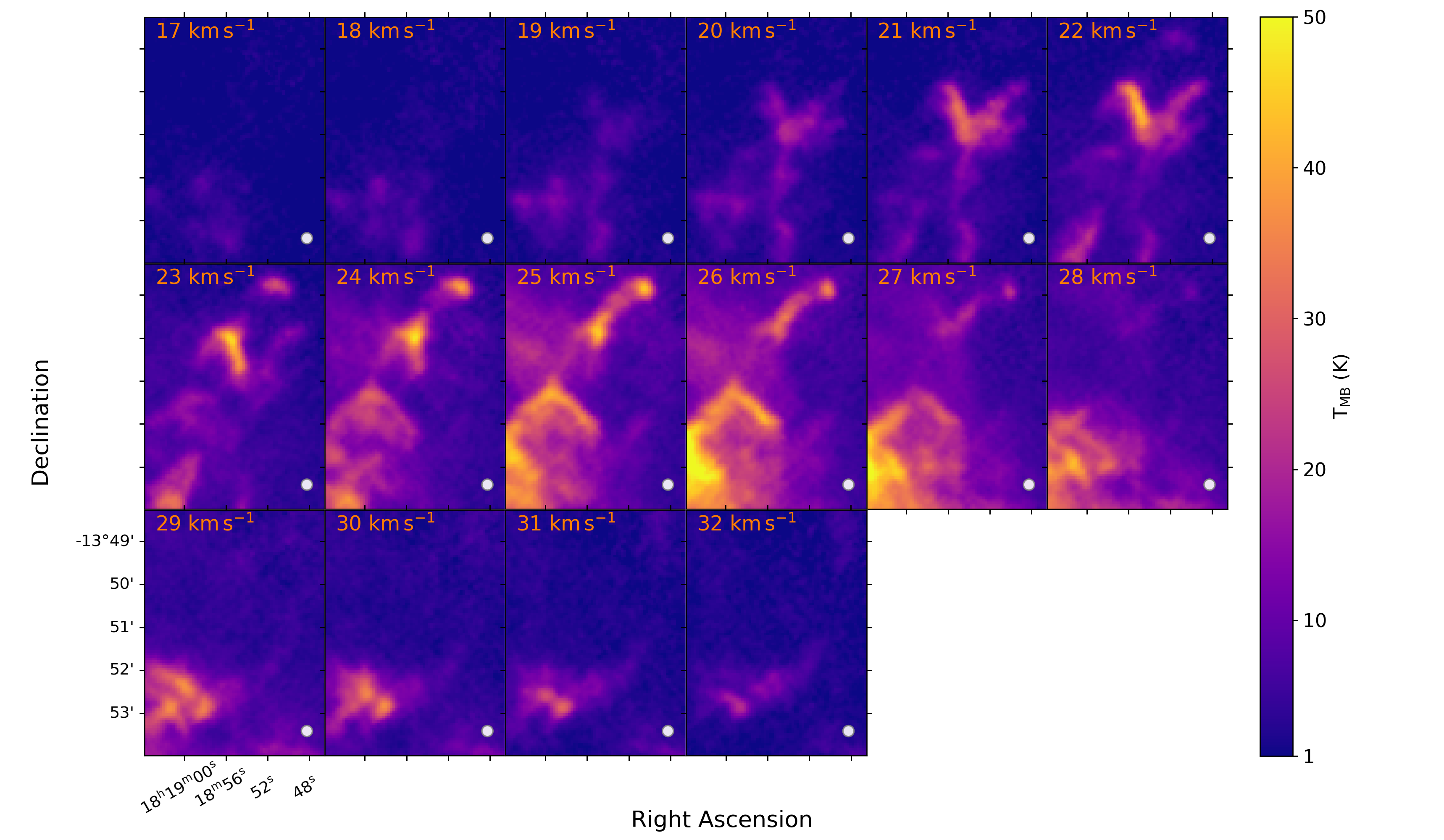}
    \caption{158~\micron\ \cii\ line channel maps binned to 1~\kms\ and centered on the Pillars. The beam is shown in the lower right corner of each map.}
    \label{fig:cii_channel_maps}
\end{figure*}

\begin{figure*}
    \centering
    \includegraphics[width=0.9\textwidth]{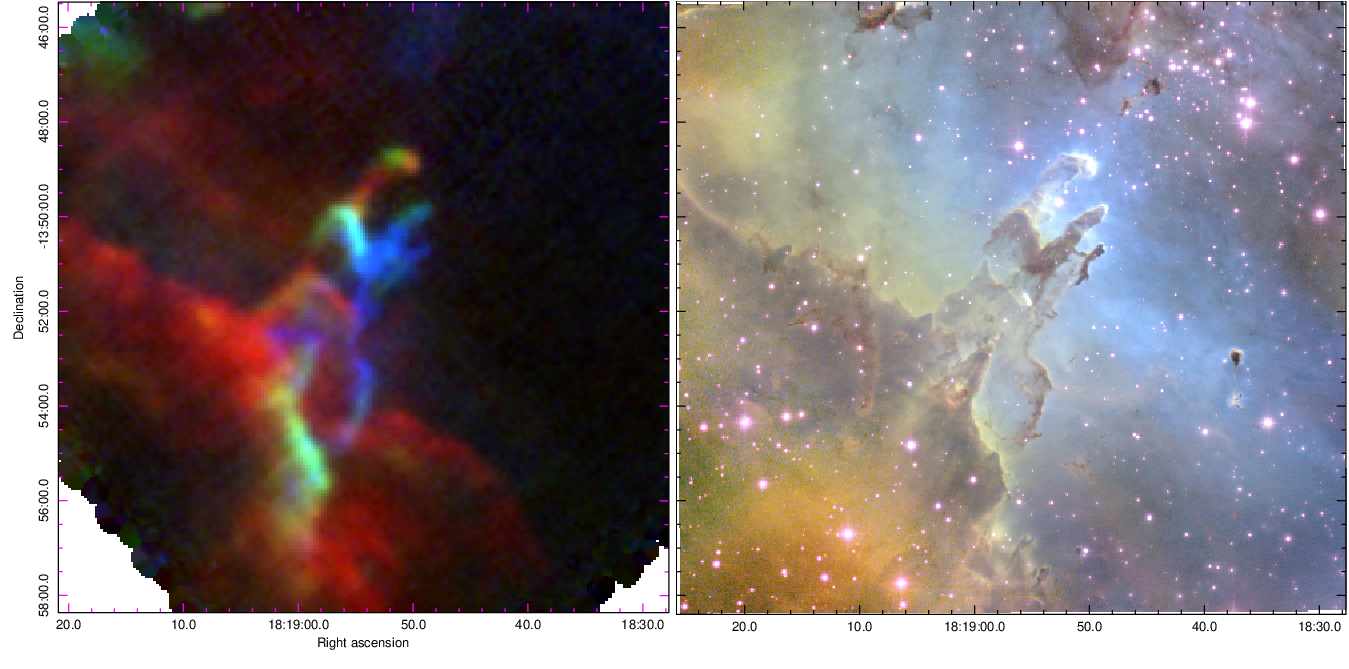}
    \caption{(\textit{Left}) Color composite using \cii\ line integrated intensities between $\vlsr = 19$--21.5~\kms\ (blue), 22--23.5~\kms\ (green), and 24--27.5~\kms\ (red). The color stretches are linear and start at 0 but have different upper limits. (\textit{Right}) NOAO optical composite prepared by T.A. Rector (NRAO/AUI/NSF and NOIRLab/NSF/AURA) and B.A. Wolpa (NOIRLab/NSF/AURA) using observations from the WIYN 0.9~m telescope at the Kitt Peak National Observatory. The image was obtained from \url{https://noirlab.edu/public/images/noao-04086} and coordinate metadata applied using Astrometry.net\textsuperscript{a} \citep{astrometrydotnet_Lang2010AJ....139.1782L}. The colors show the O\ III line at 499~nm (blue), the H$\alpha$ line at 656~nm (green), and the S\ II line at 672~nm (red).
    \small\textsuperscript{a}\url{https://astrometry.net}}
    \label{fig:cii_velocity_rgb}
\end{figure*}

Pillar 1 appears between $\vlsr=24$--26~\kms\ in the channel maps in Figure~\ref{fig:cii_channel_maps} and Pillars 2 and 3 appear between $\vlsr=20$--23~\kms.
The $\sim$10~pc scale filaments observed by \cite{2012A&A...542A.114H} and \cite{2019A&A...627A..27X} lie at similar velocities $\vlsr=20$, 22.5, and 25~\kms.
Much of the \cii\ emission in the channel maps is diffuse emission surrounding the Pillars between $\vlsr \sim 24$--27~\kms, and so the diffuse component line overlaps with and contaminates the P1 component line.
We describe in Appendix~\ref{sec:systematic-cii-background} the case and procedure for subtracting a background from \cii\ spectra towards the Pillars to handle this contamination.
The diffuse 3.3~\micron\ haze, apparent by comparing the upper left corner of Figure~\ref{fig:cii_cs_f335m} to the slightly dimmer lower right corner, may trace the same diffuse PDR as this $\vlsr \sim 25$~\kms\ \cii\ component.
The pillar system connects towards the southeast to a $\sim$10~pc scale system of emission features (only a small part of which is shown in Figure~\ref{fig:cii_channel_maps}) at $\vlsr \sim 25~\kms$ which may be the illuminated edge of the \hii\ region.
At the lower velocities of P2 and P3, a different set of emission features extend towards the southeast, away from the stars, highlighted in blue in Figure~\ref{fig:cii_velocity_rgb}.
Higher velocity channel maps $\vlsr \geq 29~\kms$ reveal a faint, ring-like feature $\sim$2~pc in diameter with a bright southern edge.
Only the western side of this ring is included in the Figure~\ref{fig:cii_channel_maps} channel maps.
Since it has a significantly different morphology than the pillar system, we deem it to be a foreground or background feature unrelated to the Pillars but likely part of the \hii\ region and a candidate for future work.

The \cii\ and \oi\ line profiles (FWHM $\sim2$--3~\kms) match the expected line width for atomic gas PDR emission.
The peak main beam temperatures $T_{\rm MB}$ of \cii\ reach $\sim$40~K towards the Pillars.
The \twcoA\ line widths $\sim2$~\kms\ and peak $T_{\rm MB}$ (reaching $\sim$80~K in the (\jmton{1}{0}) line) indicate their origin in the warm molecular gas in the PDR.
The \hcopA, \hcnA, and \csA\ line profiles are relatively thinner (FWHM $\sim 1$--2~\kms) and reach lower $T_{\rm MB}$ ($\sim$20~K in \hcopA\ and \hcnA; $\sim$10~K in \csA).
All observed line widths are likely dominated by turbulent motions.
No evidence of self-absorption in \cii\ or \oi\ is detected towards the bright Shared Base or any other location towards the Pillars (see \citealt{2020A&A...636A..16G} for a discussion of \cii\ self-absorption in other Galactic star-forming regions).

Throughout the paper, we refer to Pillars 1, 2, and 3 as P1, P2 and P3.
As was discussed in Section~\ref{sec:intro}, P1 refers to the entire structure of Pillar 1 including both P1a and P1b \citep{2015MNRAS.450.1057M}.
The ``head'' of a pillar refers to the region of the pillar nearer to the illuminating stars, and the ``tail'' or ``base'' refers to the region farther from the stars.
The ``body'' refers to the bulk midsection of the pillar.
``Along'' the body or ``parallel'' to the Pillars refers to the radial direction \wrt\ the stars, while ``across'' the body or ``transverse'' refers to the tangential direction \wrt\ the stars.
We use the proper noun ``Pillars'' to refer to the system in M16 which is the subject of this work and the common noun ``pillars'' to refer to the generic structure found in other regions as well.

\begin{figure*}
    \centering
    \includegraphics[width=\textwidth]{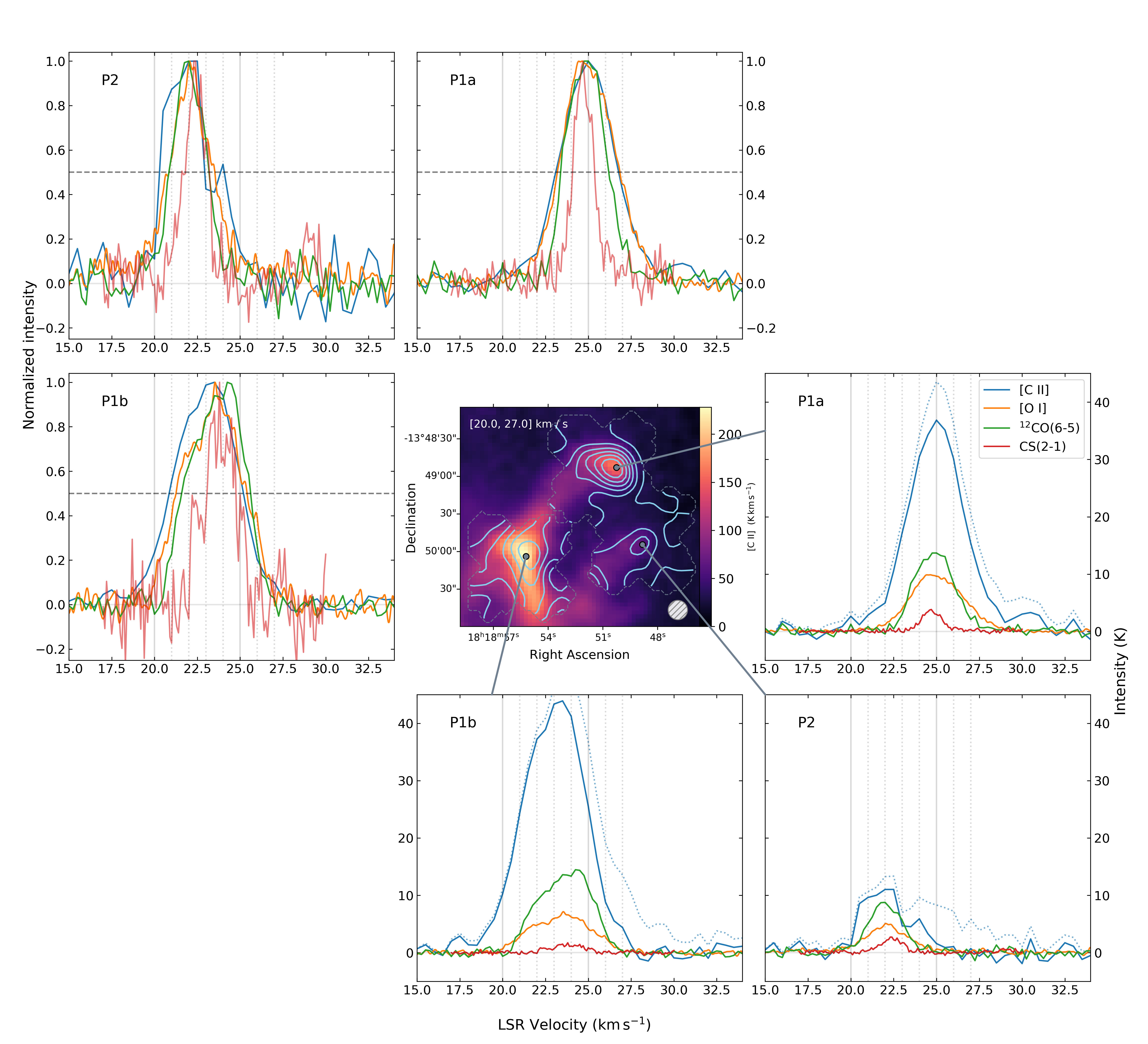} 
    \caption{Spectra of the \cii, \oi, CO(\jmton{6}{5}), and \csA\ lines towards the three brightest \oi\ positions.
    The lower-right three panels show the absolute spectra in Kelvins. The upper-left three panels show the same spectra normalized to their maxima. Half-maximum, for estimating the FWHM, is marked with dashed horizontal lines.
    The \cii\ spectra are shown after the spectral background subtraction described in Appendix~\ref{sec:systematic-cii-background}, and the unsubtracted \cii\ spectra are shown with dotted lines in the lower-right panels for context.
    Vertical lines mark every 1~\kms\ between 20--27~\kms. The \oi, CO, and \csA\ observations are all convolved to the \cii\ beam, shown in the lower right corner of the central panel.
    The central panel shows \cii\ (color) and \oi\ (contour) emission integrated between 20--27~\kms. The \oi\ contours are placed at [0, 6, 12, 18, 24, 30]~K~\kms.
    Dashed grey lines in the central panel show the area covered by these \oi\ observations. Velocities are relative to the local standard of rest (LSR).}
    \label{fig:cii_oi_spectra}
\end{figure*}

\subsection{P1}
The \cii\ and \oi\ lines, all observed molecular lines, and the 3--500~\micron\ continuum peak in brightness twice along the length of Pillar 1: towards P1a and P1b.
The peak brightness ratio of P1a to P1b is larger for denser gas tracers and smallest for \cii, in which P1b is brighter.
We show spectra of \cii, \oi, CO(\jmton{6}{5}), and \csA\ towards these brightness peaks in Figure~\ref{fig:cii_oi_spectra}.
Between the peaks, warm tracers like \cii\ and 3--160~\micron\ continuum are continuous and remain brighter than the background along the pillar, whereas the molecular lines trace the discontinuous, clumpy structure observed by \cite{2002ApJ...570..749T} in near-IR images and \cite{1999A&A...342..233W} in their radio, sub-mm, and IR observations (compare the \cii\ and \csA\ in Figure~\ref{fig:cii_cs_f335m}).

A bright rim of PAH and optical emission lies atop the molecular emission towards P1a due to direct illumination from NGC~6611 over a broad surface of neutral gas.
An embedded source in the rim \citep{2002ApJ...565L..25S, 2007ApJ...666..321I} may contribute to the 8~\micron\ emission.
At the sub-arcsecond resolution of the 3.3~\micron\ PAH image, the FUV-illuminated surfaces throughout the pillar appear as composites of the many illuminated surfaces of sub-0.01~pc clumps.
The wavy surfaces are reminiscent of the ridges and waves detected by \cite{Berne2010Natur.466..947B} on the surface of Orion's shell or by \cite{Hartigan2020ApJ...902L...1H} along the edge of Carina's Western Wall.
The gas is clumpy/porous at least down to this 0.01~pc length scale.

\subsection{P1a: Threads and Cap} \label{sec:results-p1a}

\begin{figure*}
    \centering
    \includegraphics[width=\textwidth]{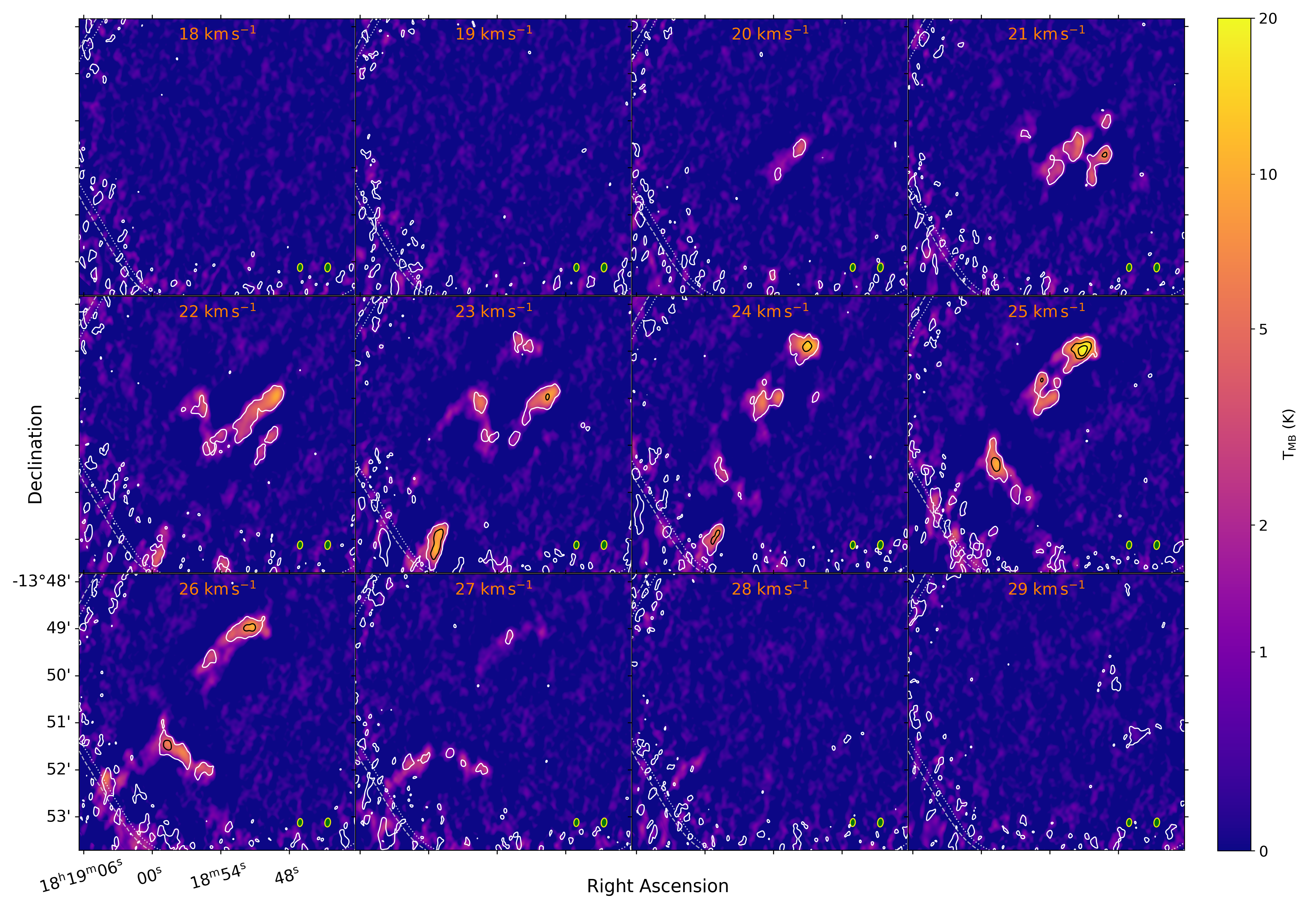}
    \caption{\hcopA\ (color) and \csA\ (contour) line channel maps binned to 1~\kms\ and centered on the Pillars. The \hcopA\ images use an arcsinh colorscale to make low-level emission more visible. The \csA\ contours mark [0.7, 4.7, 8.7]~K. The contour colors are chosen to increase contrast with the image and have no further significance. The \csA\ (left) and \hcopA\ (right) synthesized beams are shown in the lower right corner of each map. The \hcopA\ (\csA) 50\% gain contour is shown with a dashed (dotted) line.}
    \label{fig:mol_channel_maps}
\end{figure*}

\begin{figure*}
    \centering
    \includegraphics[width=\textwidth]{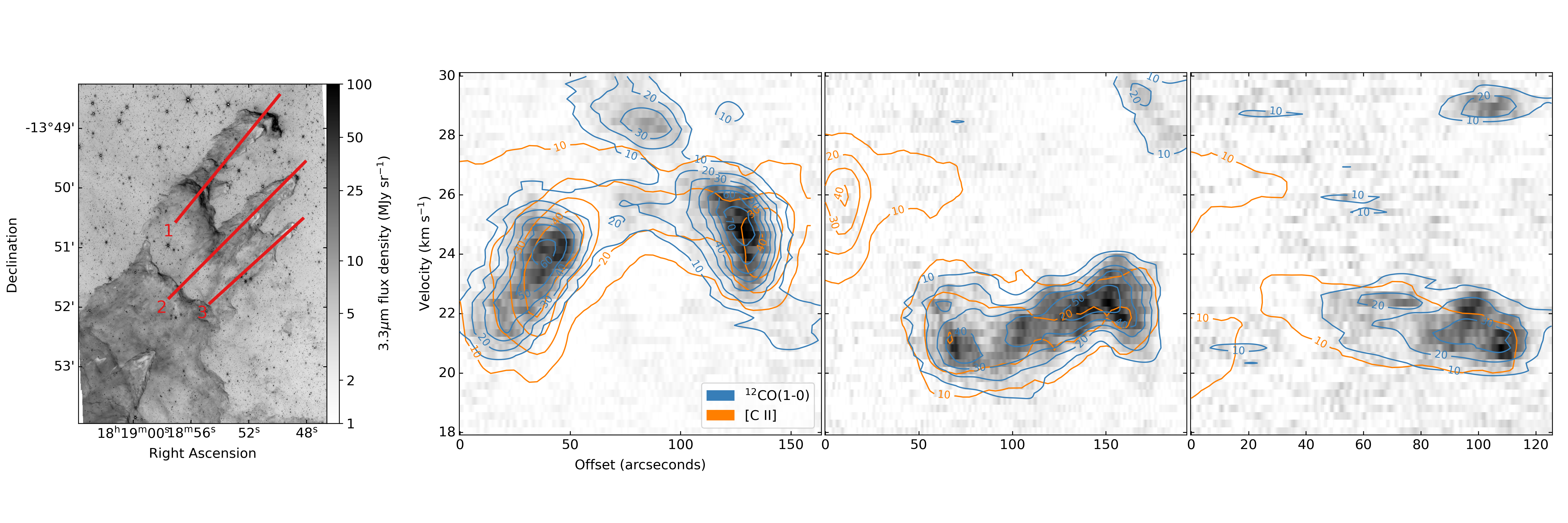}
    \caption{Position-velocity (PV) diagram along each pillar using the \cii\ and \twco\ line observations. 
    The left panel shows the F335N image with the three paths overlaid. The paths are numbered and the labels are placed at the beginning of the path.
    The next three panels show, from left to right, the PV diagrams of P1, P2, and P3 along the numbered paths. In each PV diagram, the greyscale image is the CO PV diagram at the native CO spatial resolution, while the contours show both CO and \cii\ at the \cii\ resolution. Contour labels are main beam temperatures $T_{\rm MB}$ in Kelvins. Emission at $\vlsr \gtrsim 27~\kms$ is from background features not directly related to the Pillars. Velocities are \vlsr.}
    \label{fig:pv_along}
\end{figure*}

\begin{figure*}
    \centering
    \includegraphics[width=\textwidth]{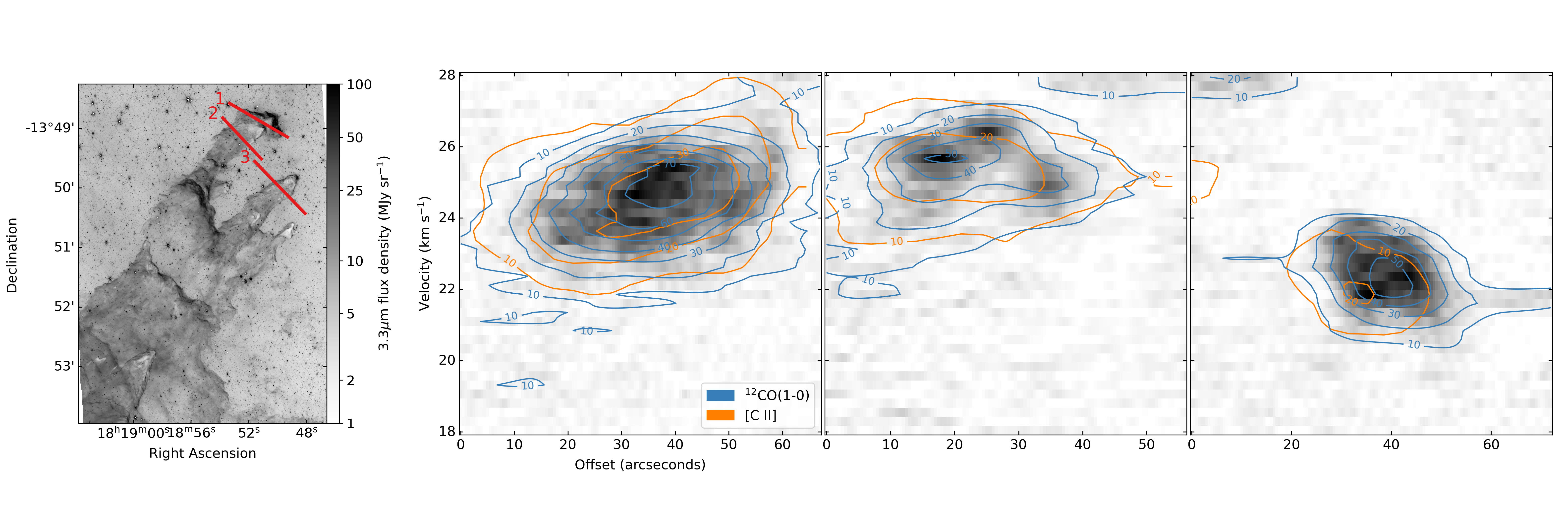}
    \caption{Same layout as in Figure~\ref{fig:pv_along}, but with PV diagrams along different paths. The first path runs across the Cap and shows its peak velocity gradient. The second path crosses the two Threads and shows the differences in \cii\ and CO velocity structure. The third path crosses the head of P2 and shows the spatial and kinematic offset of the \cii\ peak \wrt\ CO as well as the velocity gradient across the head in both lines. Velocities are \vlsr.}
    \label{fig:pv_across}
\end{figure*}

P1a appears composed of three morphologically and kinematically distinct features: the ``Cap'' oriented across the top of Pillar 1 and two filamentary structures (henceforth ``Threads''; Figure~\ref{fig:pillars_jwst}) hanging down from the head of the pillar in the NIR images in absorption (JWST F090W and F187N in Figure~\ref{fig:pillars_jwst} and the 1--2~\micron\ images by \citealt{2002ApJ...570..749T}) as well as in most molecular lines in emission.
The \hcopA\ and \csA\ line channel maps in Figure~\ref{fig:mol_channel_maps} show that the Threads are more or less parallel to each other and have similar lengths of $\sim$0.3 pc.
There are red-to-blue radial velocity gradients of order $-1$~\kms~pc$^{-1}$ along each Thread towards the pillar head (see between 70--100\arcsec\ and $\vlsr < 27~\kms$ in the position-velocity diagram along P1 in Figure~\ref{fig:pv_along}).
The Eastern Thread is spatially broader, $\sim$0.15~pc wide in CO emission just below the head of the pillar, and is redshifted by $\sim$1~\kms\ \wrt\ the thinner ($\sim$0.07~pc) Western Thread.
The Eastern Thread has a transverse gradient in radial velocity, evident in the \twcoA\ position-velocity (PV) diagrams in Figure~\ref{fig:pv_across}, redshifting to the west.
\cii\ emission towards the Threads is centered on the Eastern Thread and appears uniform across that section of the pillar rather than threaded like the molecular gas; this difference is demonstrated in Figure~\ref{fig:cii_cs_f335m} and is still evident when molecular lines are convolved to the \cii\ beam.
A detailed analysis of the velocity structure in \cii\ and \hcopA\ emission, described in Appendix~\ref{sec:geomdyn}, reveals a kinematic detection of the Western Thread in \cii\ spectra.
\cite{Bonne2023AJ....165..243B_IC63} observes filamentary features, which they termed ``legs'', extending from the IC~63 nebula away from the stars with similar velocity gradients.

The peak brightness temperatures of molecular lines towards each Thread are comparable.
This implies one or both of 1) higher column density towards the Western Thread, which in turn implies higher density assuming cylindrical symmetry, or 2) higher temperature towards the Eastern Thread.
The \ntwohpA\ line emission towards P1a around $\vlsr=23$--26~\kms\ extends slightly towards the northernmost part of the Western Thread, indicating higher density.

The Cap is a lower velocity ($\vlsr=22$--24~\kms) \cii\ component associated with P1a which extends east from the head and lies on top of the pillar like a cap.
The Cap is observed in the lines of \cii, \oi, and all but the highest critical density molecular lines.
A dark NIR feature indicating high extinction lies towards the Cap's molecular line emission, while the \cii\ and \oi\ are coincident with a bright NIR/illuminated-PAH rim spatially shifted towards the exciting stars.

There is a steep gradient in radial velocity along the Cap, with velocity increasing from east to west, in all lines in which the Cap is observed (see the first PV diagram in Figure~\ref{fig:pv_across}).
Line widths broaden significantly towards the center of the Cap where it meets the rest of P1a's associated velocity components.
If we follow the Threads northwest into the pillar head, we find that they spatially merge just as their velocity gradients lead their line profiles to blend together in velocity and become indistinguishable.
The line profiles of both Threads and the Cap blend together just northwest of the middle of P1a, where we detect the brightest molecular line and FIR continuum emission.
Detection of the \ntwohpA\ line towards this ``Merge Point'', marked in Figure~\ref{fig:pillars_jwst}, indicates cold, shielded gas.
The \cii\ emission on the eastern side of the Eastern Thread has a particularly broad line profile with a lower velocity wing.
A similar pattern is observed in \twcoA\ lines, and is present but significantly diminished in \hcopA\ when compared to \csA\ (see the spectra in Appendix~\ref{sec:measurements}).

\subsection{P1b: Base and Horns} \label{sec:p1b_results}

P1b, the lower half of P1, includes features from the Horns down to the eastern half of the Shared Base.
The complex of features is bright and continuous in warm tracers like \cii, PAH emission, and FIR dust continuum and discontinuous, particularly along the Shared Base, in molecular lines and other dense gas tracers.
The two Horns extend northwest from P1b in all tracers with spatial resolution better than $\sim$10\arcsec.
These $\sim$0.1~pc diameter clumps are particularly pronounced in the molecular line channel maps, indicating the presence of dense molecular gas, and the \ntwohpA\ and \ceighteenoA\ lines are faintly detected towards the Western Horn but not the Eastern Horn.
The Western Horn is brighter in longer wavelength ($>160~\mu$m) dust continuum, while the Eastern Horn is brighter in NIR ionized gas tracers and PDR tracers like \cii\ and PAH features, consistent with a higher molecular gas column density through the Western Horn.
A bright rim of optical, NIR, and PAH emission lies atop each Horn's molecular emission in the direction of NGC~6611.
We discuss the illumination structure of P1b further in Section~\ref{sec:pdr}.

We observe velocity gradients along each Horn: the top of the Western Horn is more blueshifted than the gas below it, while the Eastern Horn is more redshifted than the gas below it.
Interpreting these Horns as pillar-like structures along which gas is accelerated away from the stars \citep{1998ApJ...493L.113P}, the observed gradients imply that the Western Horn faces towards us and the Eastern Horn faces away as we illustrate in Figure~\ref{fig:schematic3d}.

Below the Horns lies the Shared Base, a bright \cii-emitting feature which is distinguishable from the rest of P1 by its darker appearance in the optical/NIR images (Figure~\ref{fig:cii_velocity_rgb}).
The high-resolution 3.3 and 8~\micron\ PAH emission maps show extended emission all the way across the Shared Base between P1 and P2, indicating a broad illuminated surface.
The Shared Base is the site of broad line emission in all observed lines and a strong radial velocity gradient ($\sim -3~\kms$~pc$^{-1}$) smoothly connecting P1 to P2.
The \cii\ and \twcoA\ lines contain excess low-velocity emission between $\vlsr\approx21$--22~\kms\ compared to higher critical density molecular lines like \hcopA; we show example spectra in Appendix~\ref{sec:measurements}.

\subsection{P2} \label{sec:p2_results}

Continuum images from 3--500~\micron\ show a cohesive column with a bright top.
The 3.3 and 8~\micron\ images resolve the head of P2 as a bright, $\sim$0.01~pc wide rim atop a dark clump about 0.08~pc in diameter and reveal a second similarly sized dark clump 0.4~pc below the head, about halfway down the pillar body.
This clump along P2's body is outlined at its top by thin (0.006~pc at 3.3~\micron) bright rim and coincides with a re-brightening in the 70--500~\micron\ images.

Line emission peaks between $\vlsr=22$--23~\kms\ towards the pillar head; high density tracers like \ntwohpA\ peak closer to $\vlsr=23~\kms$ while warmer, lower density tracers like \cii\ and \twco\ are blueshifted by about 1~\kms\ and all \twcoA\ line profiles have low-velocity wings.
This relative velocity shift of warm tracers to dense tracers, similar to low-velocity wings observed towards P1b and to lesser degree towards P3, persists throughout the northern half of the Pillar, but the pattern disappears as line widths increase towards the southern half.
We discuss this behavior in Section~\ref{sec:warmgasblueshift}.

Velocity gradients are detected in most lines both along and across the pillar body (Figures~\ref{fig:pv_along} and \ref{fig:pv_across}).
Gradients across P1 and P2 were also observed by \cite{2020MNRAS.492.5966S}.
Line velocities are highest towards the northeastern side of the pillar head, and decrease both to the west across the head and to the south along the body.
\cii, and \oi\ where it is observed towards the head, trace a coherent column which gets almost monotonically brighter towards the base of the pillar in integrated intensity, with only a slight local brightness maxima at the head where the PAH tracers brighten.
Molecular lines trace more substructure along P2 than \cii.
\hcopA, \hcnA, \csA\, and the CO lines trace an elongated clump towards the head, the clump towards the middle of the body, and two filametary tails below the second clump.
The \ceighteenoA\ and \ntwohpA\ lines are only detected towards the head and the mid-body clump, where all molecular line profiles are broadest.
Between these two features, molecular line emission is dim and lines are narrower (see for example the PV diagram along P2 in Figure~\ref{fig:pv_along}).

The NIR and optical \citep{1996AJ....111.2349H} images feature a bright ``wisp'' about midway along the pillar, just above the location of the clump.
The wisp seems to originate from P3 and cross the body of P2, as we see some continuous edges in the optical images both towards and off P2.
Emission from the \hcopA, \csA, and \thcoA\ lines is dimmer where the wisp overlaps with P2 between the head and the clump.
The wisp, where it passes over P2, is therefore coincident with a region of optically thinner lines of sight through P2.
The southern edge of the wisp is very close (0.05~pc or 6\arcsec, smaller than our molecular line beams) to the dark, dense clump mid-way down the pillar body.

\subsection{P3} \label{sec:p3_results}

Pillar 3, the smallest of the three main Pillars, presents in our molecular line observations as a $\sim$0.38~pc long pillar oriented roughly parallel to P1 and P2 with two tails which extend $\sim$0.34~pc in either direction at $\sim$100$^{\circ}$ angles from the body in a ``wishbone'' shape also seen in optical images.
The head of the pillar is brightest at $\vlsr\approx21.2~\kms$ in all molecular tracers.
Using the line profile modeling described in Appendix~\ref{sec:geomdyn}, we find that \cii\ is blueshifted relative to the molecular lines by $\lesssim0.3~\kms$ through the head.
\ceighteenoA\ and \ntwohpA\ are only observed towards the pillar head.
The wishbone tails are reminiscent of the ``ears'' of cometary globules simulated by \cite{1994A&A...289..559L} in 2 dimensions, which they discuss in Sections~5.2 and 5.3.1 of their paper.

\subsection{\shelf} \label{sec:shelf}
The \shelf\ and P4 (Section~\ref{sec:p4}) are detected in most molecular lines and are each associated with a compact clump of \ntwohpA\ line emission indicating enhanced gas density.
The \shelf\ is outside of the half-power beam of the CO(\jmton{1}{0}) observations, and P4 was not covered by the CO(\jmton{6}{5}) observations and lies on the edge of the \hcopA, \hcnA, \csA, and \ntwohpA\ half-power beam.

The \shelf, lying around 0.5~pc southeast of the main Pillars' tails and appearing in the same $\vlsr\approx 25$~\kms\ channel maps as P1 (Figure~\ref{fig:mol_channel_maps}), is a 0.8~pc long and 0.08~pc thick bar of molecular and \cii\ line emission oriented perpendicular to the direction of illumination.
The \shelf\ spans nearly the same width as the three-pillar system.
The integrated FIR dust emission from the \shelf\ (proxy for FUV radiation field; Section~\ref{sec:pdr}) is shifted towards NGC~6611 by $\sim$0.03~pc \wrt\ dust column density (Section~\ref{sec:physconds}), and at higher resolution, the 3.3 and 8~\micron\ PAH emission arise from the edge of the \shelf\ facing the cluster and decrease in brightness towards sites of enhanced molecular line emission along the \shelf.
We do not detect a spatial shift of \cii\ line emission \wrt\ molecular line emission.

Radial velocity increases from east to west along the \shelf\ in all lines in which it is observed.
Below the \shelf, we observe diffuse PAH and dust emission between 3--500~\micron\ as well as diffuse \cii\ emission between $\vlsr=24$--27~\kms\ (same velocity interval as the \shelf) clearly bounded to the northwest by the \shelf\ and to the northeast by another ridge of gas perpendicular to the \shelf.
This diffuse emission appears bounded to the west/southwest by a curved stream of gas, represented by the dashed line in Figure~\ref{fig:pillars_jwst}, but this stream of gas is separated by a few \kms\ from the \shelf\ in \cii, CO(\jmton{3}{2}), \hcnA, and \hcopA\ line velocity.
To the south, the diffuse emission continues into the bright, $\sim$10~pc scale feature associated with the edge of the \hii\ region.
We do not detect this diffuse emission in the molecular lines, but we do detect the \shelf, the perpendicular ridge to the northeast, and, faintly, the west/southwest stream.
The \shelf\ and northeast ridge are comparably bright in \cii, while the northeast ridge is somewhat dimmer in \twcott, much dimmer in \hcopA\ and \hcnA, and not detected in \csA.

Molecular gas column density is higher along the boundaries between the diffuse neutral gas and the ionized gas, particularly the \shelf, and all the neutral gas between the \shelf\ and northeast ridge is well illuminated.
The west/southwest stream is closer in velocity to P2 and P3, while the \shelf\ is more kinematically similar to P1 as shown by the velocity RGB image in Figure~\ref{fig:cii_velocity_rgb}.
Figures~\ref{fig:pillars_jwst} and \ref{fig:cii_velocity_rgb} show optical and NIR counterparts to the \shelf, northeast ridge, and west/southwest stream.

\subsection{P4} \label{sec:p4}

P4, a more triangular feature than the three main Pillars, has bright eastern and western edges in the 3.3 and 8~\micron\ PAH maps.
NIR continuum, like that in the central panel of Figure~\ref{fig:pillars_jwst}, shows extinction in all bands through the head and body.
From the \cii\ channel maps, we detect a radial velocity gradient along the edge of P4, from its southwestern corner ($\vlsr \sim21~\kms$) up its western side to its point ($\sim23~\kms$) and back down its northeastern corner ($\sim25~\kms$).
\cii\ line emission outlines the (upper two) triangular edges of P4 and is around 30\% dimmer towards the middle of the pillar.
The blueshifted western edge corresponds with the dark ridge in the NIR images.
We do not resolve much spatial structure in the CO(\jmton{3}{2}) observations but do detect some variability in radial velocity along the feature.
P4 lies towards the edge of the \hii\ region, which is a brighter \cii\ source with a complex line profile that is not well separated from P4's line emission.

\subsection{Summary and Geometry} \label{sec:summarygeometry}
\begin{figure*}
    \centering
    \includegraphics[width=0.9\textwidth]{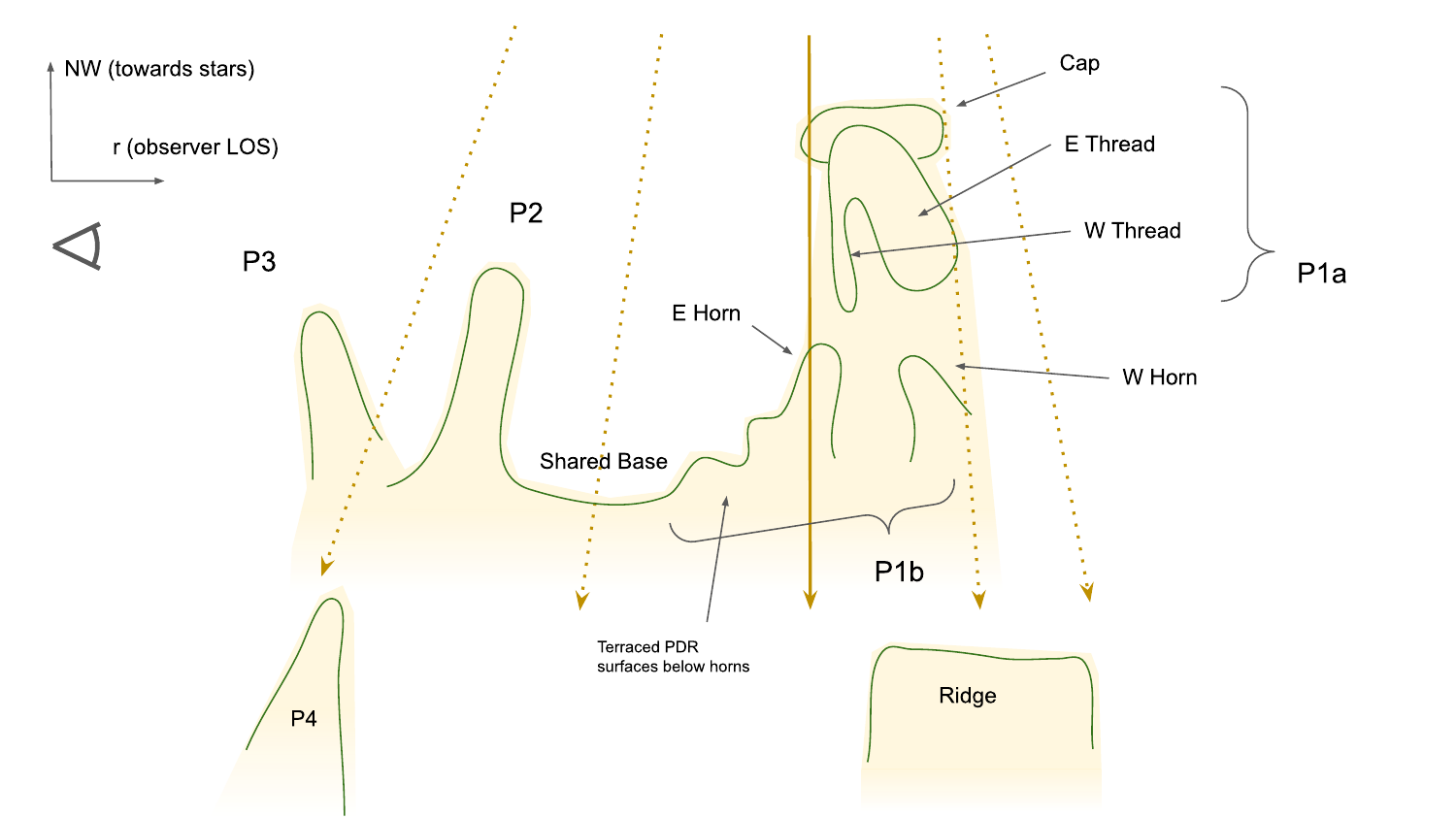}
    \caption{Schematic diagram of the pillar system as viewed from the southwest, so that the observer's line of sight runs horizontally across the figure. The green lines mark molecular gas structures and the yellow highlight marks atomic gas structure, which forms an envelope around the molecular gas (Section~\ref{sec:pdrs-p1a}). The gold rays originating from the top of the figure mark the direction of illumination, with the solid central line marking the ``perpendicular'' ray and the assumed line-of-sight position of the stars. P1a and P1b are both collections of features. The Shared Base and P1b overlap in definition, and the distinction between them is not particularly relevant because they are both approximate labels for (sets of) observed features. We estimate that the heads of P1a and P2 are separated by $\sim$1~pc along the line of sight; see Section~\ref{sec:summarygeometry} for more detail.
    }
    \label{fig:schematic3d}
\end{figure*}

The overall picture that emerges from this wealth of data is schematically represented in Figure~\ref{fig:schematic3d}.
The PAH emission traces surface structures illuminated by the stellar cluster and can be used to derive the geometry of the Pillars and their orientation relative to the stellar cluster.
The Pa-$\alpha$ emission originates mostly from diffuse ionized gas filling the \hii\ region surrounding the Pillars.
The relative strength of this line helps in placing structures such as P4 and the \shelf\ more on the near side or far side of the ionized cavity.
Peak \cii\ and molecular line velocity gradients along each pillar trace bulk motions of the gas pushed away from the stars and can indicate whether pillars are inclined towards or away from the observer \citep{1998ApJ...493L.113P, 2015MNRAS.450.1057M}.
This analysis places P1b, P2, and P3 in the foreground and, while P1b and P2 point toward the illuminating stars, P3 is backlit.
P1a is in the background and points toward the illuminating stars as well.
P1b and P2 are connected by the Shared Base, while a connection between P1a and P1b is implied by the kinematically continuous \cii\ emission between them and their on-sky projected alignment.
Warm gas tracers are blueshifted by $\sim$0.5~\kms\ towards P1b and the head of P2.
Since these pillars are both facing away from us, we interpret this to mean that the less dense outer layers are accelerated down the pillars more quickly than the dense interior layers; this is discussed in detail in Section~\ref{sec:warmgasblueshift}.
We note that we determine the sign of the pillar inclination (towards or away from the observer), but not its magnitude, from the kinematic analysis; the particular inclinations depicted in Figure~\ref{fig:schematic3d} are but one possible configuration.
\cite{2020MNRAS.492.5966S} calculates the absolute values of the inclinations to be 47$\degree$, 40$\degree$, and 40$\degree$ for P1a, P2, and P3, respectively; each may be towards or away from the observer.

The Threads' radial velocity gradients, redshifting away from the head, are consistent with material being accelerated away from the dense head along a pillar which is on the far side of the illuminating stars and points towards both the stars and the observer, consistent with the LOS position of P1a suggested by \citet{1998ApJ...493L.113P} and \citet{2015MNRAS.450.1057M}.
The Eastern Thread is redshifted \wrt\ the Western Thread; if material flows down both Threads with a similar velocity, then the Eastern Thread may be more inclined, as drawn in Figure~\ref{fig:schematic3d}, so that it has a greater projected velocity than the Western Thread.
The Cap's radial velocity gradient blueshifts away from the head, which would imply that it lies between the stars and the observer and faces away from us.
The velocity gradient may be due to unknown kinematic interactions in the pillar head, or it is possible that the head is extended along the line of sight so that parts of it are on either side of the cluster \wrt\ the observer.
This second explanation is not so unreasonable if we expect that P1's head is almost right below the cluster with very little line-of-sight offset; in this case, the line-of-sight separation between the nearest and farthest parts of the head, on opposite sides of the cluster \wrt\ us, wouldn't be that large.
We elect to place P1a directly below and only slightly behind the stars in our geometric model in Figure~\ref{fig:schematic3d}, in accordance with the second explanation for this velocity gradient.

The Shared Base is depicted in Figure~\ref{fig:schematic3d} extending a significant distance between P1 and P2.
Our analysis of column density towards the Shared Base versus average number density elsewhere implies a LOS width $\gtrsim$0.5~pc (see Sections~\ref{sec:pdrs-p1b} and \ref{sec:coldens}), and our analysis of the FUV radiation field places the heads of P1a and P2 $\sim$1~pc from each other along the LOS (see Section~\ref{sec:star_g0}).
Both of these analyses depend on the geometry and optical depth of the Pillars, and the latter is sensitive to the LOS geometry of the cluster members and extinction of FUV radiation between the cluster and the Pillars, though relative distances (such as between P1a and P2) via the latter method are not affected by uniform extinction between the cluster and all features.
Figure~\ref{fig:schematic3d} represents our analysis-based educated guess of LOS widths and separations of the Shared Base and other features, but
these uncertainties prevent us from making precise estimates.

The positions of P4 and the \shelf\ are more uncertain than those of the three main Pillars.
We place P4 closer to the observer along the LOS based on its relative darkness in Pa-$\alpha$, and the \shelf\ farther from the observer based on its relative brightness in Pa-$\alpha$ (Figure~\ref{fig:pillars_jwst}).
FIR emission towards both of these features, but particularly P4, is brighter than we expect given the projected distances of the selected stars, so the assumptions about the cluster which we use to estimate incident FUV radiation field for the three primary Pillars are not appropriate for features farther away from the cluster.
Additionally, some of P4's FIR emission might originate from 1) an embedded young stellar object (YSO) and/or 2) internal heating by that YSO, rather than solely from reprocessed FUV radiation from the main cluster.
We discuss this in more detail in Section~\ref{sec:star_g0}.

%% file: analysis_pdrs.tex
\subsection{Morphology and Geometry of the Major PDRs}

Any bright \cii-emitting feature is likely the site of a PDR, and our \cii\ observations show that M16 is rich in PDR emission.
The vast majority of the \cii\ emission is indeed from atomic gas, with a line width of $\sim$1~\kms\ rather than from ionized gas which would have a turbulence-dominated line width closer to 10~\kms\ \citep[and see Appendix~\ref{sec:systematic-cii-background} for discussion about \cii\ from the \hii\ phase]{Cuadrado2019A&A...625L...3C}.
Throughout P1 and the rest of the pillar system, sub-arcsecond resolution optical and near-IR maps trace bright rims which are $\sim$5--20 times brighter than their surroundings.
We interpret these areas, which include the Cap, Eastern Horn, and Shared Base in P1, the bright rim atop the head of P2, and parts of the head of P3 and the \shelf, to be limb-brightened edge-on PDRs.
We interpret the dimmer, ``average brightness'' emission to originate from poorly illuminated PDRs or PDRs viewed face-on or through dust extinction.

The remainder of this section highlights some of the bright PDRs throughout the Pillars with a focus on their orientation \wrt\ the observer and implications about the overall geometry of the system.
In order to discuss the geometry of some of these structures, we must consider the atomic and molecular gas column densities through them.
Column densities are discussed in depth in Section~\ref{sec:physconds}, but the results of that section inform the following discussions.

\subsubsection{P1a} \label{sec:pdrs-p1a}

Bright PDRs are associated with three distinct sets of features along the body of P1: the Cap, the Horns, and the Shared Base.
The head of P1 features a prominent PDR across the top of the Cap, observed as a bright rim in 8 \micron\ and \cii, and \oi\ sitting atop the molecular emission (Figure~\ref{fig:cii_cs_f335m}), which we are viewing edge-on \citep{2000ApJ...533L..53L}.
Below the Cap, we see relatively bright PAH emission towards a sort of ``shoulder'' atop the Western Thread, but only moderate PAH emission between the Cap and the Horns indicating that there are few edge-on illuminated surfaces towards the Threads.

Most of the \cii\ and molecular spectra towards the head of P1a are not fit well by a single Gaussian component.
Since we observe several morphologically distinct components in the channel maps presented in Section~\ref{sec:results-p1a}, we determine that P1a contains multiple components which are separated by less than their individual linewidths.
In a detailed kinematic analysis of P1a described in Appendix~\ref{sec:geomdyn}, we find that the \hcopA\ spectra towards P1a are fit well with 3 components corresponding to the Cap, Eastern Thread, and Western Thread, while the \cii\ spectra towards P1a are fit well with 2 components corresponding to the Cap and a combined-Thread component.
The \cii\ spectra towards the Threads are dominated by the Eastern Thread component, which lies at a higher velocity than the Western Thread component based on their molecular line velocities, but we detect a weak signature of the Western Thread component in the \cii\ spectra.
As the \cii\ emission does not exhibit the threaded morphology found in the molecular emission, even at matched spatial resolution, we determine that the \cii\ likely originates from a more uniform, extended envelope of atomic gas surrounding the dense molecular gas features.

Unlike the P1a spectra discussed above, the \cii\ and molecular spectra towards the Merge Point in P1a, marked in Figure~\ref{fig:pillars_jwst}, are fit well with a single component.
Based on our analysis in Appendix~\ref{sec:geomdyn}, we suggest that the Cap and two Threads are physically joined together as one component towards this position.
\ntwohpA\ line emission detected towards the Merge Point indicates cold, shielded molecular gas buried deep within the cloud.
Our proposed geometry in Figure~\ref{fig:schematic3d} envisions the Cap as a compressed rim or globule-like structure and the two Threads as legs trailing down from that cloud.

\subsubsection{P1b} \label{sec:pdrs-p1b}
The Horns, the warm gas below them, and the Shared Base are all capped with bright rims of PAH emission and form a terraced arrangement of edge-on PDRs.
These rims lie atop sites of molecular emission which are spread out in velocity.
Molecular line emission towards the Horns lies at $\vlsr=23$--25~\kms.
The bases of the Horns are connected by warm, low column density \twcoA-bright gas between $\vlsr = 23$--24~\kms\ \citep{1999A&A...342..233W}.
The Shared Base lies at $\vlsr=22$--23~\kms.
The Shared Base and the Horns are spatially and dynamically connected but are separated by about 0.13~pc and 1~\kms\ and appear as distinct features in molecular line channel maps.
These observations are consistent with the Horns and Shared Base being distinct, but connected, sites of strong PDR activity.
P1b and the Shared Base host multiple clumps of illuminated gas and behave in some ways like pillar heads.

Table~\ref{tab:columndensity} in Section~\ref{sec:physconds} lists a high atomic column density \nh\ derived from \cp\ towards the Shared Base compared to what we expect based on the PDR modeling described in Section~\ref{sec:pdr-modeling}.
Either the atomic gas density towards the Shared Base is $\sim$2 times higher than other regions or there is a long line-of-sight length through this gas structure.
We prefer the second explanation since we do not expect P1b to be the site of enhanced atomic gas density, particularly since there is little molecular gas.
If we expect the density to be closer to the median atomic gas density in Table~\ref{tab:columndensity}, then the Shared Base may extend $\gtrsim$0.5~pc along the line of sight like a ``valley'' between P1b and P2 as pictured in Figure~\ref{fig:schematic3d}.
This can explain the strong observed gradient in peak velocity along the Shared Base, as the projected gradient would be larger than the physical gradient.
The terraced arrangement of illuminated surfaces along P1b as well as the position of P1a below NGC~6611 while P1b and P2 are closer to the observer \citep{1998ApJ...493L.113P, 2015MNRAS.450.1057M} suggest that P1b continues to extend away from the observer along the LOS above the Shared Base, towards the Horns.

The emission characteristics of the Eastern and Western Horns described in Section~\ref{sec:results} indicate that the Eastern Horn is associated with more illuminated surface, either in total or facing us.
Some, but not all, of the Eastern Horn's brightness compared to the Western Horn can be explained by two geometric phenomena.
First, the Western Horn's illuminated surface must be mostly on the far side of the feature so that its $<$8~\micron\ emission is extincted by dust within the Horn.
Second, the Eastern Horn is superimposed on the body of P1, which can still be viewed through the Horn at longer wavelengths (dust continuum $\geq$70~\micron\ and the \cii\ line) and must be responsible for the Eastern Horn's brightness at those wavelengths.
This line-of-sight relationship is easiest to see in the optical images where it is clear that there is extra pillar emission surrounding the Eastern Horn.
The observed radial velocity gradients (Section~\ref{sec:results}) along the Horns suggest that the Western Horn faces towards us and the Eastern Horn faces away.

\subsubsection{P2}
All PDR tracers (PAH, 70--160~\micron, \cii) are dimmer towards P2 and P3 than towards the PDR-heavy P1.
The brightest \cii\ emission along P2 lies around $\vlsr=20$--22~\kms\ and is associated with the Shared Base.
The emission further up the body of P2 is roughly 60\% as bright as the emission towards its base and remains roughly constant in brightness even towards the head of P2.
We attribute these observations to P2 hosting a smaller illuminated PDR surface than the various locations along P1.
Towards P2, we observe PDR tracers through a limited column density and thus they appear less bright than towards P1.
Bright PAH emission is observed along the edge of the P2 head, which is not as flat in projection as the Cap in P1a and may present 1) less surface area in total and 2) a larger fraction of surface area illuminated at higher inclination (greater angle from the normal).
Below the head of P2, we see a rim of enhanced PAH emission atop the dark clump mid-way down the body and atop several cometary clumps towards the base.
P1 and P2, despite a clear difference in the brightness and abundance of PDRs along their bodies, both contain numerous small illuminated surfaces between their heads and tails.

\subsubsection{P3}
P3 is even dimmer in \cii\ than P2 (see spectra in Appendix~\ref{sec:measurements}) and is not as well resolved in \cii\ or 70~\micron\ since its projected width ($\sim$14\arcsec, 0.1~pc) is comparable to their beams.
At 3.3~\micron, the head of P3 is capped by a thin ($\approx$0.008~pc) rim of PAH emission.
The two tails of P3 which give it its wishbone shape are brighter along their cluster-facing edges in 3.3 and 8~\micron\ emission, indicating that the tails host PDRs.
As P3 and its tails are small compared to the beams of most of our observations, we will not be able to study them in as much detail as P1 and P2.

\subsubsection{\shelf}
The fact that the \shelf\ is approximately perpendicular to the direction of illumination and spans the width of the three-pillar system lying above it suggests that it may have formed in the shadow of the Pillars.
It is at least somewhat illuminated by the cluster at present, as it hosts an extended PDR surface along its 1~pc long surface facing NGC~6611.
The derived \cp\ column density towards the \shelf\ is higher than expected given its projected width ($\approx$0.1~pc), similar to what we described for the Shared Base in Section~\ref{sec:pdrs-p1b}, so it must have a LOS size of $\gtrsim$0.3~pc.
The rim of PAH emission is brighter and thicker where molecular line emission is weak along the \shelf, which could be due either to extinction through the dense clumps if the \shelf\ faces slightly away or gas density variations along the \shelf\ altering the physical PDR width if the \shelf\ is viewed nearly edge-on.

\subsubsection{P4} \label{sec:p4-pdr}
The two sides of the angular P4 both host edge-on PDRs at their surfaces.
We assume the low 1.87~\micron\ Pa-$\alpha$ emission towards P4 (green in the central panel of Figure~\ref{fig:pillars_jwst}) is due to a low foreground column density of diffuse ionized gas, indicating that P4 is relatively close to the observer within the pillar system.
The low intensity PAH emission towards the face of P4 may then originate from the illuminated far side and be extincted by dust inside the structure.
We cannot determine the orientation of P4 based on kinematics or extinction, but we orient it in Figure~\ref{fig:schematic3d} pointing away from the observer towards the main cluster.
One must consider the possibility that it faces towards the observer similar to P3.

The source driving the Herbig-Haro object HH 216 lies at the tip of P4 \citep{2004A&A...414..969A, 2007ApJ...666..321I, 2020A&A...635A.111F}.
\cite{2007ApJ...666..321I} suggest that it is one of two Class I YSOs identified towards the tip of P4, while \cite{2020A&A...635A.111F} identify a third nearby point source which may be responsible.
We discuss the possible effect of this source on the FUV radiation field at P4 at the end of Section~\ref{sec:star_g0}.

\subsection{Sources of Illumination} \label{sec:star_g0}

The brightest members of NGC~6611 are $\sim2.5$~pc from the Cap in P1a and $\sim5$~pc from P4, meaning that illumination varies by a factor of $\sim4$ along the pillar system.
We use the catalog of NGC~6611 published by \cite{1993AJ....106.1906H} to estimate the FUV radiation field $G_0$ throughout the pillar system.
The process is described in \cite{Tiwari2021ApJ...914..117T} and the software is publicly available as {\tt scoby}\footnote{The code is archived at \url{http://hdl.handle.net/1903/30441}; {\tt scoby} is also developed on \href{https://github.com/ramseykarim/scoby}{GitHub}}.
We link the spectral types of early-type NGC~6611 members published by \cite{1993AJ....106.1906H} to $T_{\rm eff}$ and ${\rm log}~g$ values using the tables of \cite{Martins2005A&A...436.1049M}, and then use the $T_{\rm eff}$ and ${\rm log}~g$ values to select stellar models from the PoWR suite \citep{powr_Sander2015A&A...577A..13S, powr_Hainich2019A&A...621A..85H}.
The PoWR stellar models provide theoretical spectra, which we combine with the catalog coordinates to estimate the FUV (6--13.6~eV) intensity at a given location from a single star assuming the projected distance and no intermediate extinction.
We estimate the total FUV intensity at the given location by summing across the contribution from all, or a subset of, stars in the catalog.
There are a handful of early-type stars within $\sim$1~pc of the Pillars in projection which cause our $G_0$ estimate to exceed 5000 Habing units towards the Pillars.
This is unreasonable given the FIR-based estimates described later in this section, so these stars must lie $\gtrsim$1~pc away along the line of sight.
We restrict our sample to stars within 2~pc in projection of the approximate cluster core $(\alpha,\,\delta)_{J2000} = (18\hh18\mm35\fs9543,\ -13\arcdeg45\arcmin20.364\arcsec)$, which includes 30 stars with types B2.5 and earlier.
We additionally restrict stars to have ${\rm log}_{10}(L_{\rm FUV} / L_{\odot}) > 4.5$ since the largest cluster members should dominate the feedback; 8 stars remain after the filtering.
We find that including all 30 stars increases $G_0$ by $\sim$50\% near the Pillars.
The FUV radiation field calculated from our final list of 8 stars (\# 161, 166, 175, 197, 205, 210, 222, 246 in the catalog of \citealt{1993AJ....106.1906H}) is $G_0 \sim 2500$ Habing units towards the heads of P1a and P2, $\sim1700$ towards P3, $\sim1500$ Habing units towards the Shared Base, and $<1000$ Habing units below the \shelf.

We can also estimate the FUV radiation field based on FIR thermal emission from dust, assuming that dust is re-radiating absorbed FUV radiation \citep{Wolfire2022ARA&A..60..247W}.
The FUV radiation field values are derived from the FIR dust emission at 70 and 160~\micron\ using the methodology described in \citealt{2016A&A...591A..40S} and corrected for a background sampled from the regions described in Appendix~\ref{sec:systematic-cii-background}.
Through this method, we estimate $G_0 \sim 2000$, 700, and 300 Habing units towards the heads of P1a, P2, and P3, respectively; and 800--1200 Habing units towards the Shared Base, the \shelf, and P4.

The stellar estimates of $G_0$ should be strict upper limits as they assume that all stars and features lie at their projected distances on the plane of the sky and that all starlight reaches every feature without extinction.
The discrepancy between the higher stellar estimate and lower FIR estimate of $G_0$ towards P2 and P3 could be a result of 1) line-of-sight separation on the order of a few parsecs, consistent with their positions on the near side of the cluster \wrt\ the observer, 2) intermediate extinction by gas/dust within the cavity, and/or 3) beam dilution by the $\sim14$\arcsec\ beam at 160~\micron, as the PDR surfaces at the heads of these two smaller pillars cover a smaller solid angle than the PDR towards the Cap in P1.
Allowing for some line-of-sight spread among the cluster members, the FIR-based estimates of $G_0$ towards the Cap, Shared Base, \shelf, and P4 are fairly consistent with the stellar estimates.
These features must lie close to their projected distances from the stars; this is most significant for the Cap, which lies fairly close in projection and so must lie approximately directly below the brightest cluster members.

The Pillars' FUV illumination is dominated by the massive cluster members near the cluster core.
The handful of early-type stars which are close in projection (within $\sim$1~pc) to the Pillars are not significant contributors to the FUV field at the Pillars and thus must be separated from the Pillars by $\gtrsim$1~pc along the line of sight, though they may still be important contributors to the FUV field below the \shelf\ and to off-axis illumination.
These near-Pillar stars may shine on parts of the pillar bodies that would otherwise be shadowed by the pillar heads \wrt\ the cluster core; that said, off-axis illumination is probably dominated by the diffuse extreme ultraviolet (EUV, $h\nu > 13.6$~eV) field created by recombinations directly to the ground state.

P3 appears not to face the bright cluster core, as explained in Section~\ref{sec:intro}.
It is possible that it faces a different star than the rest of the pillars, one which lies on the outskirts of the cluster.
P3 is the nearest of the Pillars to the observer \citep[and see schematic in our Figure~\ref{fig:schematic3d}]{2015MNRAS.450.1057M}, so a star displaced closer to the observer along the LOS could have a significant impact on the evolution of P3 while having little impact on P1 or P2.
We do not have sufficient information to identify a candidate star, but suggest that it could be among stars \# 166, 197, 205, and 210 (O8.5 V, O7 V((f)), O5 V((f*)), and B1 III, respectively), which lie $\sim$2~pc away in projection from P3 in the direction it points.
P1 and P2 are well aligned with the expected morphology and kinematics of the radiative interaction of the NGC~6611 cluster with the molecular shell; we consider it unlikely that P3's orientation away from the cluster is evidence that P3 was sculpted by a different phenomenon than the other Pillars, such as ablative Rayleigh-Taylor instability \citep{Mizuta2006ApJ...647.1151M}, given its similarity in morphology and close proximity.
Since we cannot constrain the orientation of P4, we must consider the possibility that it faces us like P3, in which case this same discussion would apply.

\subsubsection{Line-of-Sight Pillar Geometry from $G_0$}

We can roughly estimate the line-of-sight geometry of the Pillars using our two $G_0$ estimates, assuming that the catalog-based $G_{\rm 0,\,star}$ is the emitted FUV radiation field and the FIR-based $G_{\rm 0,\,dust}$ is the apparent FUV radiation field seen by the features.
The stars are assumed to be at their projected distances, lying on the plane of the sky, and all optical paths between the cluster and the features are assumed to be free of extinction.
This model attributes all differences between $G_{\rm 0,\,star}$ and $G_{\rm 0,\,dust}$ to the line-of-sight displacement of each feature from the plane on which the cluster stars lie.
To simplify the calculation, we further assume that all the stellar radiation originates from the center of the cluster using the same coordinate as earlier.
The catalog-based estimate assumes that all features are at their projected distances, so $G_{\rm 0,\,star} = L/ 4\pi r^2$, where $r$ is the projected distance from a given position to the cluster center and $L$ is the FUV luminosity of the cluster.
The FIR-based estimate measures the apparent radiation field, so $G_{\rm 0,\,dust} = L/ 4\pi (r^2 + z^2)$, where $z$ is the displacement of the feature from the cluster plane.
The ratio of the two $G_0$ estimates can be solved for the absolute value of $z$.
\begin{equation}
    \frac{G_{\rm 0,\,star}}{G_{\rm 0,\,dust}} = \frac{r^2 + z^2}{r^2} = 1 + (z/r)^2 \ \rightarrow \ 
    |z| = r \Bigg( \frac{G_{\rm 0,\,star}}{G_{\rm 0,\,dust}} - 1 \Bigg)^{1/2}
\end{equation}

The heads of P1a and P2 are estimated to lie $\sim$0--1~pc and $\sim$2--3~pc, respectively, from the plane of the cluster; higher values are obtained when the background subtraction is applied to $G_{\rm 0,\,dust}$.
The head of P3 is close to the size of the 160~\micron\ beam, so beam dilution may slightly reduce $G_{\rm 0,\,dust}$ and inflate separation; we estimate a separation of $\sim$2--5~pc from the plane.
The absolute separations for each feature are sensitive to the background subtraction, which is in turn sensitive to the 70~\micron\ opacity through the head, and also the underlying assumption of extinction-free optical paths.
The relative separation between features is less sensitive to these.
Given the assumptions and uncertainties involved, we conclude that P1a/P2 and P2/P3 are each separated on the order of $\sim$1~pc along the line of sight, so that the three primary Pillars span a few pc along the LOS.

We compare to the distances calculated by \cite{2020MNRAS.492.5966S} using peak radio continuum brightness.
LOS separations from the plane of the cluster $z = 1.8$, 2.5, and 2.8~pc for P1a, P2, and P3 are calculated as $z = D {\rm cos}(i)$ from the true distances $D = 2.6$, 3.2, and 3.6~pc and inclinations $i = 47\degree$, 40$\degree$, and 40$\degree$ presented by \cite{2020MNRAS.492.5966S}.
In both their and our estimates, P1a is nearest and P3 is farthest from the plane of the cluster.
Our estimate, from both $G_0$ and kinematic clues (Section~\ref{sec:summarygeometry}), positions P1a closer to the plane of the cluster than the estimate by \cite{2020MNRAS.492.5966S}.

Line of sight displacement $z$ can only be calculated when $G_{\rm 0,\,star} \geq G_{\rm 0,\,dust}$, which is the case for the three primary Pillars but not towards the \shelf\ or P4.
Stars $>2$~pc from the cluster core in projection may contribute more significantly to the radiation field at those two features since they are farther from the cluster core.
The LOS displacements of the \shelf\ and P4 therefore cannot be estimated using this method.
The Shared Base is extended along the LOS (Section~\ref{sec:pdrs-p1b}), so $G_{\rm 0,\,dust}$ is likely overestimated there and LOS distance cannot be determined.

At the tip of P4 lie two Class I YSOs, HH-N and HH-S, identified by \cite{2007ApJ...666..321I} in NIR and mid-IR images, and a third source of unknown type $\sim$3\arcsec\ from HH-N identified by \cite{2020A&A...635A.111F} in the NIR.
Any one of these sources may drive HH 216.
A point source appears at the tip of P4 in both the 70 and 160~\micron\ observations near the location of HH-N and the NIR point source.
The source emission is $\sim$50\% brighter than the diffuse P4 emission in both bands and likely affects the $G_{\rm 0,\,dust}$ estimate towards that position.
It is unclear whether radiation from the source heats or otherwise affects P4.
\cite{2007ApJ...666..321I} noted that the YSOs appear extincted by tens of $A_V$, but the NIR source's luminosity and extinction are not known.
If P4 is heated internally, the derived $G_{\rm 0,\,dust}$ is unsuitable for the LOS separation estimate.

\subsection{Modeling PDRs Towards the Pillars} \label{sec:pdr-modeling}

We model observations towards several PDRs associated with the Pillars to measure densities and FUV radiation fields at those locations (see \citealt{Wolfire2022ARA&A..60..247W} for an overview of PDR modeling).
We use the Wolfire-Kaufman 2020 models available in the PDR Toolbox which we access via the {\tt pdrtpy}\footnote{\url{https://dustem.astro.umd.edu}} software \citep{2006ApJ...644..283K, 2008ASPC..394..654P, Pound2023AJ....165...25P}.
This particular model set assumes a plane-parallel, face-on PDR geometry most appropriate for PDRs at the surface of non-clumpy clouds.
The observed intensities of \cii, \oi, \twco, \twcott, and \twcosf\ at selected locations towards the PDRs are integrated over the relevant velocity intervals for each PDR and convolved to the \cii\ resolution.
We do not convolve \twcott, which has a slightly larger beam.
The higher resolution of the other observations is more valuable than keeping all the beams matched.

The on-sky widths of the modeled features are close to the \cii\ beam width, so we assume that 1) the beam filling factor is 1 for all observations except \twcott, 2) there is zero emission outside the features and 3) the \twcott\ beam filling factor is the ratio of the beam areas $\Omega_{\cii} / \Omega_{\rm CO}$.
We scale the \twcott\ intensities up by the ratio $\Omega_{\rm CO} / \Omega_{\cii}$ to account for beam dilution.

\begin{figure*}
    \centering
    \includegraphics[width=\textwidth]{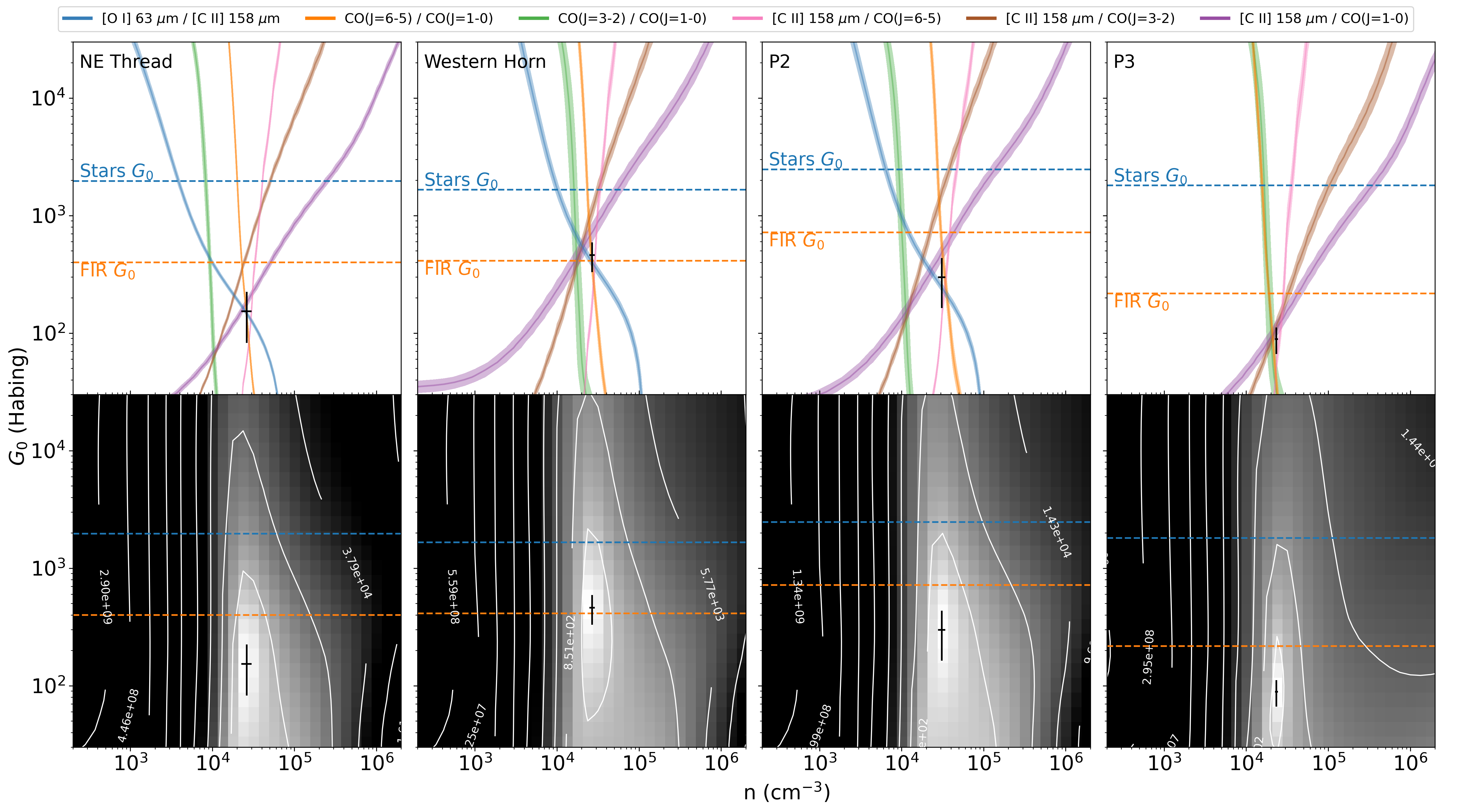}
    \caption{\textit{(Top row)} Overlay plots made using {\tt pdrtpy} which show observed line ratios towards 4 locations which are labeled in the top left corner of each plot. The black cross shows the automatically fitted solution and associated uncertainty; these values are listed in Table~\ref{tab:pdrt-fits}. \textit{(Bottom row)} The reduced $\chi^2$ associated with the automatic fit, with overlaid labeled contours.
    Radiation field values estimated via stellar catalog and FIR emission are plotted as horizontal dashed lines in each figure.}
    \label{fig:pdrt}
\end{figure*}

For pairs of these integrated intensities, the PDR Toolbox takes ratios and plots them on a grid of density $n$ and FUV radiation field $G_0$ using precalculated models.
This procedure cancels out the beam filling factor to first order.
The resulting diagrams, shown in Figure~\ref{fig:pdrt}, are called overlay plots and their usage is described in detail in \cite{Tiwari2022AJ....164..150T} and \cite{Pound2023AJ....165...25P}.
The ratios appear as lines curving across the grid.
In principle, all lines should intersect at a density and radiation field consistent with the observations.
In practice, our overlay plots generally do not converge at one specific location but rather a region of the $n$ vs. $G_0$ grid around $n\approx 2\times10^4$~cm$^{-3}$ and a few $10^2$ Habing units, which we take to be the physical conditions consistent with our observations under the Wolfire-Kaufman 2020 PDR models.
The results of the PDR Toolbox fitting routine for the four sample locations from Figure~\ref{fig:pdrt} are listed in Table~\ref{tab:pdrt-fits}.

Density is fairly well bounded under this model, with our observations fitting between $\sim$1--4 $\times10^4$~cm$^{-3}$.
We estimate this uncertainty using a contour of a few times the minimum $\chi^2$ and confirm that this region of parameter space doesn't change much between lines of sight which should have similar density and illumination.
This density estimate is consistent with the estimates made by \cite{1998ApJ...493L.113P} and \cite{2000ApJ...533L..53L} as well as broadly consistent with our estimate based on the \cii\ column density and assumed line-of-sight geometry (Section~\ref{sec:coldens-cii}; Table~\ref{tab:columndensity}).

Radiation field is much more loosely bound, as we can only place an upper limit of $G_0 \sim 1000$ Habing units and no realistic lower limit under this model.
The PDR models used here assume face-on geometry, which may produce slightly inaccurate results, so we rely on independent estimates of density and radiation field to gauge how accurate the PDR model-based estimates might be.

We can further bound the radiation field seen by the Pillars using our independent estimates of $G_0$ from stellar catalogs and FIR dust emission which we described in Section~\ref{sec:star_g0}
The stellar method, as previously discussed, yields estimates of $G_0 \sim1$--2 $\times 10^3$ Habing units in the neighborhood of the Pillars.
These estimates mark the upper end of the PDR model estimate of radiation field, so we assume that the radiation field seen by the Pillars is somewhere between 500--2000 Habing units depending on local geometry.

We are unable to make substantial claims about spatial variation in the density and radiation field based on the PDR models, as the uncertainties in our modeling techniques are greater than the local variation.
We find that the densities and radiation fields are generally self-consistent across all modeled regions towards the Pillars.

\input{table_pdrt_fits}

%% file: table_pdrt_fits.tex
\begin{deluxetable}{lcccc}
    \label{tab:pdrt-fits}
    \tablecaption{Fitted $G_0$ and $n$ to observed line ratios using {\tt pdrtpy}.}
    \tablewidth{0pt}
    \tablehead{
        \colhead{Location} & \colhead{$G_0$} & \colhead{$n$} & \colhead{$\chi^2 / {\rm dof}$} & \colhead{dof} \\
        \colhead{Name} & \colhead{(Habing)} & \colhead{(cm$^{-3}$)} & \colhead{} & \colhead{}
    }
    \decimalcolnumbers
    \startdata
        NE-thread & 150 & \nexpo{2.6}{4} & 150 & 5 \\
        W-Horn & 460 & \nexpo{2.7}{4} & 14 & 5 \\
        P2 & 300 & \nexpo{3.1}{4} & 58 & 5 \\
        P3 & 89 & \nexpo{2.3}{4} & 6.3 & 4 \\
    \enddata
    \tablecomments{The $G_0$ and $n$ values are those fitted to the observed line ratios overlayed on the top row of Figure~\ref{fig:pdrt}. These correspond to the minimum $\chi^2$ location, in color on the bottom row of Figure~\ref{fig:pdrt}, which is indicated with a black cross in all panels of that figure.}
\end{deluxetable}

%% file: analysis_conditions.tex
\subsection{Molecular and Atomic Hydrogen Column Densities} \label{sec:coldens}

Molecular and atomic gas column densities are estimated using the \thco\ and \cii\ lines, respectively, as well as with dust emission.
Column densities are summed pixel-by-pixel to estimate the total pillar masses, and line-of-sight distances are assumed to estimate densities in each gas phase.
Table~\ref{tab:columndensity} lists column and number densities and Table~\ref{tab:mass} lists pillar masses.

\subsubsection{\thcoA\ Column Density} \label{sec:coldens-co}
\begin{figure*}
    \centering
    \includegraphics[width=\textwidth]{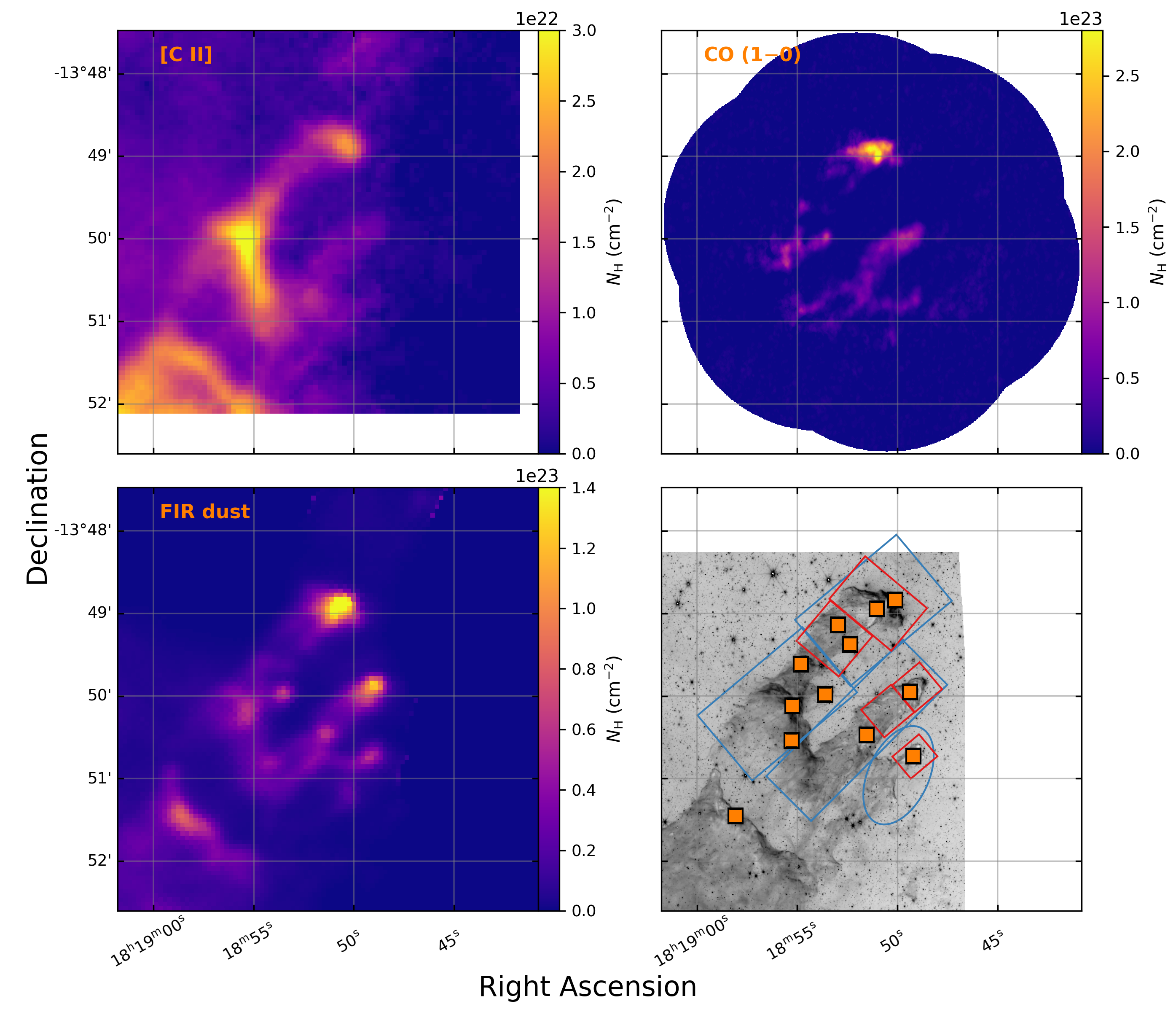}
    \caption{The top-left, top-right, and bottom-left panels show hydrogen column densities \ntot\ derived from \cp, CO(\jmton{1}{0}), and FIR dust emission. The bottom-right panel shows the JWST F335M image in greyscale as a reference. All four plots are on the same size scale and the grey gridlines are in the same place. On the bottom-right panel, blue boxes show the integration areas for the pillar masses in Table~\ref{tab:mass}, and red boxes show the same for Table~\ref{tab:lifetimes-head-neck}. Orange squares mark the locations from which column densities are sampled in Table~\ref{tab:columndensity}. In Table~\ref{tab:columndensity}, from P1a-edge to Shared-Base-Mid, locations are listed in order of decreasing declination along P1. The P2 locations are also in order of decreasing declination along P2.}
    \label{fig:columndensity}
\end{figure*}

We estimate the molecular gas column density assuming optically thin \thco\ emission following \cite{Tiwari2021ApJ...914..117T} (their Appendix~E) and \citet{MangumShirley_columndensity}.
We assume that the \jmton{1}{0} lines of \twcoA\ and \thcoA\ share the same excitation temperature along each line of sight, that \twco\ is optically thick everywhere, and that the beam filling factor is unity.
We adopt an isotopic ratio $\twcoA / \thcoA = ^{12}{\rm C} / ^{13}{\rm C} = 44.65$ based on the galactocentric radius-dependent expression given by \citet{Yan_2019} using a galactocentric radius of 6.64~kpc calculated using Equation~2 by \citet{1993A&A...275...67B}.
We convert \twcoA\ column density to molecular hydrogen column density $N({\rm H}_2)$ using the abundance ratio $\twcoA/{\rm H}_2 = 8.5\times10^{-5}$ \citep{tielens_2021}.
Hydrogen is assumed to be entirely in the molecular phase where CO is detected, so that total hydrogen column density (defined as $\ntot = \nh + 2 \nht$) is $\ntot = 2 \nht$.
Gas mass is calculated by summing \ntot\ over the boxes shown in Figure~\ref{fig:columndensity} and converting to mass using a mean molecular weight $\mu = 1.33$.

We estimate line-of-sight-averaged densities by assuming cylindrical symmetry, so that line-of-sight distance through the feature is equal to the angular width of the features, and dividing column density by that distance.
$H_2$ column densities and gas masses are listed in Tables~\ref{tab:columndensity} and \ref{tab:mass}, respectively.
Uncertainties are estimated by propagating the channel RMS noise of the \twcoA\ and \thco\ observations and, for conversion of column density into mass, the heliocentric distance uncertainty, into the procedure described above.

\subsubsection{\cp\ Column Density} \label{sec:coldens-cii}
We do not detect the \thcii\ line towards the Pillars, so we estimate an upper limit on the \cp\ column density using the \cii\ channel noise $T_{\rm rms}$ as a detection limit.
We estimate an upper limit on the \thcii\ hyperfine component at +11.2~\kms\ \wrt\ the \twcii\ line (see Table~1 in \citealt{2020A&A...636A..16G}) since it is sufficiently far from the \twcii\ line center given the observed linewidths, and we assume that the isotopic ratio $\twcp / \thcp =\ ^{12}{\rm C} / ^{13}{\rm C}$ and use the same value as in Section~\ref{sec:coldens-co}.
Equation~4 in the paper by \citet{2020A&A...636A..16G} relates the observed line brightness of \twcii\ to \thcii\ given the \twcii\ optical depth.
We find an upper limit of $\tau_{12} \lesssim 1.3$ using the brightest \twcii\ spectrum which lies towards P1b.
\cite{Bonne2023ApJ...951...39B_DR21} find a similar upper limit on \cii\ optical depth towards the DR21 ridge using the same method.

We use the method of estimating \twcp\ column density without a detected \thcii\ line described by \citet{2015A&A...580A..54O}, and we assume a constant $\Tex$ rather than a constant optical depth.
The adopted $\Tex = 107$~K is calculated using the highest \cii\ line brightness towards the Pillars and our upper limit of $\tau \leq 1.3$, and $\Tex$ is therefore a lower limit towards that location (considering the emitted radiation temperature $T_{\rm R}$ constant, $\tau$ and $\Tex$ have an inverse relationship in the optically thin and $\tau \sim 1$ regimes).
Assuming a fixed $\Tex$ across the field is equivalent to assuming that density and kinetic temperature are fixed across the map, so that only the column density influences the observed line brightness.
Kinetic temperature is determined by the density $n$ and radiation field $G_0$ according to PDR models.
Both $n$ and $G_0$ may be slightly higher at the heads of the three Pillars which contain dense molecular gas and are closer to the stars, but the column densities there are not high enough to probe $\Tex$.
$\Tex$ must therefore be sampled from the higher column density P1b location.
We will overestimate column density towards dense, highly irradiated locations like the pillar heads, and we will underestimate column density towards less dense, poorly illuminated locations along pillar bodies.
If $\Tex \sim 150$~K, column densities would be approximately halved around the map, and it is unlikely that $\Tex$ is considerably lower than 100~K.
We convert $N(\cp)$ to atomic hydrogen column density \nh\ using the abundance ratio ${\rm C}/{\rm H} = 1.6\times 10^{-4}$ \citep{Sofia2004ApJ...605..272S} and calculate column densities assuming a filling factor $\eta = 1$.
We assume hydrogen to be entirely in the atomic phase where \cp\ is present so that $\ntot = \nh$, but we note that some \cii\ emission may originate from CO-dark molecular gas \citep{Pabst2017}.

The spatially variable \cii\ background discussed in Appendix~\ref{sec:systematic-cii-background} contributes to the total column density derived in the region.
To determine just the contribution to the Pillars requires subtracting a background.
We calculate column densities with the total intensities (no background spectrum subtraction) and subtract out a column density background sampled from the same locations as the spectral background.
Within these background sample regions, we take the mean column density to be the background and the standard deviation to be a systematic uncertainty on both target and background column density due to the variability of the background.
Statistical uncertainty from the RMS channel noise is propagated through the process described above, though this source of uncertainty is reduced by the sum over pixels to obtain mass.
Background-corrected column densities are listed in Table~\ref{tab:columndensity}.
Estimates of the systematic uncertainties and the 1-$\sigma$ statistical uncertainties are given in the table's caption.

We find the highest \cp\ column density towards P1b rather than P1a despite the higher molecular gas column density towards P1a.
Mass is calculated as described in Section~\ref{sec:coldens-co}.
The atomic hydrogen mass is typically $\sim$30\% of the molecular gas mass.

\input{table_column_density.tex}
\input{table_mass.tex}

\subsubsection{Column Density from Dust Emission} \label{sec:coldens-dust}
We calculate column densities from FIR dust emission to compare to the molecular and atomic gas column densities derived from CO and \cp.
Optical depth at 160~\micron\ is calculated directly using the 70 and 160~\micron\ observations and the method described by \cite{Tiwari2021ApJ...914..117T}.
The $\geq$250~\micron\ maps have lower spatial resolution and do not resolve some Pillar features.
We obtain $\sim$14\arcsec\ resolution in our final map by using only 70 and 160~\micron.
We subtract the flux background in each band, sampled from the background regions described in Appendix~\ref{sec:systematic-cii-background} (Figure~\ref{fig:background-spectra}), prior to calculation of dust properties in order to isolate the Pillar emission.
We divide the dust optical depth at 160~\micron\ by the dust extinction cross section per H nucleus at 160~\micron, $C_{\rm ext,160} / {\rm H}$.
We take the $R_V = 3.1$ value of $C_{\rm ext,160} / {\rm H} = 1.9 \times 10^{-25}$~cm$^{-2}$/H from \cite{Draine2003ARA&A..41..241D}, following the method of \cite{Tiwari2021ApJ...914..117T}.
This yields \ntot, the column density of H nuclei or total column density, directly.
These column densities are, in theory, equal to the sum of hydrogen column densities derived from \cp, C, and CO as dust is present at all phases, but they should be dominated by the relatively larger molecular gas column.
Column densities are listed in in Table~\ref{tab:columndensity} and masses, calculated as described in Section~\ref{sec:coldens-co}, are listed in Table~\ref{tab:mass}.

We measure dust temperatures ranging from 25~K in the cold clumps, such as the pillar heads and the Western Horn, to 30~K along the lower column density pillar bodies.
Since we are using the 70 and 160~\micron\ measurements for this estimate, we are more sensitive to warm dust.
Dust temperature varies along the line of sight through the Pillars, with colder gas and dust embedded within the dense pillar heads and other clumps, shielded from radiative heating.
Assuming a constant temperature along a line of sight with real temperature variation can cause column density to be underestimated \citep{Storm2016ApJ...830..127S, Howard2019MNRAS.489..962H}.
Including the 250~\micron\ band increases sensitivity to cold dust at the expense of spatial resolution and raises the mass estimates by about 10\% (these larger estimates are not included in the table), which is within our uncertainty estimates described below.

The PACS handbook quotes a 5\% calibration uncertainty on the 70 and 160~\micron\ measurements; this dominates the total uncertainty towards the pillar heads.
We estimate an uncertainty on our flux background subtraction in each band, which is about 15\% of the background and dominates total uncertainty outside the bright pillar heads.
The PACS observations include a statistical uncertainty map; this is the smallest relative contribution to the total uncertainty, but we include it anyway since including additional sources of uncertainty is trivial using a Monte Carlo sampling method.
Using such a method, we estimate the total uncertainties on column density and mass by computing \ntot\ and $M_{\rm H}$ for 1000 realizations of the three uncertainties added to the observations and adopting the 1-$\sigma$ uncertainties to be the difference between the 84$^{\rm th}$ and 16$^{\rm th}$ percentile values divided by 2.
To these uncertainties, we add in quadrature the 15\% uncertainty from the heliocentric distance \citep{2019ApJ...870...32K}.
The median \ntot\ and $M_{\rm H}$ values are consistent with the values calculated without considering uncertainty, so we record the latter values in the tables.

The dust-derived column densities tend to be closer to the CO-derived column densities than to those from \cp, satisfying our expectation that the dust column is dominated by contribution from the molecular gas phase.
The dust-derived \ntot\ values tend to be lower than the CO-derived values, with the largest difference observed towards the bright center of P1a.
This is likely due to significant dust temperature variation along the line of sight, which causes the average temperature to be higher and the optical depth to then be underestimated.
For this reason, we take the CO values as the better gauge of molecular gas column density.
Towards the northern edge of P1a, where \cii\ is brightest, the dust-derived \ntot\ is larger than the CO-derived value; we attribute this to the line of sight passing primarily through warm atomic gas with less LOS temperature variation, as the sum of the \cp-derived and CO-derived column densities is consistent with the value from dust.

\subsection{Density} \label{sec:density}

Line-of-sight averaged densities are derived by dividing the column density measurements by estimated line-of-sight feature widths.
We estimate the angular size across each feature in Table~\ref{tab:columndensity} using the high-resolution JWST NIR images and assume cylindrical symmetry to adopt the same width along the line of sight.
We use this width for both the atomic gas and the molecular gas, even though they would be layered along lines of sight passing through molecular gas.
Densities $\sim10^5$~cm$^{-3}$ are derived from the CO- and dust-based column densities.
These densities are broadly consistent with the observed \hcopA, \hcnA, \csA, and \ntwohpA\ brightness temperatures (listed in Appendix~\ref{sec:measurements}) according to estimates from the RADEX\footnote{\url{http://var.sron.nl/radex/radex.php}} radiative transfer software \citep{radex_vanderTak2007A&A...468..627V}.
Densities $\sim10^4$~cm$^{-3}$ are derived from \cp-based column densities, which are consistent with those estimated using the PDR Toolbox.

We adopt $n_{\rm H} = \nexpo{1.8}{4}$~cm$^{-3}$ as the atomic gas density by taking the average $n_{\rm H}$ in Table~\ref{tab:columndensity}.
From this average we exclude the Shared Base and \shelf\ positions, since their line-of-sight size depends strongly on the adopted geometry which is very uncertain, and the Eastern Horn, since we cannot isolate its \cii\ emission from P1 emission behind it.
We adopt $n_{\htwo} = \nexpo{1.3}{5}$~cm$^{-3}$ as the molecular gas density by taking the average $n_{\htwo}$ towards dense features (heads, Horns, and the P2 clump) in Table~\ref{tab:columndensity}.
The molecular gas density we derive is consistent with the estimates by \cite{1998ApJ...493L.113P}, \cite{1999A&A...342..233W}, and \cite{2000ApJ...533L..53L}.

The critical density of \cii\ in the presence of either H or H$_2$ ranges from 3--$\nexpo{5}{3}$~cm$^{-3}$ at $\sim$100~K, so our adopted atomic gas density implies that the kinetic temperature $T_{\rm K}$ should be similar to the excitation temperature $\Tex$.
Recalling that our $\Tex$ is a lower limit due to optical depth, we place a lower limit on atomic gas $T_{\rm K} > 100$~K.
The derived molecular gas densities are greater than the critical density of \twco\ by a factor of $\sim50$, so the radiation temperature $T_{\rm R}$, assumed equal to the observed $T_{\rm MB}$, of the optically thick  \twco\ line should be similar to the kinetic temperature of the gas it probes.
The kinetic temperature of the warm, outer layers of molecular gas probed by \twcoA\ must therefore be $T_{\rm K} \sim 50$--70~K for most of the pillar bodies and reach $T_{\rm K} \sim 100$~K towards brightly illuminated regions in P1a and P1b.
The derived molecular gas density is close to the critical densities $\sim10^5$~cm$^{-3}$ of \hcopA\ and \hcnA\ and slightly below the critical densities $\sim\nexpo{3}{5}$~cm$^{-3}$ of \ntwohpA\ and \csA.
The \hcopA\ and \hcnA\ lines are optically thick according to RADEX given their assumed abundance \citep{tielens_2021} and the CO-derived column density \nht, so their observed $T_{\rm MB} = T_{\rm R} \sim 15$--20~K likely traces $T_{\rm K} \sim 20$~K molecular gas towards P1a.

%% file: table_column_density.tex
\begin{deluxetable*}{lcc|cc|ccc|cc}
    \label{tab:columndensity}
    \tablecaption{Column and number densities.}
    \tablewidth{0pt}
    \tablehead{
        \colhead{} & \colhead{} & \colhead{} &
        \colhead{\cp} & \colhead{} &
        \colhead{CO} & \colhead{} & \colhead{} &
        \colhead{Dust} & \colhead{} \\
        \colhead{Location} & \colhead{Width} & \colhead{} &
        \colhead{\ntot} & \colhead{$n_{\hone}$} &
        \colhead{\ntot} & \colhead{$\sigma_{\rm stat}$} & \colhead{$n_{\htwo}$} &
        \colhead{\ntot} & \colhead{$\sigma_{\rm stat}$} \\
        \colhead{Name} & \colhead{(\arcsec)} & \colhead{(pc)} &
        \colhead{($10^{21}$ cm$^{-2}$)} & \colhead{(cm$^{-3}$)} &
        \colhead{($10^{21}$ cm$^{-2}$)} & \colhead{($10^{21}$ cm$^{-2}$)} & \colhead{(cm$^{-3}$)} &
        \colhead{($10^{21}$ cm$^{-2}$)} & \colhead{($10^{21}$ cm$^{-2}$)}
    }
    \decimalcolnumbers
    \startdata
        P1a-edge & 17 & 0.14 &
        20.9 & \nexpo{4.7}{4} &
        31 & 8 & \nexpo{3.5}{4} &
        59 & 12 \\
        P1a-center & 45 & 0.38 &
        13.9 & \nexpo{1.2}{4} &
        286 & 21 & \nexpo{1.2}{5} &
        139 & 24 \\
        E-Thread & 20 & 0.17 &
        8.2 & \nexpo{1.6}{4} &
        31 & 9 & \nexpo{3.0}{4} &
        22 & 7 \\
        W-Thread & 10 & 0.08 &
        3.8 & \nexpo{1.4}{4} &
        29 & 9 & \nexpo{5.5}{4} &
        24 & 9 \\
        E-Horn & 13 & 0.11 &
        14.8 & \nexpo{4.4}{4} &
        80 & 14 & \nexpo{1.2}{5} &
        22 & 6 \\
        W-Horn & 14 & 0.12 &
        1.7 & \nexpo{4.6}{3} &
        125 & 15 & \nexpo{1.7}{5} &
        54 & 15 \\
        Shared-Base-E & 33 & 0.28 &
        29.2 & \nexpo{3.4}{4} &
        59 & 13 & \nexpo{3.4}{4} &
        59 & 10 \\
        Shared-Base-Mid & 22 & 0.19 &
        18.0 & \nexpo{3.1}{4} &
        $\leq$6 & 6 & ... &
        24 & 7 \\
        P2-head & 18 & 0.15 &
        5.0 & \nexpo{1.1}{4} &
        114 & 14 & \nexpo{1.2}{5} &
        95 & 22 \\
        P2-clump & 9 & 0.08 &
        6.3 & \nexpo{2.7}{4} &
        65 & 11 & \nexpo{1.4}{5} &
        59 & 16 \\
        P3-head & 13 & 0.11 &
        3.6 & \nexpo{1.1}{4} &
        67 & 11 & \nexpo{9.8}{4} &
        58 & 22 \\
        \shelf & 13 & 0.11 &
        19.2 & \nexpo{5.7}{4} &
        ... & ... & ... & 
        63 & 13
    \enddata
    \tablecomments{
    Column and number densities derived from \cp, CO, and dust emission measurements; the species is indicated above each column density column.
    Column densities are sampled towards each location, shown in Figure~\ref{fig:columndensity}, and given in units of $10^{21}$~cm$^{-2}$.
    The standard conversion between column density and $A_V$ is $A_V = \ntot / 1.9 \times 10^{21}~{\rm cm}^{-2}$ \citep{Bohlin1978ApJ...224..132B_NH_AV}, so the column densities listed in the table can be divided by 1.9 to estimate $A_V$.
    Column densities derived using the \cii\ line have statistical uncertainty $\sigma_{\rm stat} = 0.6$ (table units) except for location Shared-Base-E, which has $\sigma_{\rm stat} = 0.7$, and an additional systematic uncertainty of 1.5 or 1.7 (table units) for the northern and southern background samples, respectively (see Appendices~\ref{sec:measurements} and \ref{sec:systematic-cii-background}), which estimate the spatial variability of the \cp\ background (Section~\ref{sec:coldens-cii}).
    Atomic and molecular gas number densities $n_{\hone}$ and $n_{\htwo}$ are estimated using the \cp\ and CO measurements of column density by dividing \nh\ and \nht, respectively, by the assumed LOS width of each feature.
    We assume $\ntot = \nh$ in the atomic gas and $\ntot = 2\nht$ in the molecular gas in order to estimate density and pressure in each phase independently; however, some of the \cii\ emission could arise from CO-dark molecular gas.
    On-sky angular widths are estimated using the JWST images and are adopted as the LOS widths through the features assuming cylindrical symmetry.
    }
\end{deluxetable*}

%% file: table_mass.tex
\begin{deluxetable}{c ccc|cc|cc|c}
    \label{tab:mass}
    \tablecaption{Atomic and molecular gas mass.}
    \tablewidth{0pt}
    \tablehead{
        \colhead{Feature} &
        \colhead{\cp} & \colhead{} & \colhead{} &
        \colhead{CO} & \colhead{} &
        \colhead{dust} & \colhead{} &
        \colhead{} \\
        \colhead{Name} &
        \colhead{M$_{\rm H}$} & \colhead{$\sigma_{\rm stat}$} & \colhead{$\sigma_{\rm sys}$} &
        \colhead{M$_{{\rm H}_2}$} & \colhead{$\sigma_{\rm stat}$} &
        \colhead{M$_{\rm H}$} & \colhead{$\sigma_{\rm tot}$} &
        \colhead{M$_{\rm tot}$}
    }
    \decimalcolnumbers
    \startdata
        P1a &
        24 & 1 & 7 &
        79 & 3 & 
        83 & 23 &
        103 \\
        P1b & 
        41 & 2 & 8 & 
        37 & 2 & 
        64 & 20 &
        78 \\
        P2 & 
        34 & 2 & 7 & 
        70 & 3 &  
        95 & 33 &
        103 \\
        P3 & 
        4.2 & 0.6 & 2.7 & 
        14 & 1 &  
        15 & 7 &
        18 \\
    \enddata
    \tablecomments{Gas mass $M$ derived for the pillars from \cii, \thco, and dust emission.
    Masses are given in $M_{\odot}$ and rounded; sums are calculated from exact numbers.
    Statistical and systematic uncertainty estimates are given where applicable.
    Total masses are considered to be the sum of atomic gas mass from \cp\ and the molecular gas mass from CO.
    }
\end{deluxetable}

%% file: analysis_pressure.tex
\subsection{Pressure Balance} \label{sec:pressures}
We estimate pressures for the ionized, atomic, and molecular gas phases of gas and comment on the possibility of pressure equilibrium.
Pressures are listed in Table~\ref{tab:pressure}.
We follow \cite{1999A&A...342..233W} in using the presence of pressure equilibrium between the ionized and molecular gas as a proxy for whether or not the ionization-driven shock has already been driven through the pillar heads.

\subsubsection{Ionized Gas}
Our observations do not probe directly the physical conditions of the ionized hydrogen (\hii) phase, so we take values $n_{\rm e}\sim 1120$--$1800$~cm$^{-3}$ and $T_{\rm e} \approx 8000$--10,000~K from optical ion line studies of the pillar surface by \cite{2006MNRAS.368..253G} and \cite{2015MNRAS.450.1057M}.
\cite{2000ApJ...533L..53L} note that temperature may be higher near the PDR-\hii\ interface due to photoelectric heating, a harder radiation field, and high density.
We estimate thermal pressure in the ionized gas flowing from the surface of the pillar $P_{\rm i}/k_{\rm B} = 2n_{\rm e}T = 1.8$--\nexpo{3.6}{7}~K~cm$^{-3}$.
We neglect any sources of non-thermal pressure in the ionized gas near the pillar surface; magnetic fields should not be important in the ionized phase due to low densities and, while a turbulent linewidth was detected in the ambient \hii\ region (far from the pillar surface, where the photoevaporative flow does not dominate gas kinematics) in M16 by \cite{Higgs1979AJ.....84...77H}, the radial striations evident in the optical images presented by \cite{1996AJ....111.2349H} indicate that the photoevaporative flow near the pillar surface is not turbulent.

\subsubsection{Molecular Gas}
\cite{2018ApJ...860L...6P} discuss the pressures originating within the molecular gas phase.
We follow their discussion and recalculate pressures using our observations of the molecular gas.
We determined $n_\htwo = \nexpo{1.3}{5}$~cm$^{-3}$ in Section~\ref{sec:density}.
Molecular gas temperature should be coupled to dust temperature $T_d$ at densities $n\gtrsim10^5$~cm$^{-3}$ \citep{1999A&A...342..233W}.
We find $T_{\rm d}\sim25$~K towards the dense gas (Section~\ref{sec:coldens-dust}), comparable to the value $T\sim20$~K found by \cite{1999A&A...342..233W} using 350--2000~\micron\ photometry, which is more sensitive to cold dust but less sensitive to warm dust than the 70 and 160~\micron\ photometry used for our estimate.
Thermal pressure in the molecular gas is $P_{\htwo,\,{\rm therm}}/k_{\rm B} = n_\htwo T \approx \nexpo{3.2}{6}$~K~cm$^{-3}$.
We estimate the pressure from turbulent support using the molecular line velocity dispersion $\sigma_{\rm obs} \approx 0.6$~\kms\ (thermal velocity dispersion, $\sigma_{\rm therm} \sim 0.1~\kms$, contributes less than 2\% of $\sigma_{\rm obs}$) to be $P_{\htwo,\,{\rm turb}}/ k_{\rm B} = \rho \sigma^2 = (2 n_\htwo \mu m_\hone) \sigma^2 \approx \nexpo{1.5}{7}$~K~cm$^{-3}$ ($\mu = 1.33$), consistent with pressure derived from the line widths observed by \cite{1999A&A...342..233W}.

\cite{2018ApJ...860L...6P} estimate the plane-of-sky magnetic field strength $B=170$--320~$\mu$G within the Pillars, calculated using the Davis-Chandrasekhar-Fermi method \citep{Davis1951PhRv...81..890D, Chandrasekhar1953ApJ...118..113C} and therefore assuming that the turbulent velocity field is isotropic, which implies magnetic pressure $P_{\htwo,\,{\rm B}}/k_{\rm B}\approx0.9$--\nexpo{3.0}{7}~K~cm$^{-3}$.
The factor of 3 range in their pressure value seems to be driven mostly by variation in the molecular line widths observed by \cite{1999A&A...342..233W}; our observations indicate that some of this line width can be attributed to line-of-sight confusion rather than turbulent velocity dispersion and we observe single-component line widths near the lower end of their range.
We use the Equation~1 in the paper by \cite{2018ApJ...860L...6P} to estimate $B_{\perp} \sim 320~\mu$G and $P_{\htwo,\,{\rm B}}/k_{\rm B} \approx \nexpo{3.0}{7}$~K~cm$^{-3}$ from our observed molecular gas velocity dispersion $\sigma_v \approx 0.6$~\kms\ along lines of sight with limited line-of-sight confusion.
The magnetically-dominated total pressure within the molecular gas is $P_\htwo/k_{\rm B} \approx \nexpo{4.8}{7}$~K~cm$^{-3}$.

The total pressure in the molecular gas is slightly higher than the pressure in the ionized gas.
The molecular and ionized gas phases are likely in pressure equilibrium, with the extra pressure supporting the molecular gas against self-gravity.
We assume that the ionization-driven shock has already passed through the molecular gas.

We observe a velocity gradient across the head of P2, which could indicate an additional mode of rotational support in the pillar head.
For comparison, we estimate a rotational ``pressure'' $P_{\htwo,\,{\rm rot}} = \rho \omega^2 r^2$ where $\omega r \approx 0.5$~\kms\ based on the CO line velocities (see the PV diagram in Figure~\ref{fig:pv_across}), finding $P_{\htwo,\,{\rm rot}}/k_{\rm B} \approx 10^7$~K~cm$^{-3}$ (not included in our total pressure).
This is comparable to turbulent pressure and less than magnetic pressure ($\nexpo{3}{7}$~K~cm$^{-3}$, see above), and could contribute $\sim$20\% of the total support to the pillar head, but we would expect the head to be extended in the plane of rotation across the pillar if that were so.
Since we do not observe oblateness, we find it unlikely that rotational support is a significant support mechanism in the pillar heads.
Pillar rotation is further discussed by \cite{2020MNRAS.492.5966S}.

\subsubsection{Atomic Gas}
We determined the density of the warm, atomic gas to be $n_{\hone} \approx \nexpo{1.8}{4}$~cm$^{-3}$.
The lower limit on the $\Tex$ of the \cii\ line indicates a kinetic temperature $T_{\rm K} > 100$~K (Section~\ref{sec:density}), and temperatures up to $250$~K are consistent with the approximate $n$ and $G_0$ in the PDR models \citep{Pound2023AJ....165...25P}.
This results in a thermal atomic gas pressure $P_{\hone} /k_{\rm B} = n_{\hone} T \sim 2$--\nexpo{5}{6}~K~cm$^{-3}$.
The atomic gas phase is likely hydrostatically supported by turbulence similar to the molecular gas.
We remove the thermal velocity dispersion of \cii\ $\sigma_{\rm therm} \sim 0.5~\kms$ given the assumed $T_{\rm K}$ from the observed \cii\ velocity dispersions $\sigma_{\rm obs} \sim 1.1$--1.4~\kms\ to find $\sigma_{\rm turb} = ( \sigma_{\rm obs}^2 - \sigma_{\rm therm}^2)^{1/2} \sim 1.0$--1.3~\kms.
Adopting the turbulent velocity dispersion $\sigma_{\rm turb} \approx 1.3$ yields a turbulent pressure of $P_{\hone,\,{\rm turb}}/k_{\rm B} \sim \nexpo{5}{6}$~K~cm$^{-3}$ in the atomic gas.
This is an upper limit since the photoevaporative flow velocity $\sim 0.5$~\kms\ projected along the line of sight should contribute to the observed width of the \cii\ line, but we cannot disentangle these contributions in our observations.
We discuss the impact of this flow velocity on the photoevaporative lifetime of the Pillars in Section~\ref{sec:pillar_lifetimes}.
Ultimately even the highest estimate of turbulent pressure in the atomic gas allowed by the observations is an order of magnitude smaller than the total pressures in the ionized and molecular phases in Table~\ref{tab:pressure}.

It is likely that ionization-driven shocks have already passed through the Pillars, so the atomic gas should be in pressure equilibrium with the ionized and molecular gas phases surrounding it.
However, our estimate of the total atomic gas pressure falls significantly short of the ionized gas pressure by $P/ k_{\rm B} \sim\nexpo{1}{7}$~K~cm$^{-3}$.
The missing pressure is likely from the magnetic field in the atomic gas whose strength must be $B\sim 200\mu$G.
\cite{Pellegrini2007ApJ...658.1119P, Pellegrini2009ApJ...693..285P} estimate comparable magnetic field strengths in the PDRs in M17 and the Orion Bar and determine that the magnetic field is important to the structure, particularly the width, of the PDR.
We don't have an independent estimate of the field strength in the atomic layer, but scaling the molecular phase $B$ field by the ratio of the molecular to atomic gas densities gives a field strength of $B_{\perp} \sim 85~\mu$G and a corresponding $P_{\rm B}/k_{\rm B} \sim \nexpo{2}{6}$~K~cm$^{-3}$, which is insufficient for pressure equilibrium.

We conclude that ambipolar diffusion, the flow of neutrals past ions, has weakened the magnetic field in the molecular gas faster than in the atomic gas because the molecular gas has a lower ionization fraction.
The pillar heads are still magnetically supported now, but the magnetic field in the molecular gas must have been larger in the past.
The ambipolar diffusion timescale $\tau_{\rm AD} \sim 10^6$~yr in the molecular gas is similar to the age of the cluster, as we demonstrate in the next section.
Carbon is ionized in the atomic gas, so the ionization fraction there is $\chi_{\rm ion} \sim 10^{-4}$ and $\tau_{\rm AD} \sim 10^9$~yr.

\subsection{Ambipolar Diffusion and Molecular Cloud Collapse}
\cite{Mouschovias1991ApJ...373..169M} determined that the reduction in magnetic support due to ambipolar diffusion could precipitate the collapse of molecular cores into stars.
We estimate the possibility of such a collapse in the Pillars before they are photoevaporated.
Critical mass under magnetic field support can be expressed as
\begin{equation} \label{eq:mcrit}
    M_{\rm c,B} = 9.7 \Bigg( \frac{R}{0.1~{\rm pc}} \Bigg)^2 \Bigg( \frac{B}{100~\mu{\rm G}} \Bigg) M_{\odot}
\end{equation}
The magnetic critical masses, calculated using the assumed radii in Table~\ref{tab:lifetimes-head-neck}, for the heads of P1a, P2 and P3 are 83, 21, and 10~$M_{\odot}$, respectively, while the estimated H$_2$ masses are 72, 14, and 6~$M_{\odot}$.
For all the pillar heads, but particularly P1a whose geometry is complex (see Section~\ref{sec:results-p1a}), the $R^2$ dependence carries significant uncertainty into the critical mass determination from our assumptions about line-of-sight width, but we can conclude that all the heads are nearly critical.
\cite{1994A&A...289..559L} present calculations to determine the gravitational stability of magnetically supported globules in Section~5.9 of their paper, and these suggest that the pillar heads are unstable given their current magnetic field strengths.

For magnetically subcritical clouds, gravitational collapse is controlled by the process of ambipolar diffusion of the supporting magnetic field.
The ambipolar diffusion timescale is
\begin{equation} \label{eq:tauAD}
    \tau_{\rm AD} \approx \nexpo{2}{5} \Big( \frac{\chi_{\rm ion}}{10^{-8}} \Big) ~ {\rm yr}
\end{equation}
\begin{equation} \label{eq:chiion}
    \chi_{\rm ion} \approx \nexpo{2}{-7} \Bigg( \frac{10^4~{\rm cm}^{-3}}{n_\htwo} \Bigg)^{1/2}
\end{equation}
where the degree of ionization $\chi_{\rm ion}$ is calculated for a typical cosmic ray ionization rate $\zeta_{\rm CR} = \nexpo{3}{-17}$~s$^{-1}$ \citep{Elmegreen1979ApJ...232..729E}.
For the adopted $n_\htwo = \nexpo{1.3}{5}$~cm$^{-3}$, we find $\chi_{\rm ion} \sim \nexpo{6}{-8}$ and $\tau_{\rm AD} \sim 1$~Myr which is similar to both the age of the cluster and Pillars as well as the estimated photoevaporative lifetime of the Pillars (Section~\ref{sec:pillar_lifetimes}).
The models by \cite{Bergin_1999}, which use physical conditions similar to those observed towards the pillar heads, indicate that $\chi_{\rm ion} < 10^{-8}$ is more typical of these massive cores; this would mean ambipolar diffusion weakens the magnetic field more quickly than expected.
Adopting this lower ionization fraction would yield $\tau_{\rm AD} \sim 0.2$~Myr, smaller than both the age of the system and the photoevaporative lifetime.
It is possible that any of the pillar heads may collapse to form stars before they are photoionized as ambipolar diffusion weakens the magnetic fields supporting them against gravitational collapse.
Protostars have been observed towards the heads of P1a and P2, so fragmentation and local collapse may have already taken place \citep{2002ApJ...565L..25S, 2007ApJ...666..321I, 2007ApJ...654..347L}.

The interplay between gravity, magnetic fields, and turbulence in context of cloud collapse is a complicated one which we simplify in our discussion here; see the review by \cite{Hennebelle2019FrASS...6....5H}.
The ambipolar diffusion timescale is slow compared to the dynamic timescales of clouds.
Quicker alternatives for overcoming magnetic support are available, such as turbulence or other methods of locally enhancing the mass-to-magnetic-flux ratio \citep{Vazquez-Semadeni2011MNRAS.414.2511V, Bailey2014ApJ...780...40B}.
If magnetic support were overcome more quickly, the magnetically supported and critical Pillar heads would be even more likely to fragment and collapse into stars before they evaporate.

\begin{deluxetable}{lcccc}
    \label{tab:pressure}
    \tablecaption{Pressures in the molecular, atomic, and ionized gas phases.}
    \tablewidth{0pt}
    \tablehead{
        \colhead{Component} & \colhead{$P_{therm}$} & \colhead{$P_{turb}$} & \colhead{$P_B$} & \colhead{$P_{tot}$}
    }
    \decimalcolnumbers
    \startdata
Molecular gas & 3 & 15 & 30 & 48 \\
Atomic PDR gas & 2--5 & 5 & 2* & 9--12 \\
Ionized flow & 18--36 & & & 18--36 \\
    \enddata
    \tablecomments{Pressures in the molecular, atomic, and ionized gas near P1a.
    Pressures are expressed as $P/k_{\rm B}$ in units of ($10^6$~K~cm$^{-3}$).
    Total pressures are calculated by summing across lower and upper estimates, respectively. The value ranges are intended as approximate estimates of our uncertainty in the pressures due to uncertain, or multiple, estimates of density or other properties. The atomic gas magnetic pressure (*) in this table is calculated by scaling down the magnetic field in the molecular gas by the ratio of mass densities in the two phases, but we suspect it is much larger as discussed in the text.}
\end{deluxetable}

\subsection{Constraints on Pillar Age} \label{sec:pressure-age}
Simulations by \cite{2001MNRAS.327..788W} indicate that the shock can pass through the dense, molecular pillar head in $10^5$~yr while the ionization front takes $\sim 0.5 \times 10^6$~yr to photoevaporate the pillar, comparable to the $\sim10^6$~yr age of the system.
Our observations are consistent with a shocked, equilibriated pressure structure, in which the shock passed through long ago and left the ionization front behind.
This relieves us of the tight upper limit of $\sim10^5$~yr on the pillar age required for unshocked pillar heads \citep{1999A&A...342..233W, 2001MNRAS.327..788W}.
Our estimated photoevaporation timescales $\sim10^6$~yr (Section~\ref{sec:pillar_lifetimes}) and the NGC~6611 cluster age $\sim\nexpo{2}{6}$~yr also favor the post-shock, $\sim$10$^6$-yr-old Pillars scenario.
See also the discussion of timescale tension by \cite{2001MNRAS.327..788W}.

%% file: discussion.tex
\subsection{Warm Gas Dynamics} \label{sec:warmgasblueshift}
Along lines of sight towards the Shared Base and the head of P2, the peak \cii, \oi, and CO line velocities are blueshifted by $\sim$0.5~\kms\ \wrt\ the denser molecular gas tracers.
The \cii\ line, but not the CO, is similarly blueshifted towards the head of P3.
This is inconsistent with a photoevaporative flow from an illuminated surface that is mostly facing away from us, but is consistent with gas flowing down these pillars away from the star (towards us in projection along the LOS).

Simulations by \cite{1994A&A...289..559L} predict that less dense, outer layers of gas move down the pillar bodies more quickly than denser gas (see their Figure 4c--f).
These simulations also predict blue line wings towards the head of a globule (oriented away from us, as P1b and P2 are) at two different evolutionary stages: a wing due to compressed gas on the far side of the head moving towards the observer during the first $\sim$0.1~Myr while the cloud is collapsing into a globule, and a wing during the stable cometary phase at an age of $\sim$1~Myr as small clumps are ejected and accelerated along the globule.
As we mention in Section~\ref{sec:p3_results}, the wishbone tails of P3 resemble the ``ears'' which emerge in these simulations; comparisons could be drawn between the later stages of these ears and the Threads.
Observational comparisons could be made between the Threads, ears, and tails and the ``legs'' observed by \cite{Bonne2023AJ....165..243B_IC63} towards IC~63.

Since \cii, \oi, and \twcoA\ should all trace warmer, outer layers of gas, the observed velocities can be explained if these layers are moving faster down the pillar than the colder, denser gas deeper in the pillar heads.
We, in our \cii\ and molecular lines, and \cite{1998ApJ...493L.113P} in CO and \cite{2015MNRAS.450.1057M} in optical ion lines, have observed radial velocity gradients along the Pillars which are thought to trace gas accelerated away from the star as it flows down the pillar.
Flow of gas away from the head down the body is predicted in simulations of certain pillar configurations by \cite{2010MNRAS.403..714M} and of cometary globules by \cite{1994A&A...289..559L}.
To explain our blueshifted outer layers gas, we need only require that the phenomenon acts more strongly on less dense gas, accelerating it more quickly near the head so that we see a relative velocity shift there but not further down the pillar body where all phases have been accelerated.
The driving force is unknown, but we suggest a few possibilities.
This could be a shearing flow generated by the action of stellar photons on the slanted pillar structure.
The action is transmitted in the warm PDR surface layers, so they respond first.
Deeper self-gravitating layers will be slowly carried along through shearing action.
Other candidates for the driving force could be the rocket effect or shocks driven by the ionization front, perhaps in combination with surface geometry.
Stellar winds, while a good candidate for transferring momentum radially away from the exciting stars, do not reach the surface of the pillars.
The striations detected in optical images by \cite{1996AJ....111.2349H} indicate that photoevaporated material flows away from the ionization front uninterrupted for at least a few tenths of a parsec, so the winds must terminate against the \hii\ region further away towards the stars (see the diagram in the left panel of Figure~19 in the paper by \citealt{2013MNRAS.435...30W}).

Towards the Eastern Thread, the \cii\ and \twcoA\ lines are blueshifted \wrt\ \csA\ and other molecules.
P1a faces towards, not away from, the observer, so this velocity shift is not well explained by these mechanisms.
The velocity shift towards the Eastern Thread may be due to an entirely different phenomenon than the velocity shifts towards the Shared Base, P2, and P3.

\subsection{Pillar Lifetimes} \label{sec:pillar_lifetimes}
We estimate the photoevaporative lifetime of the pillars based on our observations.
We adopt the equation for mass loss rate $\dot M = A v_f \rho_\hone$ from \cite{Gorti2002ApJ...573..215G}, where $A$ is the total evaporative surface area, $v_f$ is the photoevaporative flow velocity from the surface, and $\rho_\hone$ is the mass density at the base of the flow.

\input{table_lifetime}

Warm atomic gas should flow along the pressure gradient through the PDR towards the ionization front at a $v_f$ comparable to, but no greater than, the sound speed in the atomic gas $c_\hone = (P_{\hone,therm}/\rho_\hone)^{1/2} \sim1~\kms$.
The observed \cii\ velocity dispersion gives an upper limit $v_f \lesssim \sigma_{\rm obs} \approx 1.4~\kms$, though we attribute most of the observed velocity dispersion to turbulence (Section~\ref{sec:pressures}).
We adopt $v_f \sim 0.5~\kms$.

Mass density at the base of the flow $\rho_\hone = \mu m_\hone n_\hone$, where we use $\mu = 1.33$ and the atomic gas density $n_\hone = \nexpo{1.8}{4}$~cm$^{-3}$ from Section~\ref{sec:density}.
Evaporative surface area $A$ is estimated based on the on-sky sizes of the Pillars in high-resolution optical and NIR images.
In order to better understand the evolution of the Pillars, we estimate the evaporative lifetimes of the heads and bodies separately.
Evaporation of low column density regions towards the pillar bodies is of particular interest, since we detect significant \cii\ line and PAH emission from these regions.
We define the pillar ``necks'' as the lower column density segments of the body between the head and any dense clumps along the body.
We use the lack of detected \ntwohpA\ emission to distinguish the neck from the head and rest of the body.
For P1a, the neck comprises both Threads, and for P2 the neck extends between the pillar head and the clump.
We do not resolve P3 well enough to define a neck.
The boxes used to integrate the head and neck masses are shown in the rightmost panel of Figure~\ref{fig:columndensity}.
We model pillar heads as hemispheres and necks as cylinders, and we select a single typical radius for both components of each pillar based on the angular sizes in Table~\ref{tab:columndensity}.
The height of the neck is the length of the box used to integrate its mass.
The adopted dimensions and corresponding mass loss rates are listed in Table~\ref{tab:lifetimes-head-neck}.
The surface area is the only variable changed from region to region, as the number densities in Table~\ref{tab:columndensity} do not vary by more than a factor of a few and the total \cii\ velocity dispersion only varies by $\sim$50\%.
Column densities calculated using \cp\ and CO are summed pixel-by-pixel within the boxes to find the atomic and molecular gas masses, and the total mass of each feature is considered to be the sum of the mass in both phases.
The photoevaporative lifetime of each feature is the mass divided by the mass loss rate.
The masses and photoevaporative lifetimes are listed in Table~\ref{tab:lifetimes-head-neck}.

The pillar heads will evaporate in $\sim$1--2~Myr by our estimate, consistent with the estimate by \cite{2015MNRAS.450.1057M} of $\sim$3~Myr for these pillars as well as the estimate by \cite{2013MNRAS.435...30W} of $\sim$2~Myr for the pillars surrounding NGC~3603.
This is similar to the $\sim$2~Myr age of the NGC~6611 system, so the Pillars likely formed with roughly a few times as much mass as they have now and have photoevaporated over the last $\sim$Myr down to their present-day masses.
As we discussed in Section~\ref{sec:pressures}, the photoevaporative lifetime is at least as large as $\tau_{AD}$, so the heads of the Pillars may still collapse to form stars before they are completely photoevaporated.

The pillar necks appear to evaporate on a shorter timescale of $\sim$0.1~Myr due to their much larger surface areas.
If we have overestimated their mass loss rates, the culprit could be a diminished flow velocity which is obscured in the observed velocity dispersion by the more dominant turbulent velocity dispersion.
In the case of Pillar 1, the flow velocity would have to be diminished by a factor of $\sim$10.
We do observe thinner (by $\lesssim $0.5~\kms) line profiles towards the Threads in P1a and the neck of P2, so this order of magnitude decrement in flow velocity between the head and the neck is possible and must be observationally confirmed with, for example, velocity resolved radio recombination lines or optical ion lines.
$G_0$, measured from FUV emission, is diminished only by $\sim$60\% from the Head to the Neck of P1a, and by $\sim$30\% in P2.

We briefly consider the possibility that the necks do evaporate $\sim$10x more quickly: either they will evaporate and cause the dense clumps towards the heads to appear as cometary globules like those observed towards Cygnus~X \citep{2016A&A...591A..40S} or at the 0.01~pc scale elsewhere in M16, or they are being replenished by bulk gas flows from the head like those discussed in Section~\ref{sec:warmgasblueshift}, which may hasten the dissipation of the pillar heads by a factor of $\sim$2.
The meaningful difference between the two evolutionary scenarios is whether the Pillars are observed as globules or coherent pillars for the last $\sim$1~Myr of their lives; the work of \cite{2016A&A...591A..40S} suggests the former.

%% file: table_lifetime.tex

\begin{deluxetable*}{lcccccccc}
    \label{tab:lifetimes-head-neck}
    \tablecaption{Pillar head and neck masses, mass loss rates, and photoevaporative lifetimes.}
    \tablewidth{0pt}
    \tablehead{
        \colhead{Feature} & \colhead{R(, H)} & \colhead{R(, H)} & \colhead{A} & \colhead{$M_{H}$} & \colhead{$M_{H_{2}}$} & \colhead{$M_{\rm tot}$} & \colhead{$\dot M$} & \colhead{$t$} \\
        \colhead{Name} & \colhead{(\arcsec)} & \colhead{(pc)} & \colhead{(pc$^{-2}$)} & \colhead{($M_{\odot}$)} & \colhead{($M_{\odot}$)} & \colhead{($M_{\odot}$)} & \colhead{($M_{\odot}$~Myr$^{-1}$)} & \colhead{(Myr)}
    }
    \decimalcolnumbers
    \startdata
        P1a Head &
        20 & 0.17 & 0.18 &
        15 & 72 & 86 &
        54 & 1.6 \\
        P1a Neck &
        20, 38 & 0.17, 0.32 & 0.34 &
        8 & 7 & 15 &
        103 & 0.1 \\
        P2 Head &
        10 & 0.08 & 0.04 &
        2 & 14 & 16 &
        14 & 1.2 \\
        P2 Neck &
        10, 28 & 0.08, 0.24 & 0.13 &
        3 & 21 & 24 &
        38 & 0.6 \\
        P3 Head &
        7 & 0.06 & 0.02 &
        1 & 6 & 7 &
        7 & 1.0 \\
    \enddata
    \tablecomments{Pillar heads are modeled as hemispheres and necks as cylinders; the R(, H) columns give radius and, if applicable, height. The adopted height for each pillar neck matches the height of the box used to sum over column density.
    Mass values are rounded to the nearest whole number.
    Adopted atomic gas density at the base of the photoevaporative flow is $n_{\rm H} = \nexpo{1.8}{4}$~cm$^{-3}$. Adopted flow velocity is 0.5~\kms.
    }
\end{deluxetable*}

%% file: conclusion.tex
We observed the Pillars of Creation in M16 in velocity-resolved \cii\ and \oi\ for the first time and detected PDRs which are spatially and kinematically associated with the molecular gas.
We combine these data with observations of APEX, CARMA, BIMA, \textit{Herschel}, \textit{Spitzer}, and JWST to obtain a multi-layer view of these gas structures.

The largest pillar, P1, is the site of the brightest emission from both molecular gas and PDR tracers.
It is separated into two substructures, P1a and P1b, which roughly correspond to its head and its base, respectively.
These substructures are somewhat separated along the line of sight, though still linked by kinematically continuous \cii\ emission.
In P1a, the molecular tracers CO, \hcnA, \hcopA, and \csA\ reveal two dense filamentary structures which we term the ``Threads'' running parallel to the pillar body and terminating in a dense ``Cap'' at the head of the pillar.
Our kinematic modeling indicates that these three distinct components are connected towards a region of high column density in the pillar head.
Below the Threads past a region of low column density lies P1b, which is linked to the base of P2 by a bright, high column density PDR called the Shared Base.
P2, a $\sim$1~pc long pillar which lies closer to the observer than P1, features a dense clump at its head and another halfway down its body.
P2 and P3 are dwarfed in brightness by the PDRs towards P1.
We include in our studies a handful of nearby PDR host structures, namely the \shelf\ and P4, which are kinematically related to the Pillars.

This wealth of data allows us to develop a geometric model for the structure of this pillar system, summarized by the schematic in Figure~\ref{fig:schematic3d}.
Velocity gradients along the Pillars indicate that P1a and P3 point towards the observer while P2 points away.
The relative strength of diffuse foreground ionized gas emission places P1a farthest from the observer, and P2 and P3 on the near side \wrt\ the bright O stars.
P2 and P3 are darker optical sources and so their illuminated sides must face away from the observer.
All of this is consistent with the conclusions of \cite{1998ApJ...493L.113P} and \cite{2015MNRAS.450.1057M}.
We establish from kinematic and column density information that P2 and P1b are connected by the Shared Base, a PDR which is significantly extended along the line of sight.

The \cii\ line traces the same parsec-scale spatial and dynamic patterns as the molecular lines, exhibiting the same velocity gradients along P1 and P2 (P3 is too dim in \cii\ and not well enough resolved to confirm or rule out a gradient).
At smaller scales, \cii\ and \oi\ emission is generally smoother and more continuous than molecular emission, particularly towards the Threads in P1a, so we conclude that a PDR layer can manifest as a common envelope around dense features like the Threads.
The \cii\ and \twco\ lines are blueshifted by $\sim$0.5~\kms\ w.r.t. the rest of the molecular lines towards the Shared Base and the heads of P2 and P3.
We suggest that this is caused by the same process as the observationally established acceleration of gas away from the stars down the bodies of pillars \citep{1998ApJ...493L.113P, 2015MNRAS.450.1057M} and that the process acts more strongly on the less dense gas in the outer layers of pillars.
These motions are similar to those simulated by \cite{1994A&A...289..559L} and \cite{2010MNRAS.403..714M} and may be a shear flow of the outer layers down the surface of the pillars.

We study the illumination based on both cataloged cluster members and FIR dust emission and constrain the FUV intensity to $G_0 \sim 10^3$~Habing units towards the Pillars.
The head of Pillar 1 is characterized by a relatively high $G_0 \sim 2000$~Habing units and good agreement between the stellar and FIR estimates, indicating that it is directly illuminated and the projected distance to the stars is close to the true distance.

We determine the average molecular gas density in the Pillars to be $\sim\nexpo{1.3}{5}$~cm$^{-3}$ and the average atomic gas density to be $\sim\nexpo{1.8}{4}$~cm$^{-3}$.
The integrated column densities towards the Pillars yield total masses of 103, 78, 103, and 18 $M_{\odot}$ for P1a, P1b, P2, and P3 respectively.
The atomic gas is typically $\sim$25\% of the total mass, with the exception of P1b which contains about half its mass in the atomic phase.
The masses of the Pillars are concentrated in the heads and occasional clumps along the bodies, such as the Horns atop P1b and the clump halfway down P2.
These clumps are illuminated, as we detect PDR emission from their cluster-facing surfaces.

Based on pressures derived from our number densities, we assume that the pillars are in pressure equilibrium with the ionized gas around them and that the PDRs associated with the Pillars are magnetically supported.
In addition, the magnetic pressure of the dense molecular gas seems to be supporting the Pillar heads and clumps against gravity, but we infer that ambipolar diffusion is reducing the magnetic field support over a $\sim$1~Myr timescale.

We estimate that the heads of each pillar will be photoevaporated by radiation from NGC~6611 in $\sim$1--2~Myr.
Whether they remain coherent pillars or evolve into free-floating globules as they evaporate may indicate whether or not flows of gas from the head into the body are a significant mass loss pathway compared to photoevaporation.
Improved understanding of the evolutionary phases of pillars will reveal their relationships to cometary globules and other categories of observed features.

The Pillars' column shapes are to some extent a shadow of the concentrated mass at their heads, but it is clear from our observations and analysis that there are pre-existing density enhancements all along their bodies.
The structure and orientation of the Pillars are affected by the pre-existing structure in the molecular cloud which is sculpted by the radiation field when massive stars turn on.

%% file: measurements.tex
\begin{figure*}
    \centering
    \includegraphics[width=\textwidth]{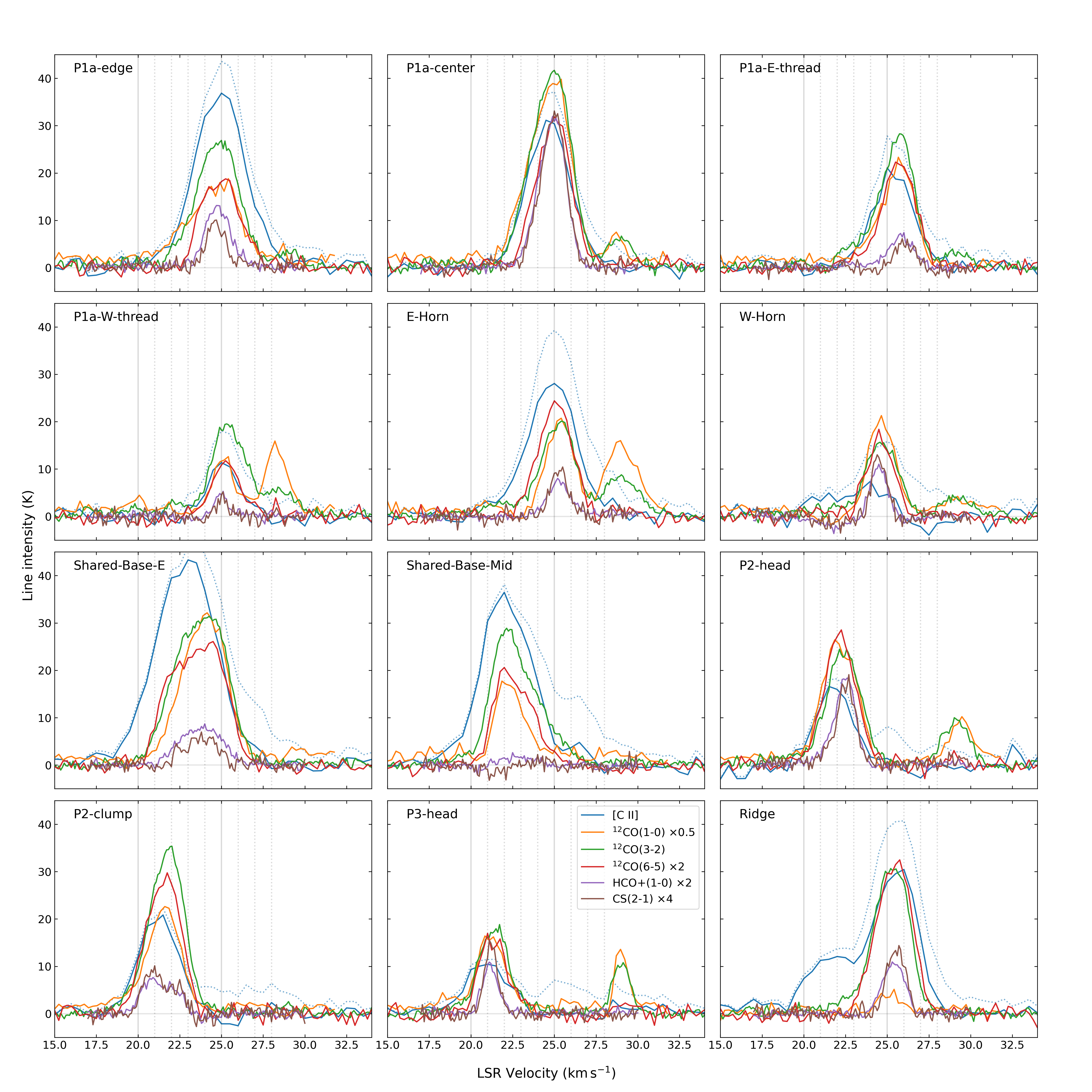}
    \caption{Spectra observed towards the pillars in several lines. All data are convolved to the \cii\ resolution, except CO(\jmton{3}{2}) which is shown at its native resolution. Locations are those shown in the leftmost panel of Figure~\ref{fig:columndensity} and listed by name in Tables~\ref{tab:columndensity} and \ref{tab:measurements}. The \cii\ line spectra are shown with a dotted line before and a solid line after subtracting out the background as described in Appendix~\ref{sec:systematic-cii-background}; the second column in Table~\ref{tab:columndensity} lists whether the northern or southern background was subtracted. Vertical lines mark every 1~\kms\ between 20--28~\kms. Several lines/locations show additional components separated from the peak emission, such as CO at $\vlsr \sim 29$~\kms\ towards multiple positions or \cii\ at $\vlsr \sim 20$--23~\kms\ towards the \shelf; all such components originate from background features identified in channel maps.}
    \label{fig:spectra}
\end{figure*}
\begin{figure*}
    \centering
    \includegraphics[width=\textwidth]{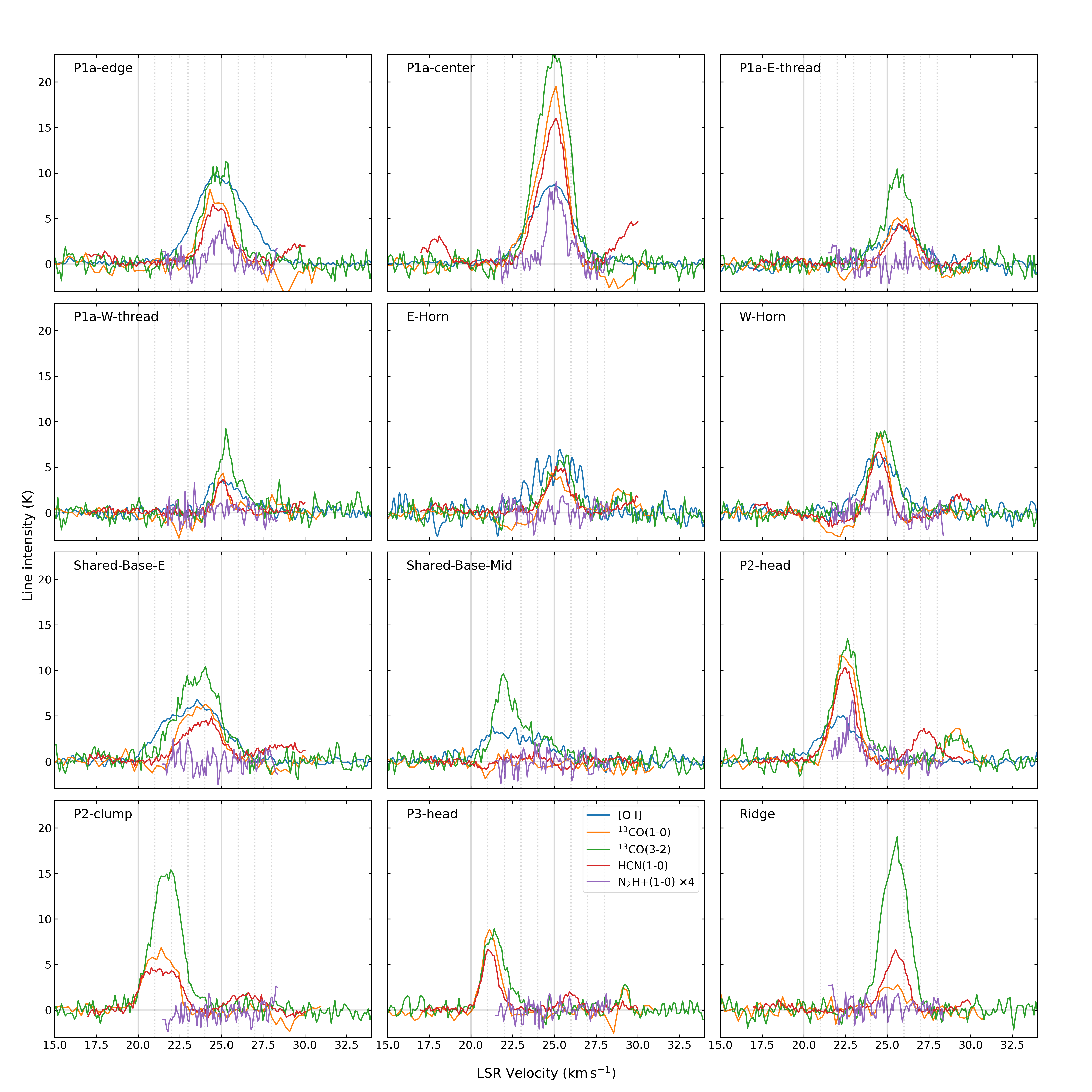}
    \caption{Same as Figure~\ref{fig:spectra} showing different species. All data except \thcott\ are convolved to the \cii\ resolution. The line intensity axis spans a smaller range here. Multiple satellite lines appear in the \hcnA\ spectra. The \ntwohpA\ line is the satellite line (J, F1, F) $ = $ (1--0,~0--1,~1--2). The CO emission around $\vlsr \sim 29$~\kms\ is from a background feature.}
    \label{fig:spectra2}
\end{figure*}

Table~\ref{tab:coordinates} lists coordinates and Table~\ref{tab:measurements} lists measured main beam temperatures for the 12 locations for which we give column and number densities in Table~\ref{tab:columndensity}. Their positions are shown in the leftmost panel of Figure~\ref{fig:columndensity}.
The Table~\ref{tab:measurements} measurements are all given at their native resolutions.
Figures~\ref{fig:spectra} and Figure~\ref{fig:spectra2} show most line spectra towards these locations.
The spectra are all shown convolved to the \cii\ beam (15.4\arcsec), except for the CO(\jmton{3}{2}) lines which are shown at their native resolutions ($\sim$20\arcsec).

\begin{deluxetable*}{lllc}
    \label{tab:coordinates}
    \tablecaption{Sample location information.}
    \tablewidth{0pt}
    \tablehead{
        \colhead{Location} & \colhead{Right Ascension} & \colhead{Declination} & \colhead{} \\
        \colhead{Name} & \colhead{(hh:mm:ss)} & \colhead{(dd:mm:ss)} & \colhead{Background}
    }
    \decimalcolnumbers
    \startdata
P1a-edge        &  18\hh18\mm50\fs1147 & -13\arcdeg48\arcmin50.286\arcsec & N \\
P1a-center      &  18\hh18\mm51\fs0331 & -13\arcdeg48\arcmin56.830\arcsec & N \\
P1a-E-thread    &  18\hh18\mm52\fs9684 & -13\arcdeg49\arcmin08.407\arcsec & N \\
P1a-W-thread    &  18\hh18\mm52\fs3583 & -13\arcdeg49\arcmin22.366\arcsec & N \\
E-Horn          &  18\hh18\mm54\fs8134 & -13\arcdeg49\arcmin36.870\arcsec & S \\
W-Horn          &  18\hh18\mm53\fs6043 & -13\arcdeg49\arcmin58.878\arcsec & S \\
Shared-Base-E   &  18\hh18\mm55\fs2294 & -13\arcdeg50\arcmin07.362\arcsec & S \\
Shared-Base-Mid &  18\hh18\mm55\fs2857 & -13\arcdeg50\arcmin32.292\arcsec & S \\
P2-head         &  18\hh18\mm49\fs3745 & -13\arcdeg49\arcmin57.146\arcsec & N \\
P2-clump        &  18\hh18\mm51\fs5274 & -13\arcdeg50\arcmin28.378\arcsec & N \\
P3-head         &  18\hh18\mm49\fs2048 & -13\arcdeg50\arcmin43.792\arcsec & N \\
\shelf\         &  18\hh18\mm58\fs0793 & -13\arcdeg51\arcmin27.241\arcsec & S \\
    \enddata
    \tablecomments{Coordinates are J2000. Background column lists whether the northern or southern background sample was used for column density (Section~\ref{sec:coldens-cii}) and spectral (Appendix~\ref{sec:systematic-cii-background}) subtraction.}
\end{deluxetable*}

\begin{deluxetable*}{lrrrrrrrrrrrr}
    \label{tab:measurements}
    \tablecaption{Peak T$_{\rm MB}$ for each line.}
    \tablewidth{0pt}
    \tablehead{
        \colhead{Location Name} & \colhead{\cii} & \colhead{\oi} & \colhead{CO(1-0)} & \colhead{$^{13}$CO(1-0)} & C$^{18}$O(1-0) & \colhead{CO(3-2)} & \colhead{$^{13}$CO(3-2)} & \colhead{CO(6-5)} & \colhead{\hcnA} & \colhead{\hcopA} & \colhead{\csA} & \colhead{\ntwohpA}
    }
    \decimalcolnumbers
    \startdata
P1a-edge        &  43.6 &   13.4 &        43.8 &         12.0 &          0.7 &        26.9 &         11.2 &          9.5 &    6.1 &     6.2 &   2.2 &        1.8 \\
P1a-center      &  37.1 &   11.5 &       104.2 &         27.5 &          3.3 &        41.7 &         23.1 &         19.5 &   22.6 &    22.1 &  12.8 &        3.6 \\
P1a-E-thread    &  27.8 &    7.1 &        51.0 &         13.9 &          0.5 &        28.3 &         10.4 &         13.1 &    6.6 &     5.6 &   3.2 &        1.6 \\
P1a-W-thread    &  18.1 &    7.5 &        52.8 &         18.3 &          0.9 &        19.5 &          9.2 &          5.1 &    6.9 &     4.9 &   2.8 &        1.1 \\
E-Horn          &  39.3 &    7.6 &        81.1 &         16.3 &          0.4 &        20.1 &          6.3 &         16.7 &    9.1 &     7.1 &   6.4 &        1.3 \\
W-Horn          &  15.8 &   10.8 &        82.9 &         28.7 &          1.5 &        15.7 &          9.1 &         11.5 &   12.8 &    10.7 &   7.9 &        2.2 \\
Shared-Base-E   &  47.9 &    8.6 &        80.7 &         11.7 &          1.3 &        31.4 &         10.5 &         17.0 &    6.5 &     6.1 &   2.9 &        1.3 \\
Shared-Base-Mid &  38.1 &    6.0 &        33.7 &          5.4 &          0.6 &        28.9 &          9.6 &         11.5 &    1.7 &     2.0 &   1.9 &        1.2 \\
P2-head         &  18.2 &    8.0 &        68.6 &         20.9 &          1.8 &        24.5 &         13.5 &         17.3 &   14.6 &    14.1 &   7.0 &        3.7 \\
P2-clump        &  22.2 &\nodata &        47.8 &         17.4 &          1.3 &        35.4 &         15.4 &         15.9 &    7.3 &     7.0 &   5.3 &        1.3 \\
P3-head         &  11.5 &\nodata &        59.2 &         26.6 &          1.3 &        18.8 &          8.9 &          9.7 &   11.8 &    10.3 &   8.1 &        1.4 \\
\shelf\         &  40.7 &\nodata &     \nodata &      \nodata &      \nodata &        30.6 &         19.0 &         20.4 &    5.7 &     5.5 &   2.5 &        1.4 \\
    \enddata
    \tablecomments{Peak main beam temperatures in Kelvins from lines at their native resolutions, which are listed alongside RMS temperatures in Table~\ref{t-mapsummary}. The positions are shown in the leftmost panel of Figure~\ref{fig:columndensity}. Since these are calculated as the maximum value from each spectrum, a measurement 1 or 2 times the RMS noise in Table~\ref{t-mapsummary} should be regarded as a potential nondetection; the spectra in Figures~\ref{fig:spectra} and \ref{fig:spectra2} provide additional context. The \ntwohpA\ measurements are from the satellite line (J, F1, F) $ = $ (1--0,~0--1,~1--2) since the brightest lines at the center of the bandpass overlap (see Section~\ref{sec:observations}). The \shelf\ is outside the half-power level of BIMA's primary beam, so the CO(\jmton{1}{0}) measurements there are not given. The listed \cii\ measurements have not had any background subtracted from them.}
\end{deluxetable*}

%% file: systematic_background.tex
\begin{figure*}
    \centering
    \includegraphics[width=0.8\textwidth]{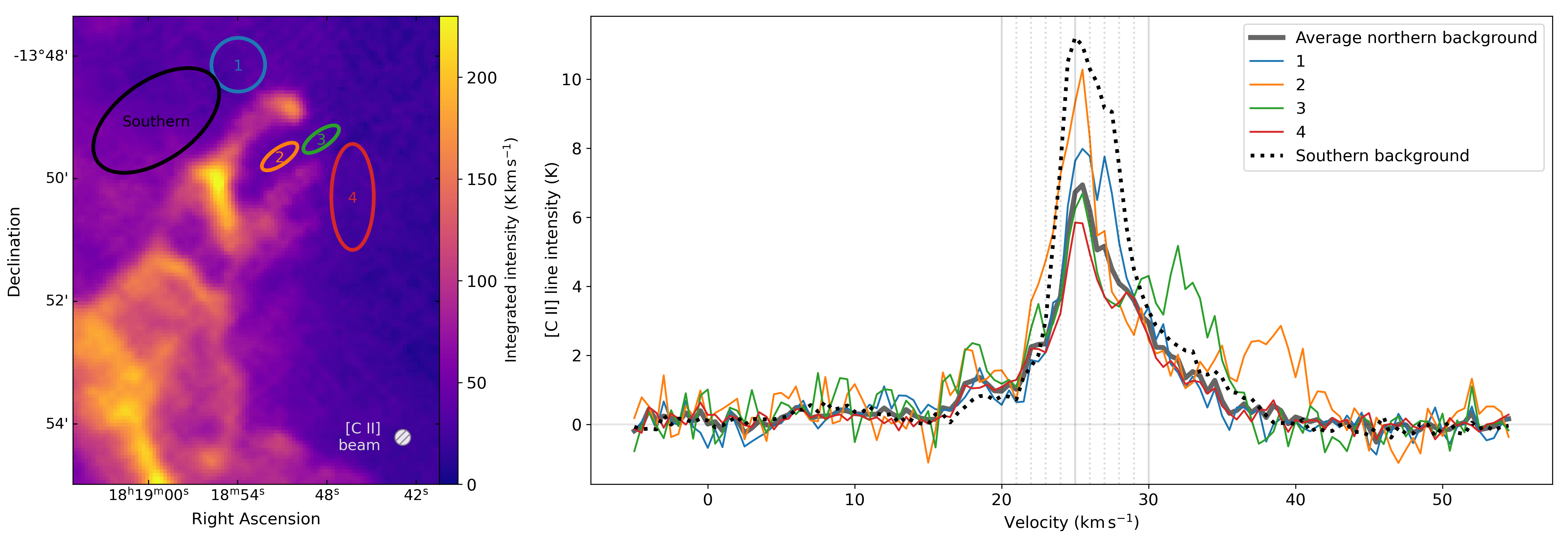}
    \caption{\textit{(Left)} The \cii\ integrated intensity between $\vlsr=18$-27~\kms, the same interval shown earlier in Figure~\ref{fig:moment_panels}, is shown in color. The four northern background regions (numbered) and the single southern background region are outlined.
    \textit{(Right)} The mean spectrum from within each individual background region on the left panel, labeled accordingly. The ``Average'' background spectrum drawn in the bold, solid line is the average across all pixels in the numbered northern regions and does not include the southern region. Vertical lines mark every 1~\kms\ between 20--30~\kms.}
    \label{fig:background-spectra}
\end{figure*}

Key to understanding the \cii\ line spectra through the Pillars is understanding the atomic gas that surrounds the them.
Foreground and background elements may appear in the spectra towards the Pillars, and must be acknowledged and accounted for in any meaningful interpretation of Pillar spectra.
Around the velocity of P1, $\vlsr \sim 25~\kms$, the surrounding area is also bright in \cii, particularly east of the Pillars as seen in the channel maps in Figure~\ref{fig:cii_channel_maps}.
The spatial extent of this component indicates that $\vlsr \sim 25~\kms$ may be a region-scale ``systematic velocity'' related to the bulk velocity of the clouds from which the star cluster was born.

We investigate the background \cii\ spectrum around the Pillars by integrating spectra within several handpicked, parsec-scale regions shown in Figure~\ref{fig:background-spectra}, selected for their proximity to the Pillars and, for the four northern regions, lack of distinct morphological features in the channel maps.
The southern region is selected for its proximity to P1b and inclusion of the diffuse feature which we suspect lies in the foreground/background of P1b and other ``southern'' features.
The regions are kept at least one beam (15\arcsec) away from the optical/NIR edges of the Pillars, which ensures that the regions don't overlap with the Pillars and include Pillar emission while remaining close to the Pillars so that the background spectra are relevant to the Pillars.
The \cii\ spectra from within these regions include significant emission around $\vlsr \sim 25~\kms$, demonstrating that the Pillars are not the only source of emission in those channels.

The peaks of the background line profiles are thin ($\sim$1--3~\kms), which indicates that most of this emission originates from the atomic gas \citep{Cuadrado2019A&A...625L...3C}.
The ionized \hii\ phase contributes $\sim$10\% of the total \cii\ emission on large ($\sim$500~pc) scales \citep{Tarantino2021ApJ...915...92T}.
\cii\ line profiles from the diffuse, ambient \hii\ region should have $\sim$10~\kms\ widths reflective of the turbulent velocity dispersion observed by \cite{Higgs1979AJ.....84...77H} towards the M16 \hii\ region in lines of H and He.
The background \cii\ line profiles have wings that could arise from a wide, low-intensity component originating from the ambient \hii\ region.
This will be explored in greater detail in a future study.
The characteristic PDR velocity $\vlsr \sim 25$--26~\kms\ of the region surrounding the Pillars may originate from pre-cluster dynamics and poses a unique challenge for the analysis of P1, which lies almost exactly at this velocity.

On a more local scale around the Pillars, this diffuse emission is even brighter towards the southern end of P1, near P1b.
Channel maps between $\vlsr \sim 24$--28~\kms\ indicate structure immediately east of P1b, as it is not clear if the structure is related to the Pillar system.
This feature has a wider line profile than the northern background samples, as we see in the comparison in Figure~\ref{fig:background-spectra}.

In order to mitigate adverse effects of this background on our analysis of the Pillars, we conduct a background sampling and subtraction in order to help isolate \cii\ emission from the Pillars themselves.
We use the average of all pixels in four northern regions, drawn with the bold, solid line on the right panel of Figure~\ref{fig:background-spectra}, to correct ``northern'' targets such as P1a, P2, and P3.
We use the average from within the southern region to correct spectra towards P1b and the \shelf; see the Background column of Table~\ref{tab:coordinates}.
These corrections are made for all \cii\ spectra we show going forward and particularly throughout the kinematic analysis in Section~\ref{sec:geomdyn} where the shape of the line is important to our analysis.
We do not account for optical depth while subtracting background spectra; we find in Section~\ref{sec:coldens-cii} that \cii\ has an optical depth $\lesssim 1$, so adverse effects of background subtraction on line shape should be minimal.
We do not make a channel-by-channel spectral subtraction while calculating \cp\ column densities in Section~\ref{sec:coldens-cii}.

Some distinct components are identified in the background spectra from Regions \#2 and \#3 around 30--40~km/s in Figure~\ref{fig:background-spectra}.
Referencing the channel maps in Figure~\ref{fig:cii_channel_maps}, these can be associated with a morphologically distinct north-to-south strip of redshifted gas which does not appear directly associated with the Pillars.
The influence of this redshifted component is not seen in the averaged background spectrum, and the feature itself has very little overlap with P1, so it should not adversely affect our background subtraction.

The background subtraction proves to be useful in determining whether characteristics of the \cii\ line profile should be associated with the Pillars.
The molecular and (unsubtracted) \cii\ spectra through the pillar bodies, and particularly through the head of P1 in Figure~\ref{fig:spectra}, differ noticeably in that the \cii\ line profile has more emission at higher velocities than the other tracers.
No shift is detected in the \cii\ line center w.r.t. the molecular lines, so the asymmetry in the line profile must be limited to the low-intensity line wings.
If we were to associate all \cii\ emission towards the head of P1 with the pillar itself, then we would conclude that the PDR layer has a significantly different kinematic signature than the deeper molecular gas layers, thus resulting in an extended red tail in the \cii\ spectrum.
However, when we apply the \cii\ background subtraction, we find that the redshifted \cii\ tail disappears and the \cii\ spectrum looks much more like the molecular lines, suggesting that the dynamics of the PDR and molecular layers are not dissimilar.
The background identification and subtraction is therefore necessary in order to identify Pillar-related spectral characteristics and avoid ascribing background emission to the Pillars.

Jackknife tests using combinations of these and other background samples confirm that the choice of background samples doesn't upset results, i.e. change the profile and velocity of lines too much, but reveal limitations in our ability to make specific and precise claims about background-subtracted \cii\ line intensities since the background varies throughout the system (see the velocity shift between Regions \#1 and \#2 in Figure~\ref{fig:background-spectra}).

%% file: kinematic_modeling.tex
P1a is composed of three morphologically and kinematically distinct components, making it the most complex structure in the pillar system.
A 3D geometrical model, informed by an in-depth kinematic analysis of P1a, is necessary to contextualize other physical characteristics derived from our observations.
We fit Gaussian line profiles to observed \cii\ and molecular line spectra in order to quantify, or at least approximate, characteristics of the observed profiles.
In advance of this kinematic modeling, we subtract channel-by-channel from \cii\ spectra the background identified near the Pillars.
We describe the identification and correction in Appendix~\ref{sec:systematic-cii-background}.
Since we only model spectra towards P1a, we only use the northern background sample described in Appendix~\ref{sec:systematic-cii-background}.

\subsection{Kinematic Modeling Towards P1a} \label{sec:appdx-kin-modeling}
We fit 1, 2, 3, and 4 component models to each \hcopA\ and \cii\ spectrum in a $\sim1$' box surrounding P1a ($\sim$500 \cii\ spectra, $\sim$10,000 \hcopA\ spectra).
For each line, all spectra were modeled using the same model template and initial conditions.
The strength of this unsupervised pixel grid fit is that real, physically meaningful patterns may emerge from this large number of modeled spectra.
We visualize the fitted models in (RA, Dec, Velocity) space by plotting each pixel's fitted component line centers.
There are 1--4 fitted components per pixel depending on the model template and the number of components with sufficiently large amplitudes \wrt\ the noise.
We plot these components using a 2D histogram by projecting along Dec to view the RA-Velocity plane; due to the components' spatial orientations, we find the RA-Velocity projection easier to interpret than the Dec-Velocity projection and so we limit our discussion to the former.
These histograms are similar in orientation and meaning to PV diagrams.

\begin{figure*}
    \gridline{\fig{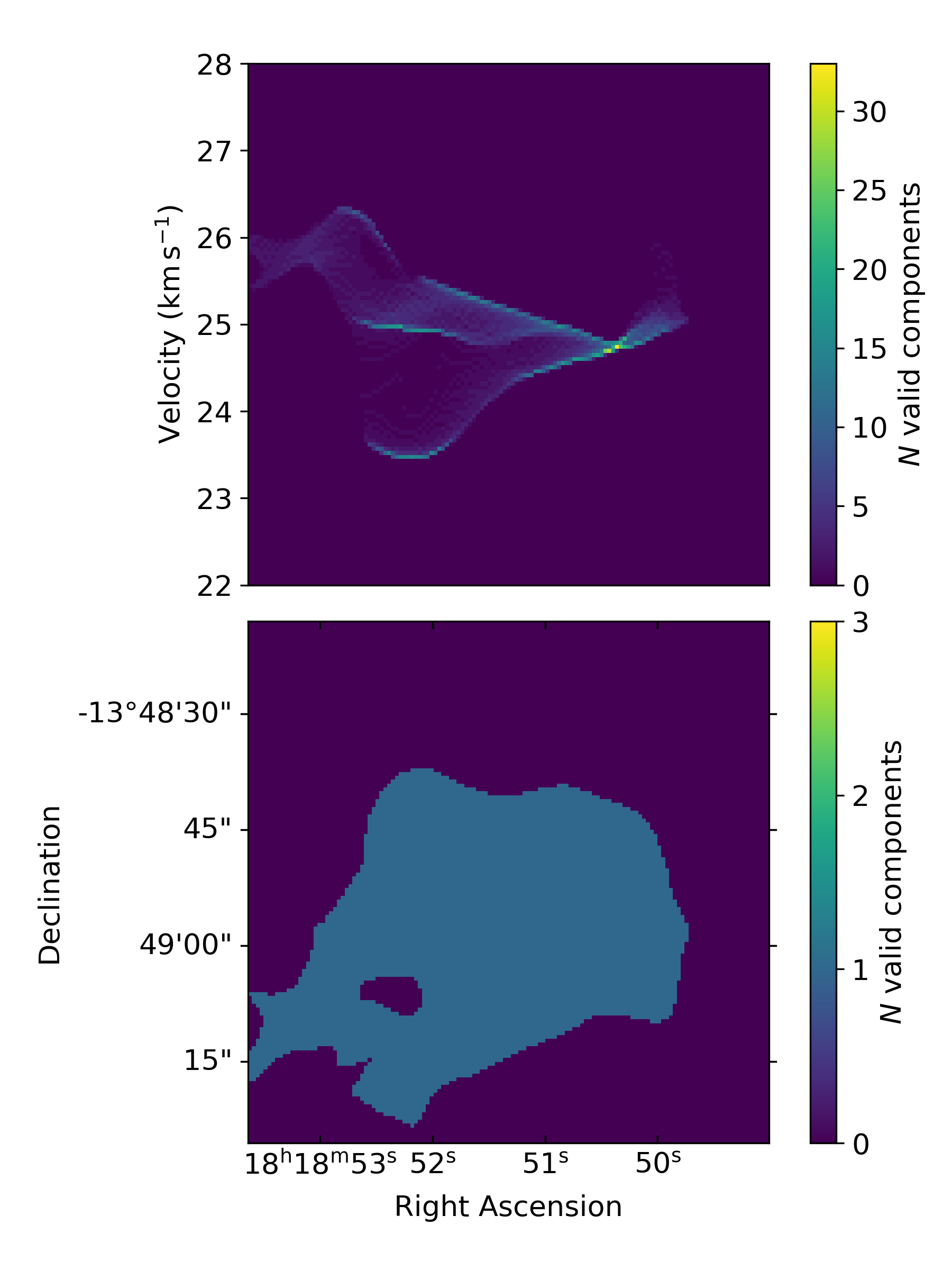}{0.3\textwidth}{\hcopA\ 1 component}
              \fig{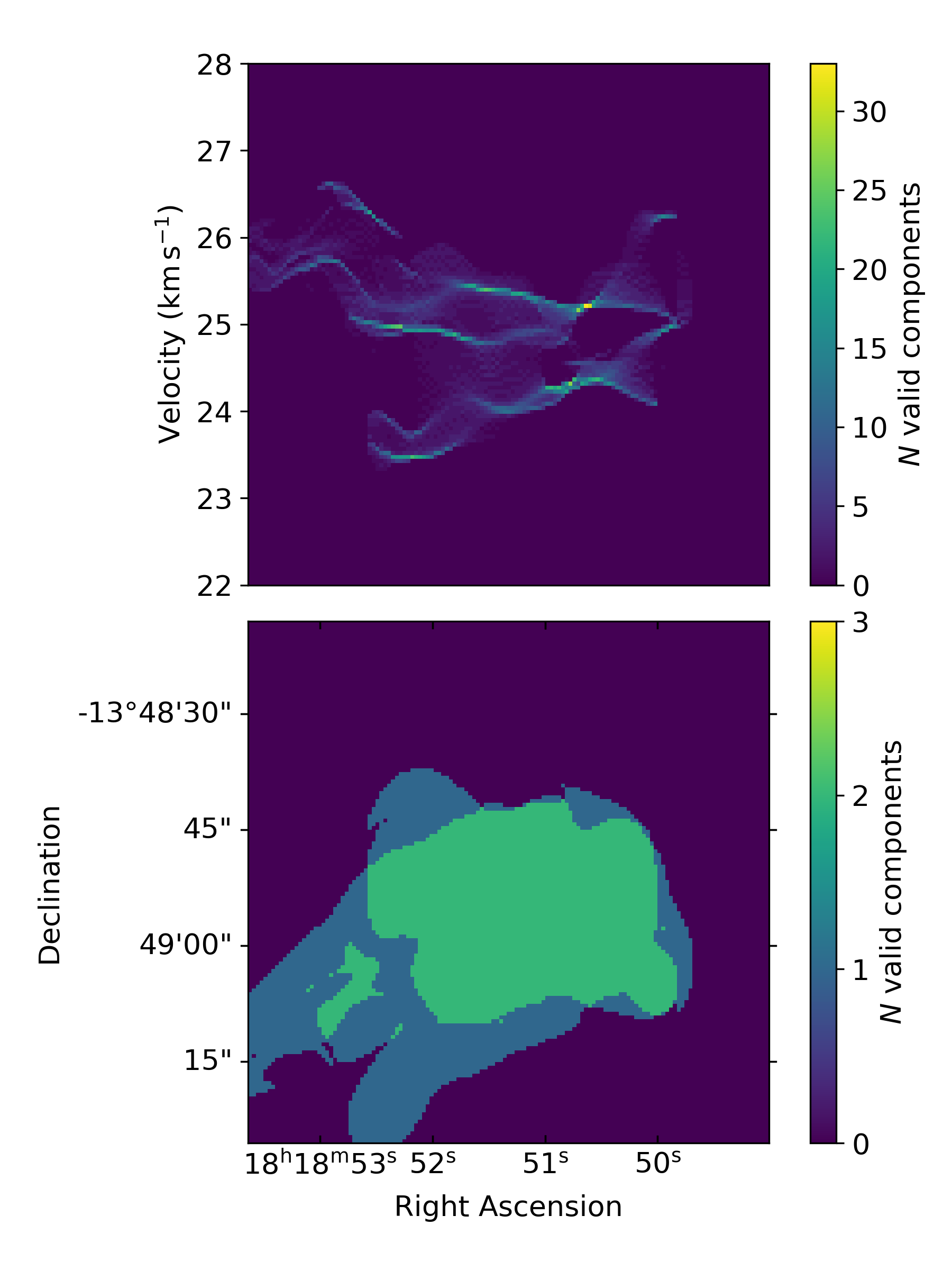}{0.3\textwidth}{\hcopA\ 2 component}
              \fig{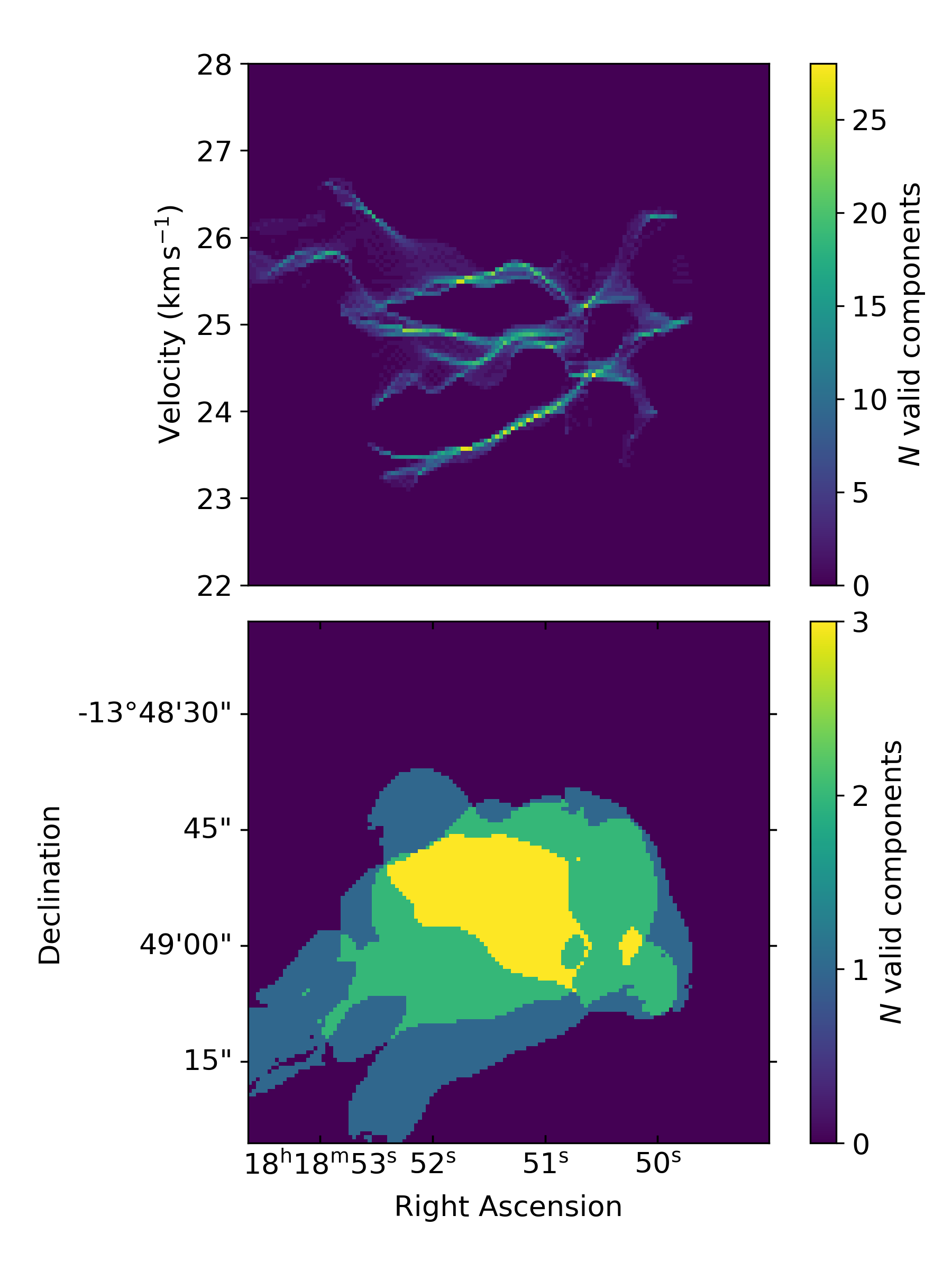}{0.3\textwidth}{\hcopA\ 3 component}}
    \gridline{\fig{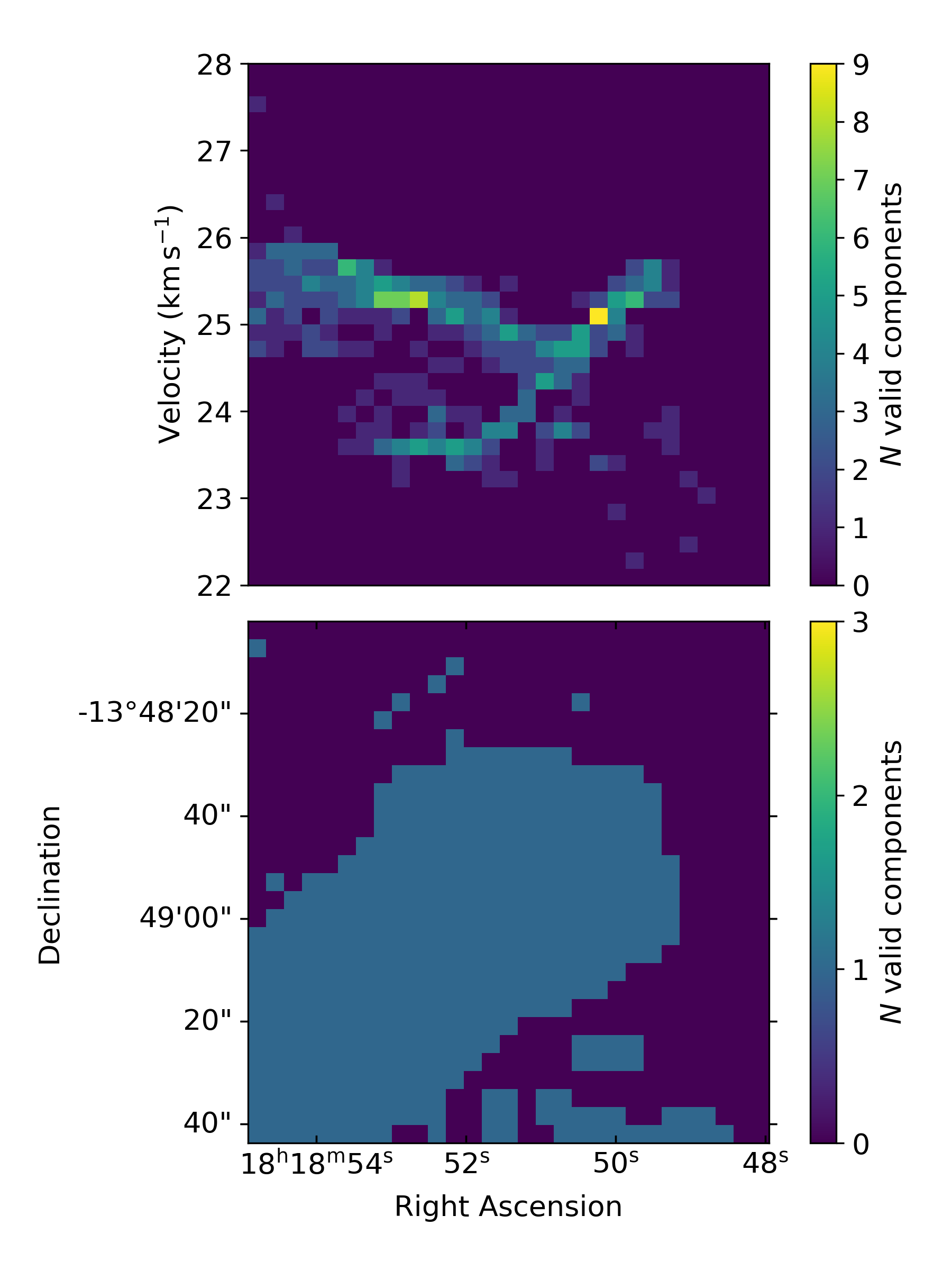}{0.3\textwidth}{\cii\ 1 component}
              \fig{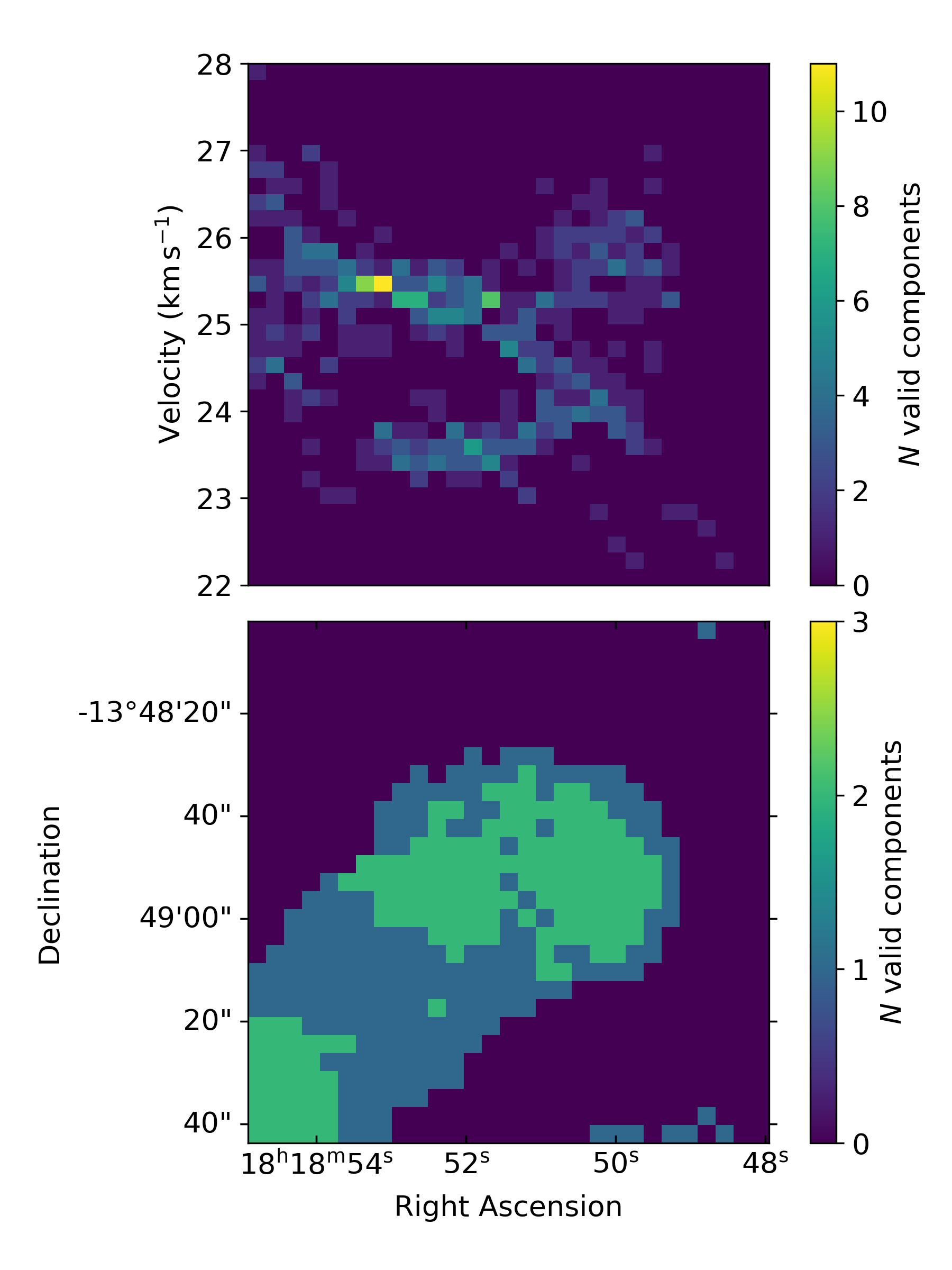}{0.3\textwidth}{\cii\ 2 component}
              \fig{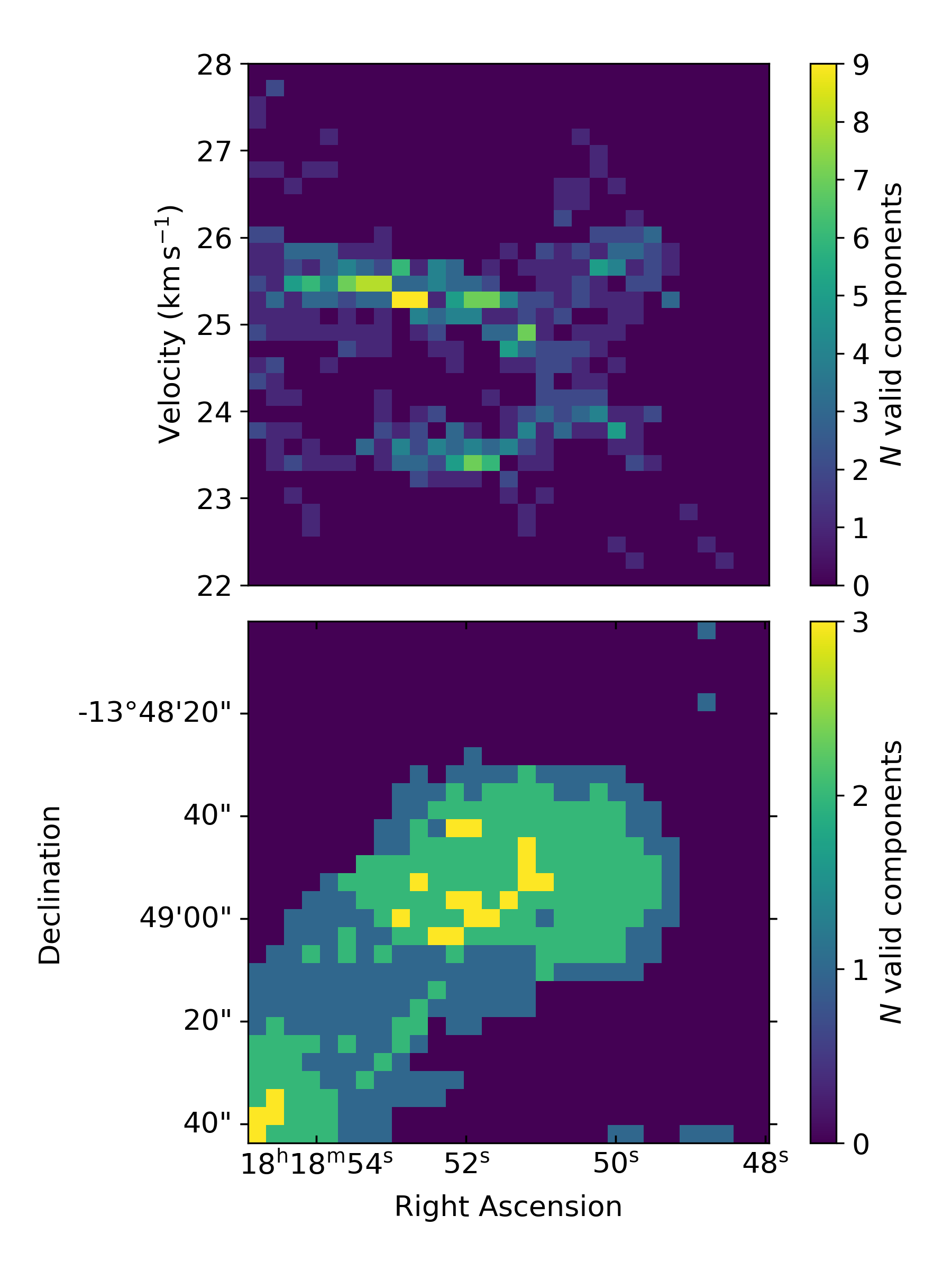}{0.3\textwidth}{\cii\ 3 component}}
    \caption{RA-Velocity projections of the 1, 2, and 3 component line center solutions. 4 component model results are not shown here but are not significantly different than the 3 component results.}
    \label{fig:unsupervised_solution_projections}
\end{figure*}

These projections, shown in Figure~\ref{fig:unsupervised_solution_projections}, all illustrate a spread of components towards the east which move kinematically closer together towards the Merge Point in the west, where they appear to merge together in the \cii\ series and the \hcopA\ 1-component projection.
In the \hcopA\ model series, three line center groupings are observed towards the eastern side of the image: one at $\vlsr \approx 25.5$~\kms, one at 24.5--25~\kms, and one at 23.5~\kms\ in the east with a strong gradient towards higher velocity to the west.
These velocities and their gradients are consistent with those of the two Threads and the Cap which we observed in the channel maps of most lines.
These groupings are observed in the 1, 2, and 3 component figures, and these groupings do not fundamentally change or disappear even when a 4th component is made available.
In the \cii\ model series, we only observe two groupings at $\vlsr \approx 23.5$~\kms\ and 25.5~\kms, and as with \hcopA\ this pattern persists even when more than two components are allowed.

The \hcopA\ results show a shift in behavior towards the western part of the head, near the Merge Point in Figure~\ref{fig:pillars_jwst}, between 1 and 2 allowed components.
In the 1-component models, we observe the kinematic merging of the high- and low-velocity component groupings into a single group around $\vlsr\approx 24.7~\kms$.
When we make a second component available to the \hcopA\ line models, we see the three eastern velocity groups converge towards the same velocity but then dramatically shift into a two-component grouping towards the Merge Point where the 1-component models converged.
This pattern persists even as 3 and 4 components are available to the \hcopA\ spectra.
This is peculiar for two reasons: first, it is at odds with the picture of several components merging together which we see in the \hcopA\ single component model and all the \cii\ models; and second, it relies on an abrupt shift in component velocity coupled with the abrupt disappearance of an entire component.
We suspect that, rather than a physical phenomenon, this is a sort of computational ``phase shift'' within the solution space of the model fit in which the result abruptly transitions from one solution to another between nearby pixels.
A more detailed investigation of the affected spectra is required in order to tell if the components really do merge and the \hcopA\ ``phase shift'' is really just a computational artifact, or if the situation too complex for us to make such a claim.
For a few locations towards the head, we perturbed the models' initial conditions and examined the fitted models to check whether our results are robust and whether we come to the same conclusions as described above.
Through this more supervised method, we find that all results from the unsupervised method are sound except those towards the northwestern corner of the head around the Merge Point, where the solution space phase shift occurs.
We find that the 2-component solutions towards that location are not unique and that a single component works comparably well.

We conclude that \hcopA\ line spectra towards the eastern and southern parts of the head can be decomposed into three distinct components in all the molecular lines, and the components are consistent from line to line.
The \cii\ spectrum towards these locations can be decomposed into two components which generally correspond to the highest and lowest velocity molecular gas components (the Eastern Thread and the Cap).
\hcopA\ line spectra towards the Merge Point can be decomposed into 1--3 components with similar reduced $\chi^2$ and so a single component is the least complex solution.
The three components merge into one towards the Merge Point, as we see in the single-allowed-component panel in Figure~\ref{fig:unsupervised_solution_projections}, and become too indistinct to fit separately.

\subsection{Geometry of P1a} \label{sec:appdx-kin-geometry}
We interpret these results to originate from a pillar head composed of three distinct molecular gas components, the two Threads and the Cap, embedded in a warm PDR gas envelope.
We observe the two Threads as spatially separate entities south of the pillar head and we can distinguish the components in the eastern and southern regions of the head in velocity, even though they are spatially blended.
Due to its strong velocity gradient, the Cap is the most kinematically separated from the rest of the head to the east.
At that location, we observe the widest line profiles which we are able to decompose into three components.
These spectra can also be fit with two components, where one represents the Cap and the other wider component represents a merged-Thread component, but they cannot be well fit with a single component.

These three morphologically distinct components spatially overlap towards the head and kinematically overlap towards the northwest part of the head, indicating that they are physically merged towards the northwest.
The gradients of the components are similar in absolute value but not in sign and are organized radially around the Merge Point.
Line profiles in every line towards the Merge Point are consistent with a single emitting component, while line profiles towards other locations of the head cannot be modeled by a single component.
We detect \ntwohpA\ and \ceighteenoA\ emission and high \thcoA\ column density towards the Merge Point, and all lines peak in brightness near that region.
The \ntwohpA\ detection implies sufficiently cold, shielded gas in addition to high column density, which is more likely towards a physical intersection of clouds rather than a projected stack of clouds.

Component interactions and overlays like this one may be responsible for the broad \cii\ line profiles observed towards other regions.
\cite{Ossenkopf2013A&A...550A..57O} describe ``macro-turbulent'' motions, which cause a greater portion of the \cii\ line to be optically thick and create a broader line profile than expected for opacity-broadening.
Our line profile analysis and high spatial resolution observations reveal the individual components towards P1a, but they are likely obscured by geometry or insufficient spatial resolution elsewhere.

The \cii\ emission towards P1a shares broad kinematic characteristics like the gradients along the Threads with the molecular line emission as we see in Figure~\ref{fig:unsupervised_solution_projections}, but some key differences noted in Section~\ref{sec:results-p1a} lead us to conclude that the PDR gas forms an extended envelope around the molecular gas.
We do not detect the threaded morphology just south of the pillar head in \cii\ or \oi\ (see Figure~\ref{fig:cii_cs_f335m}).
This is not solely due to the coarser spatial resolution of the fine structure line observations because smoothing the CO to the same resolution preserves the structure.
The \cii\ emission along the pillar body towards the Threads is spatially broad, $\sim$0.3~pc, and is centered on the Eastern Thread; from a purely spatial standpoint, the \cii\ emission appears to trace only the Eastern Thread.
We do not detect the same transverse velocity gradient across the Eastern Thread in \cii\ as we do in the molecular lines.
The PV diagrams in Figure~\ref{fig:pv_across} show that \cii\ shifts to lower velocities instead, such that the \cii\ line peaks at similar velocities as the molecular lines towards both the Eastern and Western Threads.
We interpret this as a kinematic detection of \cii\ emission associated with the Western Thread.
Modeled \cii\ central velocities show that, while the Eastern Thread has a bulk molecular gas velocity $\sim$1~\kms\ higher than the Western Thread, the \cii\ emission only shifts by $\sim$0.3~\kms\ between the same locations.
The \cii\ PV diagrams are spatially and kinematically broad and uniform, and the molecular line PV diagrams show more structure with more extreme velocities and narrower line widths.